\documentclass[symmetry,article,accept,pdftex,moreauthors]{Definitions/mdpi} 
\newcommand{\ethicalapproval}[1]{%
\vspace{6pt}\noindent{\fontsize{9}{11.2}\selectfont\textbf{Ethical Approval:} {#1}\par}}

\firstpage{1} 
\pubvolume{1}
\issuenum{1}
\articlenumber{0}
\pubyear{2022}
\copyrightyear{2022}
\externaleditor {\textls[-5]{{Academic Editors: Jan Awrejcewicz and Igor V. Andrianov}} 
}
\datereceived{28 January 2022 } 
\dateaccepted{21 March 2022} 
\datepublished{} 
\hreflink{\textls[-5]{https://doi.org/}} 

\Title{Mapping Long-Term Natural Orbits about Titania, a Satellite \linebreak of Uranus}

\TitleCitation{Mapping Long-Term Natural Orbits about Titania, a Satellite of Uranus}

\Author{Jadilene Xavier 
 $^{1,}$*\orcidA{}, Antônio {Bertachini} 
 Prado $^{2,3}$\orcidB{}, Silvia {Giuliatti} 
 Winter $^{1}$\orcidC{} and Andre Amarante $^{4}$\orcidE{}}

\AuthorNames{Jadilene Xavier, Antônio Bertachini Prado, Silvia Giuliatti Winter and Andre Amarante}

\AuthorCitation{Xavier, J.; Prado, A.B.; Winter, S.G.; Amarante, A.}

\address{%
$^{1}$ \quad Grupo de Din\^amica Orbital e Planetologia (GDOP), São Paulo State University (UNESP), \linebreak Guaratinguet\'a 12516-410, 
 SP, Brazil; giuliatti.winter@unesp.br\\
$^{2}$ \quad Academy of Engineering, Peoples’ Friendship University
 of Russia (RUDN University), 6 Miklukho-Maklaya Street, \linebreak 117198 Moscow, Russia; antonio.prado@inpe.br\\
$^{3}$ \quad \textls[-15]{Postgraduate Division, National Institute for Space Research (INPE), São José dos Campos 12227-010, SP, Brazil}\\
$^{4}$ \quad Federal Institute of Education, Science and Technology of São Paulo (IFSP), São José dos Campos 12227-010, 
 SP, Brazil; andre.amarante@unesp.br}

\corres{Correspondence: jadilene.rodrigues@unesp.br}

\abstract{Close polar and circular orbits are of great interest for the exploration of natural satellites. There are still no studies in the literature investigating orbits around Titania, the largest satellite of Uranus. In this work, we present results of a set of numerical simulations carried out to obtain long-duration orbits for a probe around Titania. Through an expansion of the gravitational potential up to second order, the asymmetry of the gravitational field due to Titania's coefficient $C_{22}$, the zonal coefficient $J_2$, and the gravitational perturbation of Uranus is considered. The analysis of lifetime sensitivity due to possible errors in the values of $J_2$ and $C_{22}$ is investigated using multiple regression models. Simulations were performed for different eccentricity values, and lifetime maps were constructed. The results show that low-altitude and near-circular orbits have longer lifetimes due to the balance between the disturbance of Uranus and the gravitational coefficients of Titania. The results also show that non-zero values of the longitude of periapsis ($\omega$) and longitude of the ascending node ($\Omega$) are essential to increase the lifetime up to eight times compared to cases where $\omega= \Omega=0^\circ$. We also show that an orbit with eccentricity $10^{-3}$ is the most affected by errors in the values of $J_2$ and $C_{22}$.}


\begin{document}




\section{Introduction}
\label{intro}

Uranus and its satellites are one of the least explored systems in our solar system. The only mission sent to visit Uranus took place in 1986 with Voyager 2. A mission to explore the Uranus system would be extremely important to understanding the formation and evolution of the solar system.

The main information about the Uranus system comes from observations made from Earth and data sent by space telescopes, but some important questions still remain unanswered. The distances of the ice giant planets are still an enigma, as some models suggest they should be closer to the Sun. Their high obliquity and low luminosity are intriguing. In addition to all these issues, information about Uranus' satellites would also be of great importance, as they have irregular surfaces and there is evidence of oceanic formation \cite{Jarmak2020,Cartwright2021}. According to \cite{Hofstadter2019}, NASA is working in partnership with ESA to plan possible missions to the ice giants during the period 2024--2037. These possible missions can help to understand the origin and evolution of this system.

Titania, Uranus' largest natural satellite, has a surface with flaws that could indicate past or still active seismic activity. In addition, there are indications that its surface has a mixture of water and ice with carbon dioxide \citep{Cartwright2015,Cartwright2018,Fletcher2020}.
An orbiter around Titania could investigate, for example, whether this satellite is or once was an ocean world, or even determine recent endogenous geological activity \cite{Cartwright2021}.

The best way to study this natural satellite is to send a probe around it. Therefore, it is necessary to look for the best orbits around Titania. In the literature, we still do not have references to the study of orbits around Titania that include perturbations due to its non-sphericity. Using maps that show the orbital duration, we present in this work a study to find the best conditions to obtain longer lifetimes of a probe around Titania. Here, Titania is at the center of the system and is perturbed by Uranus. Due to the asymmetry caused by Titania's ellipticity, the coefficient $C_{22}$ is analyzed, as well as $J_2$, which generates a vertical symmetry in the gravitational field around the satellite. 

Lifetime mapping is a technique used to find the best initial conditions for orbits as a function of their duration. The map provides the best values of the orbital elements for which orbits last the longest. With this, it is possible to extract significant values of the semi-major axis ($a$), eccentricity ($e$), inclination ($I$), argument of periapsis ($\omega$), longitude of ascending node ($\Omega$), and mean anomaly ($M$) to analyze the trajectory of a space vehicle. 

This technique is used in the work of \cite{Gomes2016} to study a hypothetical system, with masses and distances similar to the Earth-Moon system. In this work, the authors looked for the longevity of a probe around a natural satellite with high eccentricity. The study was done through a second-order expansion of the potential of the perturbing planet, assuming that it is in an elliptical orbit. The results indicated that the high variation of the eccentricity and inclination significantly reduce the useful life of the probe. The authors also showed that this high eccentricity variation is responsible for a 7--8\% decrease in the probe lifetime.

These maps were used in \cite{Cardoso2017} to analyze the best orbits for a probe around Callisto. In this study, the authors used the double-averaged method to investigate the importance of hypothetical eccentricity values for the perturber, Jupiter. Callisto's non-sphericity was considered and an analysis of the effect of its gravity coefficients was presented. Lifetime maps were used to find the best initial conditions for long-life orbits. In addition, the authors highlighted the importance of adopting specific values for $\omega $ and $\Omega $ in order to increase the useful life of the probe. The results obtained by them showed that for this system, the contributions due to Callisto gravitational coefficients $J_2$ and $C_{22} $ are small.

The double-averaging method was also used to study another Galilean satellite in \cite{Carvalho2012}. The dynamics of orbits around Europa were analyzed considering the perturbation of Jupiter in an elliptical orbit and the gravity coefficients $J_2$, $J_3$, and $C_{22}$ of Europa. The study used two processes to find long-duration orbits: the double-average and single-average methods. The authors concluded that even with not eliminating the coefficient $C_{22}$, the simple-average process was more realistic and can be used to find orbits that reach 80 days of life. Orbits up to 200 days of duration have also been found using this method, however, the eccentricity increases slightly in this case. Orbits around Europa were studied again in \cite{Carvalho2012_2}. In this work, the authors found a set of long-lived frozen orbits with initial conditions given by $a = 1850$~km, $e = 0.09$, $I = 40^\circ$, $\omega = 270^\circ $, and $\Omega = 90^\circ$.

Frozen orbits are of great interest for projecting future space missions. These orbits are characterised by keeping the variation of eccentricity, inclination, and argument of periapsis constant. For these orbits to be found, the argument of periapsis must be fixed at $90^\circ$ or $270^\circ$ and the short period elements eliminated. These elements are responsible for causing large variations in eccentricity and inclination due to the dependence on the term $C_{22}$.

In \cite{Tao2018}, frozen orbits around the Moon were investigated considering the Earth perturbation and the effects of the coefficients $J_2$ and $C_{22}$. The authors performed a closed-form transformation between osculating elements and averaged elements in order to find initial osculating elements for frozen orbits. This transformation was done only for circular orbits. The results showed that it was necessary to transform the average elements into osculating elements in order to add the $C_{22}$ coefficient, as this term caused large variations in high-altitude orbits and can totally change the behaviour of the orbits.

Orbits that keep their eccentricity, inclination, and argument of periapsis constant over a period resemble frozen orbits. In \cite{Lara2003} three families of periodic lunar orbits were found. The results showed that these orbits are very sensitive to the initial conditions and that these conditions have small differences with respect to the frozen orbits. Families of frozen orbits around the Moon were also found in the work of \cite{Abdelsalam2016}. The authors eliminated the short period terms and considered the oblateness of the Moon to order 4 ($J_4$). For eccentric frozen orbits, all families had a critical inclination around $64^\circ$. In the case of very low quasi-circular lunar orbits, several families with different critical inclinations were obtained, although they quickly disappeared due to small changes in the semi-major axis and eccentricity.

Despite causing large variations in some orbital elements, such as inclination, eccentricity, and argument of periapsis, the term $C_{22}$ has an important role in the dynamics of orbits around natural satellites. The combination of this term with the zonal term $J_2$ is able to attenuate the effects caused by the perturbation due to the third body acting as a ``protection mechanism'' to prolong/increase the lifetime of the desired orbit \cite{Carvalho2018,Saedeleer2006}. In \citep{Tzirt2009} the combination of this coefficient with the $J_2$ term was also investigated in orbits around the Moon. The results showed that the values of $J_2$ and $C_{ 22}$ strongly affected the near-critical inclination and the libration amplitude of the argument of periapsis. This work also showed that these effects were greater when added to the contribution of the rotation rate.

In several works found in the literature, the perturbation due to the third body was considered when it comes to orbits around natural satellites. The effects of this perturbation were mainly observed in the eccentricity of the orbit around the observed satellite. These high eccentricity values achieved by the probe's orbit are related to the effect of the Kozai--Lidov mechanism. This mechanism was studied for the first time in \cite{Kozai1962,Lidov1962}. This phenomenon causes the argument of periapsis to oscillate at a constant value. Whereas the argument of periapsis oscillates around a specific value, the eccentricity and inclination suffer periodic oscillations (e.g., see \cite{Naoz2016,Naoz_2017} and the references therein).

The structure of this work is organised as follows: in Section \ref{sec:model} the mathematical model adopted to analyze the system is described, as well as the methodology. In Section \ref{sec:result} we discuss the results obtained. In Section \ref{sec:element} we show the isolated effect of each perturbation through the analysis of the orbital elements of each specific case. In this same section, we also present the mechanism for obtaining long-time orbits through the evolution of the orbital elements. In Section \ref{sec:grav} we present an estimate of the probe's lifetime when affected by possible errors in the values of Titania's gravitational coefficients. Finally, in Section \ref{sec:final}, we present our conclusions.

\section{Mathematical Model}\label{sec:model}

The system addressed in this work consists of a central body (Titania), a space probe, and a perturber (Uranus). We assume that the perturber body is in a Keplerian orbit around Titania with a radius of $25.362 \times 10^{3}$ km and mass $8.68 \times 10^{25}$ kg, (\url{https://ssd.jpl.nasa.gov/}), accessed on April 05, 2021. The orbital parameters are given in Table  \ref{tab:1}. To make an accurate investigation of the problem, we consider the main gravity coefficients ($J_2$ and $C_{22}$) of Titania, regarding the oblateness and ellipticity of the body. These terms are the most important in the gravitational field, after the Keplerian term. Previous work \cite{Tao2018,Tzirt2009,Tzirt2010,Scheeres2006} shows that these coefficients are the most relevant to the pertubative effects caused on a probe by a satellite and, in most cases, the most significant to model the irregular shape of a body. It is important to note that these two coefficients are the only coefficients available in the literature. As our focus is to analyze orbits very close to the surface of this satellite, these will be the main terms to be considered in this work. Titania has a mass of $ 35.27 \times 10^{20} $ kg and a radius of $ 788.9 $ km, (\url{https://ssd.jpl.nasa.gov/}), accessed on April 05, 2021. The other parameters of Titania are shown in Table \ref{tab:2}. The equations of motion are described according to \cite{Scheeres2006}:

\begin{equation}
\ddot{\Vec{r}}= - \dfrac{G(M_T+m)\Vec{r}}{\Vec{r}^3} + GM_U \left(\dfrac{\Vec{r}_U-\Vec{r}}{|\Vec{r}_U-\Vec{r}|^3}-\dfrac{\Vec{r}_U}{\Vec{r}_U^3}\right) + \Vec{P}_T + \Vec{P}_U 
\end{equation}

\begin{equation}
\begin{split}  
P_{Tx} = & - \dfrac{G(M_T+m) J_2x}{2r^5}\left[3-15\left(\dfrac{z}{r}\right)^2\right]\\
& + \dfrac{3 G(M_T+m) C_{22}x}{r^5} \left[2- \dfrac{5(x^2-y^2)}{r^2}\right] 
\end{split}
\end{equation}

\begin{equation}
\begin{split}
P_{Ty} =  & - \dfrac{G(M_T+m) J_2y}{2r^5}\left[3-15\left(\dfrac{z}{r}\right)^2\right] \\ 
& - \dfrac{3 G(M_T+m) C_{22}y}{r^5} \left[2 + \dfrac{5(x^2-y^2)}{r^2}\right] 
\end{split}
\end{equation}

\begin{equation}
\begin{split}
P_{Tz} =  & - \dfrac{G(M_T+m) J_2z}{2r^5}\left[9-15\left(\dfrac{z}{r}\right)^2\right] \\
& - \dfrac{15 G(M_T+m) C_{22}z}{r^7}(x^2-y^2) 
\end{split}
\end{equation}
\begin{equation}
\begin{split}
P_{Ux} =  &  \dfrac{GM_U (x_U-x)}{|\Vec{r}_U-\Vec{r}|^3} \bigg[\dfrac{ J_2}{(r_U-r)^2}\left(\dfrac{15}{2}\dfrac{(z_U-z)^2}{(r_U-r)^2}-\dfrac{3}{2}\right) \\
& +\dfrac{J_4}{(r_U-r)^4}\left(\dfrac{315}{8}\dfrac{(z_U-z)^4}{(r_U-r)^4}-\dfrac{105}{4}\dfrac{(z_U-z)^2}{(r_U-r)^2}\right)\bigg] 
\end{split}
\end{equation}

\begin{equation}
\begin{split}
P_{Uy} =  &  \dfrac{GM_U (y_U-y)}{|\Vec{r}_U-\Vec{r}|^3} \bigg[\dfrac{ J_2}{(r_U-r)^2}\left(\dfrac{15}{2}\dfrac{(z_U-z)^2}{(r_U-r)^2}-\dfrac{3}{2}\right)\\
& +\dfrac{J_4}{(r_U-r)^4}\left(\dfrac{315}{8}\dfrac{(z_U-z)^4}{(r_U-r)^4}-\dfrac{105}{4}\dfrac{(z_U-z)^2}{(r_U-r)^2}\right)\bigg] 
\end{split}
\end{equation}

\begin{equation}
\begin{split}
P_{Uz} =  &  \dfrac{GM_U (z_U-z)}{|\Vec{r}_U-\Vec{r}|^3} \bigg[\dfrac{ J_2}{(r_U-r)^2}\left(\dfrac{15}{2}\dfrac{(z_U-z)^2}{(r_U-r)^2}-\dfrac{9}{2}\right)\\
&+\dfrac{J_4}{(r_U-r)^4}\left(\dfrac{315}{8}\dfrac{(z_U-z)^4}{(r_U-r)^4}-\dfrac{175}{4}\dfrac{(z_U-z)^2}{(r_U-r)^2}+\dfrac{75}{8}\right)\bigg], \\
\end{split}
\end{equation}

\noindent where $\vec{P}_T$ and $\vec{P}_U$ are the expansions of the gravitational potential up to second order for Titania and up to fourth order for Uranus, respectively; $P_{Tx}$, $P_{Ty}$, and $P_{Tz}$ are components of the $P_T$ vector and $P_{Ux}$, $P_{Uy}$, and $P_{Uz}$ the components of the $P_U$ vector; $m$ is the mass of the space probe, $M_T$ the mass of Titania, and $M_U$ the mass of Uranus. The terms $\vec{r}$ and $\vec{r}_U$ are the radius vector of the probe and Uranus, respectively. 

The investigation of orbits around natural satellites with a non-uniform mass distribution has been the subject of several studies in recent years. Studies \cite{Cardoso2017, Scheeres2006, Lara2006, Carvalho2010, Carvalho2012, Carvalho2018} emphasise the importance of considering the oblateness and ellipticity of the body to be orbited by the probe. However, it is understood that analyzing the influence of higher oblateness coefficients of the disturbing body can provide a much more accurate study of the orbits.

Uranus is very far from Earth, approximately 3 billion km, and the time to send signals from Earth is on the order of 2.8 h; therefore it takes approximately 5.6 h to send a command and receive a signal back. Because of this time lag, orbital maneuvering is a logistical problem. An important reason to search for long-term natural orbits in which to place a space probe is because they do not require frequent orbital maneuvers, thereby saving fuel and simplifying the orbit control. For the sake of comparison, we did a single test considering the highest order gravity coefficients $ J_2 $ and $ J_4 $ of the disturbing body. This implementation was made through an adaptation of the Mercury package \citep{Chambers1999}, which was used in our numerical simulations.

\begin{table}[H]
  \caption{Parameters of Titania with respect to Uranus.}
 \label{tab:1}
\newcolumntype{C}{>{\centering\arraybackslash}X}
\begin{tabularx}{\textwidth}{CC}
\toprule
   {\bf Parameter}         & {\bf Value} \\
  \midrule
  Semi-major axis (km)    &  $435.8 \times 10^{3}$     \\
  Eccentricity            &  $1.18 \times 10^{-3}$   \\  
  Inclination     ($^\circ$)   &  $10^{-1}$                    \\
  Argument of periapsis ($^\circ$)&  $1.64 \times 10^{2}$  \\
  Longitude of ascending node ($^\circ$) & $1.67 \times 10^{2}$    \\
  Mean anomaly  ($^\circ$)     & $2.05 \times 10^{2}$       \\
 \bottomrule
\end{tabularx}
\noindent{\footnotesize{JPL. Website: \url{https://ssd.jpl.nasa.gov/}.}} Accessed on April 05, 2021.
 \end{table}
 
 \vspace{-6pt}
 
  \begin{table}[H]
  \caption{Gravity coefficients of Titania and Uranus.}
 \label{tab:2}
\newcolumntype{C}{>{\centering\arraybackslash}X}
\begin{tabularx}{\textwidth}{CCCC}
\toprule
 {\bf Body} &  \boldmath{$J_2$} & \boldmath{$J_4$} & \boldmath{$C_{22}$}  \\
\midrule
Uranus $^{1}$  &  $3.34343 \times 10^{-3}$ & $ -2.885 \times 10^{-5}$ & -\\

Titania $^{2}$ & $1.13 \times 10^{-4}$  & - & $3.38 \times 10^{-5}$   \\
  \bottomrule
\end{tabularx}
\vspace{-12pt}
\noindent{\footnotesize{\textsuperscript{1} 
 \cite{French2012}; \textsuperscript{2} \cite{Chen2014}.}}
\end{table}

Our main goal is to study circular orbits with high inclinations. Circular orbits are better for observations as eccentric orbits have a lower perigee and are more susceptible to collisions. Highly inclined orbits take advantage of the central body's rotation to observe the entire natural satellite. However, we understand that analyzing other values of eccentricities can contribute to a more complete approach to the problem, giving more choices for the mission designers. Therefore, we considered the following eccentricity values for the probe's orbit: $e=0$, $e=10^{-1}$, $e=10^{-2}$, $e=10^{-3}$, and  $e=10^{-4}$. In order to analyze those orbits very close to the surface of Titania, we adopted a semi-major axis interval ($a$) ranging from 810 km to 2500 km for each value of the eccentricity. The inclination is in the range between $75^\circ$ and  $90^\circ$. 

We considered two cases regarding the argument of periapsis and the longitude of the ascending node ($\omega$ and $\Omega$, respectively). First, we consider $\omega =\Omega = 0^\circ$ and, after that, we choose specific values in regions with longer lifetimes, to explore best results. 

All cases were simulated for 1000 days, equivalent to approximately $10^4$ orbital periods for the closest orbit to Titania's surface and $10^3$ revolutions for the farthest orbit. For our purposes, this time is adequate, sufficient to analyze the orbital evolution of a space probe and collect and send data for analysis. The lifetime of the orbit is defined by the time that the space probe remains in orbit around Titania without colliding with the central body. Therefore, the maps of $a_i$ versus $I_i$ show the lifetime of the probes numerically simulated up to 1000 days. Our main goal is looking for longer lived orbits around Titania under the effects of its gravitational coefficients and Uranus.

In order to analyze the influence of the gravity coefficients of Titania and the third body, we constructed life maps considering two cases. Point mass models are considered without taking into account the gravity coefficients of both bodies,with gravity coefficients for both bodies, Titania and Uranus, and including $J_2$ and $C_{22}$ of Titania. ``Difference maps'' (third body---$J_2$, $C_{22}$ from Titania) are also presented. The case in which we analyzed the $J_2$ and $J_4$ coefficients showed insignificant changes in lifetimes compared to the case where these coefficients are not included. Therefore, we will adopt Uranus as a point of mass.  
Our purpose with these maps is to find regions where the orbits are affected by the gravity coefficients of Titania and the third body. These effects are in opposite directions, and the orbits can last longer without collision or escape. The time evolution of the orbital elements $a$, $e$, $I$, $\omega$, and $\Omega$ of the orbits with a long lifetime are shown Section \ref{sec:element}. 

\section{Analyzing the Results}
\label{sec:result}

The  results show diagrams $a \times I$ considering: (i) only the contribution of the disturbing body (Uranus), modelled as a point of mass; (ii) the gravity coefficients of Titania ($ J_2 $ and $ C_{22} $), and (iii) the more complete model, where both effects are considered. In the figures, the color bar represents the lifetimes of the orbits. We constructed ``difference maps'' by subtracting the lifetime of the orbits obtained when Uranus is considered. The lifetime of the orbits when considering both effects, the gravitational effects of Uranus, and the 
gravity coefficients of Titania, are presented in Section \ref{sec:result}.

It should be noted that, in these figures showing the differences in the lifetimes, a positive sign indicates those values where only the presence of  Uranus prolonged the probe's life, whereas negative values indicate that the gravity coefficients of Titania helps to naturally control the third-body perturbations Uranus, thus increasing the lifetime of the orbits. 

\begin{figure}[H]
  \fbox{\includegraphics[width=9.2cm]{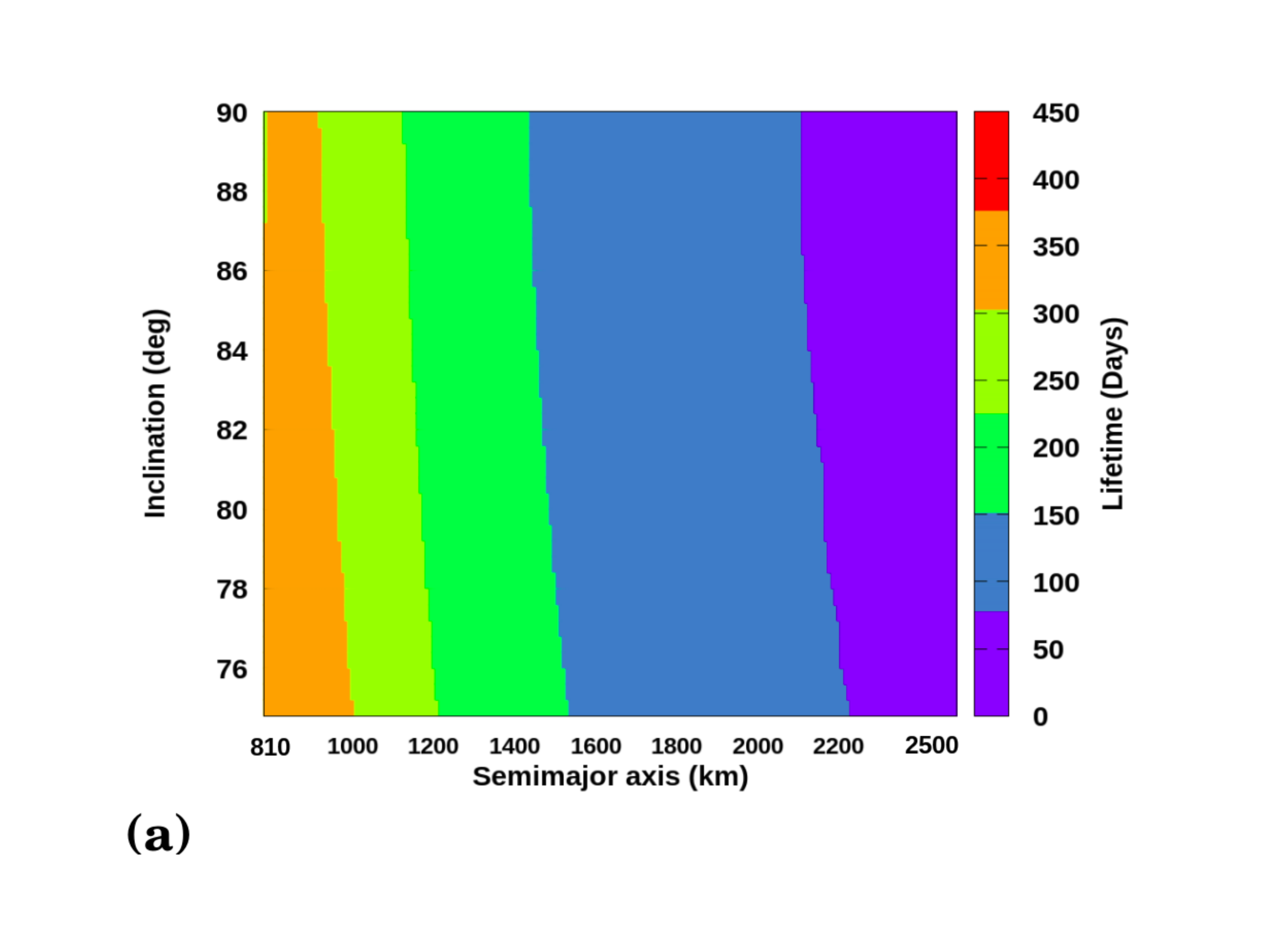}} \\ \fbox{\includegraphics[width=9.2cm]{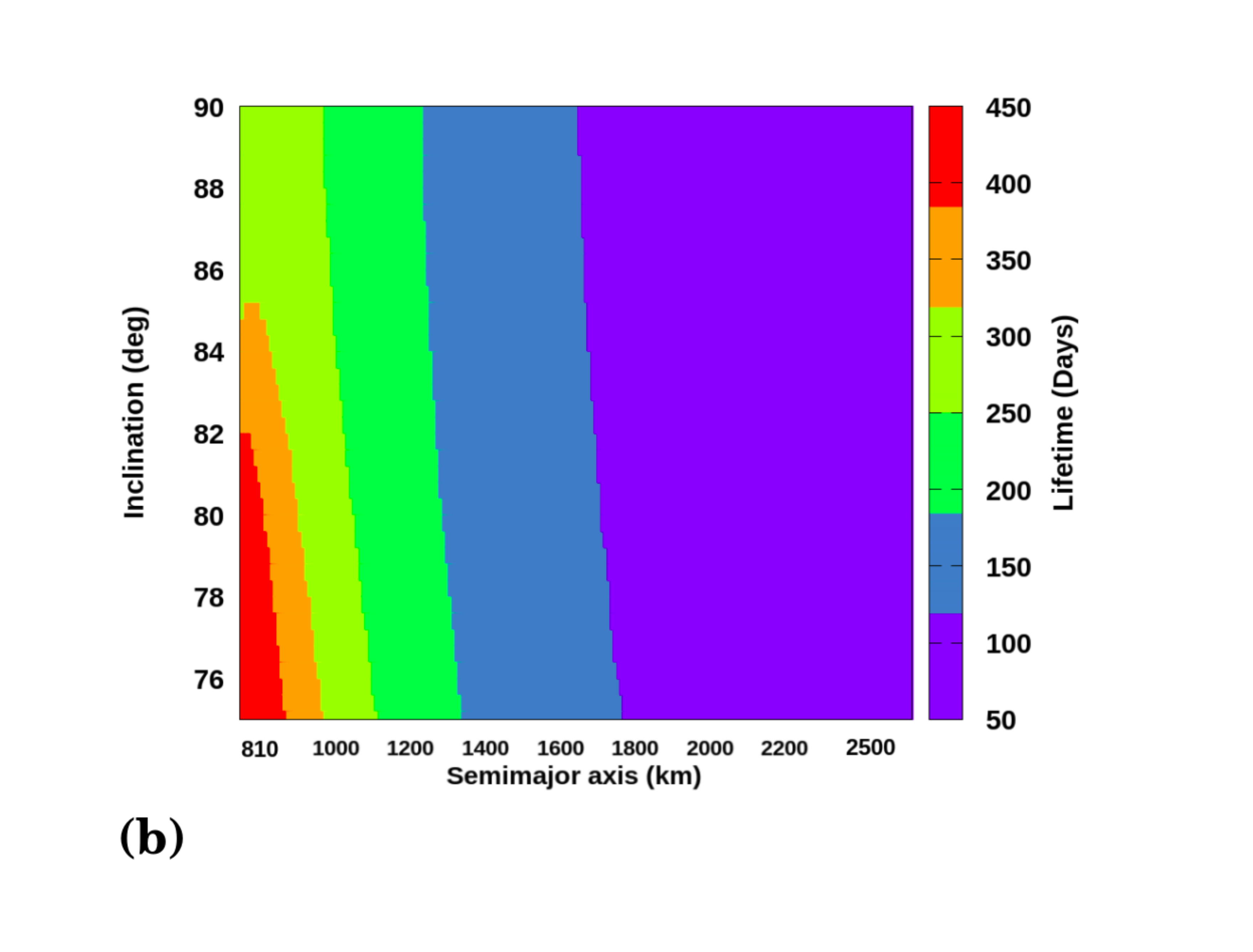}} \\
  \fbox{\includegraphics[width=9.2cm]{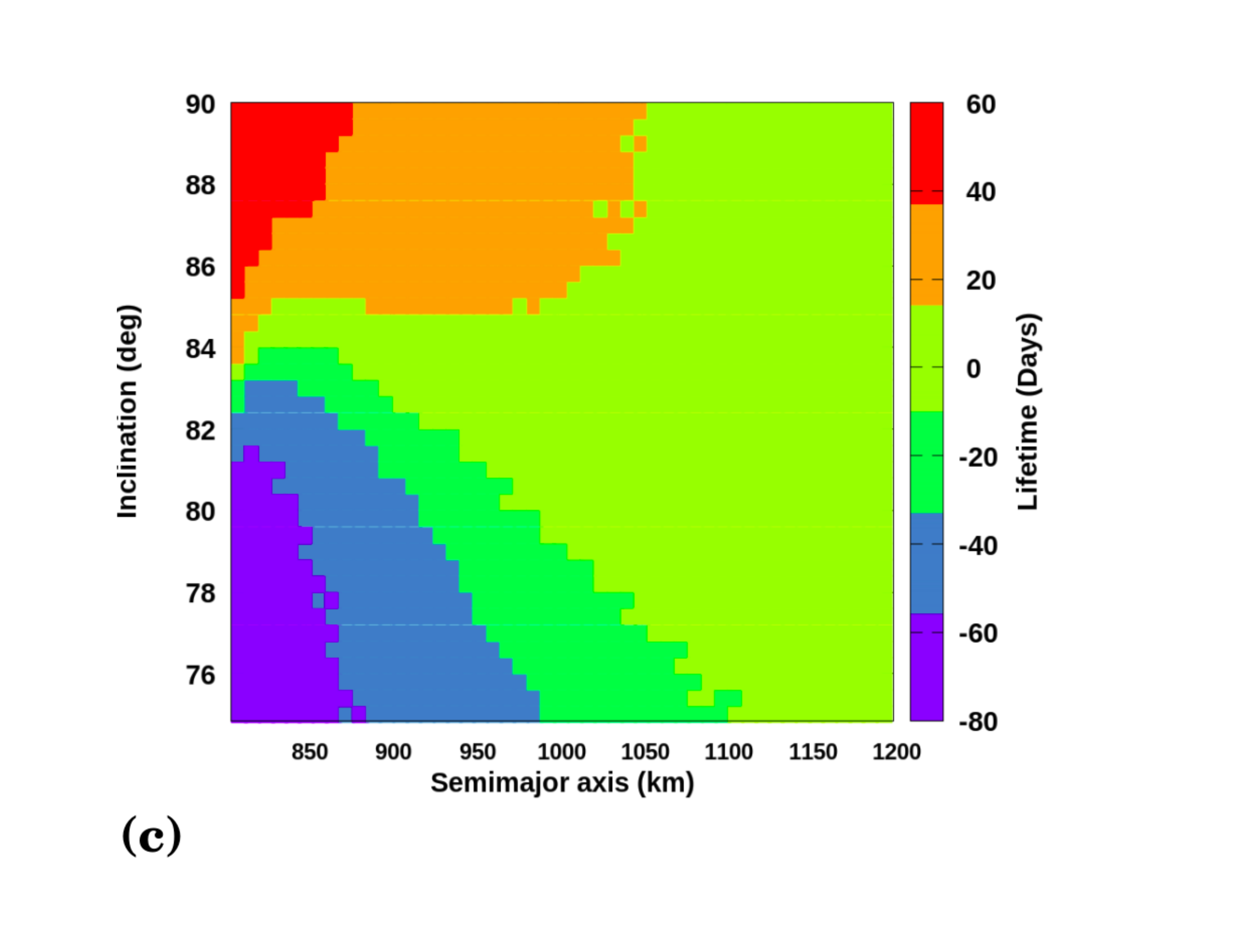}}
\caption{Diagram of $a$ versus $I$ for $e=0$: (\textbf{a}) considering only the effects from Uranus; (\textbf{b}) including
the third-body and the effects of $J_2$ and $C_{22}$ of Titania; (\textbf{c}) lifetime differences (third body $- J_2$ and $C_{22}$ of Titania). Initial values are \emph{a} = 810--1200~km, \emph{I} = 7--90\textdegree, $\omega=0^\circ$, and $\Omega=0^\circ$.}
\label{fig:1}
\end{figure}

Figure \ref{fig:1} presents a map where we considered only the third body (a) and the third-body and the gravity coefficients of Titania (b). For the case where only the gravity coefficients of Titania are considered, Figure \ref{fig:1}b shows an increase of 50 days in the life of the probe for orbits closer to the surface of Titania. Orbits with lifetimes up to 450 days are observed in the same altitudes and inclinations. Lifetimes in the range 100--300 days for all values of inclinations occur for the semi-major axis in the range 1000--2200 km (Figure \ref{fig:1}a), and $a$ = 1000--1800 km (Figure \ref{fig:1}b) (shown in light green, green, and blue squares). 

We show lifetime ``difference maps'' in Figure \ref{fig:1}c. The maximum value of the semi-major axis is $a = 1200$ km; after that, there are no significant changes in the orbits' lifetime because the space probe is too far from Titania. Its irregular shape is unable to balance the third-body perturbations from Uranus. Figure \ref{fig:1}c shows that the largest differences in lifetime appear for values of  $a$ = 810--1000~km and \emph{I} = 75--83\textdegree~(blue region), where the equilibrium between the gravity coefficients of Titania and Uranus occurs. We found increases in the lifetime on the order of 60 days, which is not negligible. In addition, those orbits are closer to Titania, which helps in the observation of this body. It is noted that the most negative numbers occur for lower values of inclinations and semi-major axes. In these regions, the third body perturbations are smaller, and the gravity coefficients dominate the dynamics and reduce the duration of the orbits.

For $ e = 10^{-4} $, the results present similar behaviour compared to $e = 0$. The simulations were performed using the same initial conditions used in Figure \ref{fig:1}, except for the eccentricity, which now equals $ e=10^{-4} $.

We found an equal maximum lifetime when the third body and the $J_2$ and $C_{22}$ coefficients of Titania are considered. However, when considering only Uranus as a perturber, the region with the longest orbit life ranges in the interval \emph{a} = 810--900~km and \emph{I} = 75--80\textdegree. In this region, the longest lifetime is approximately 450 days. When including the gravity coefficients of Titania, this region increases to \emph{a} = 810--1000~km and  \emph{I} = 75--84\textdegree. Another fact is that the number of orbits with a lifetime between 300--350 days decreases when the gravity coefficients of Titania are considered.

The results showed that the orbits in the range $a$ = 810--1050~km and $I$ = 75--84\textdegree~have increased by 60 and 20~days. Orbits with $a$ = 810--900~km and {\em I} = 85--90\textdegree~have their lifetime extended by approximately 40--80 days, thanks to the effect caused by the equilibrium of the perturbations.

\begin{figure}[H]
\fbox{\includegraphics[width=9.2cm]{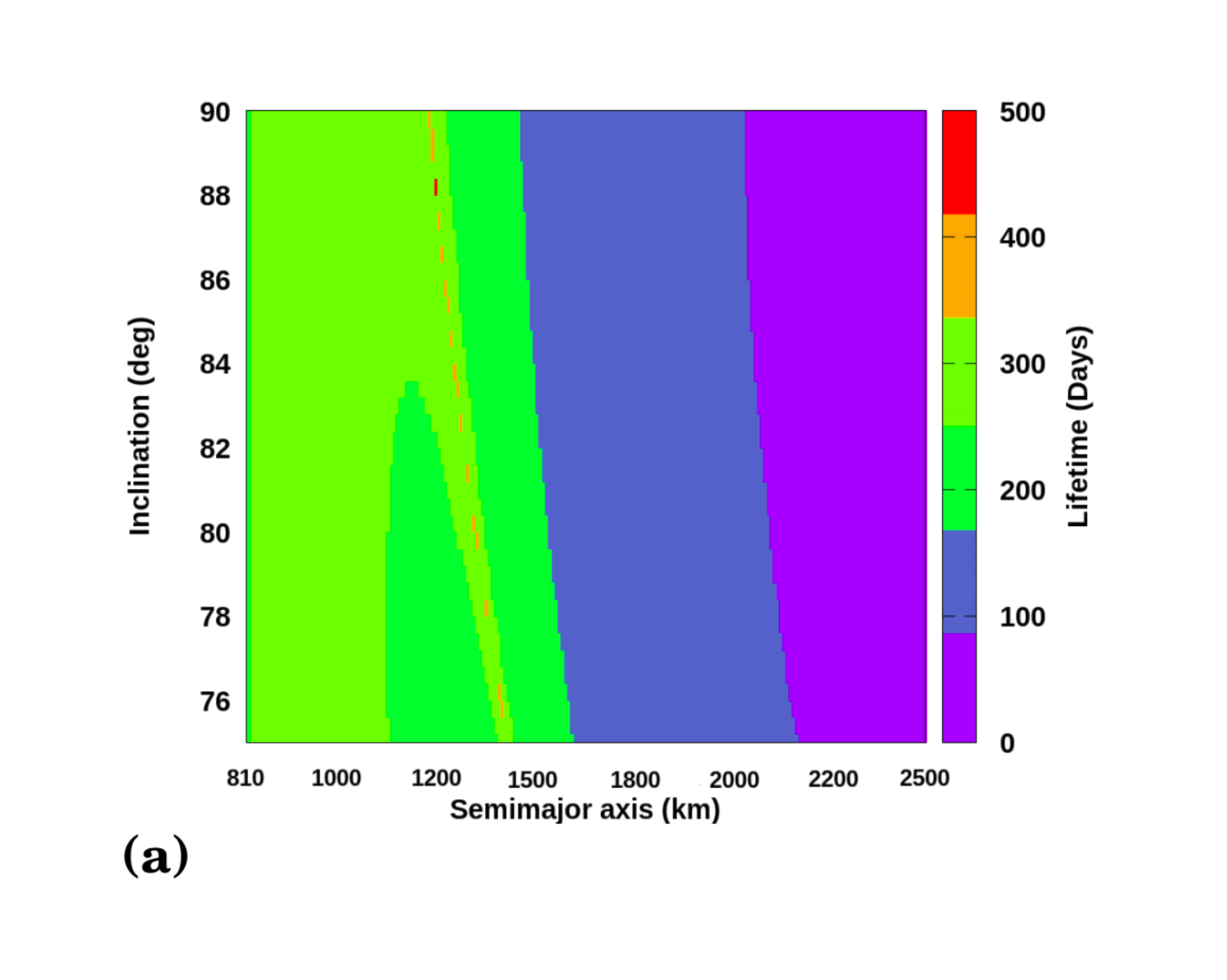}}\\
\fbox{\includegraphics[width=9.2cm]{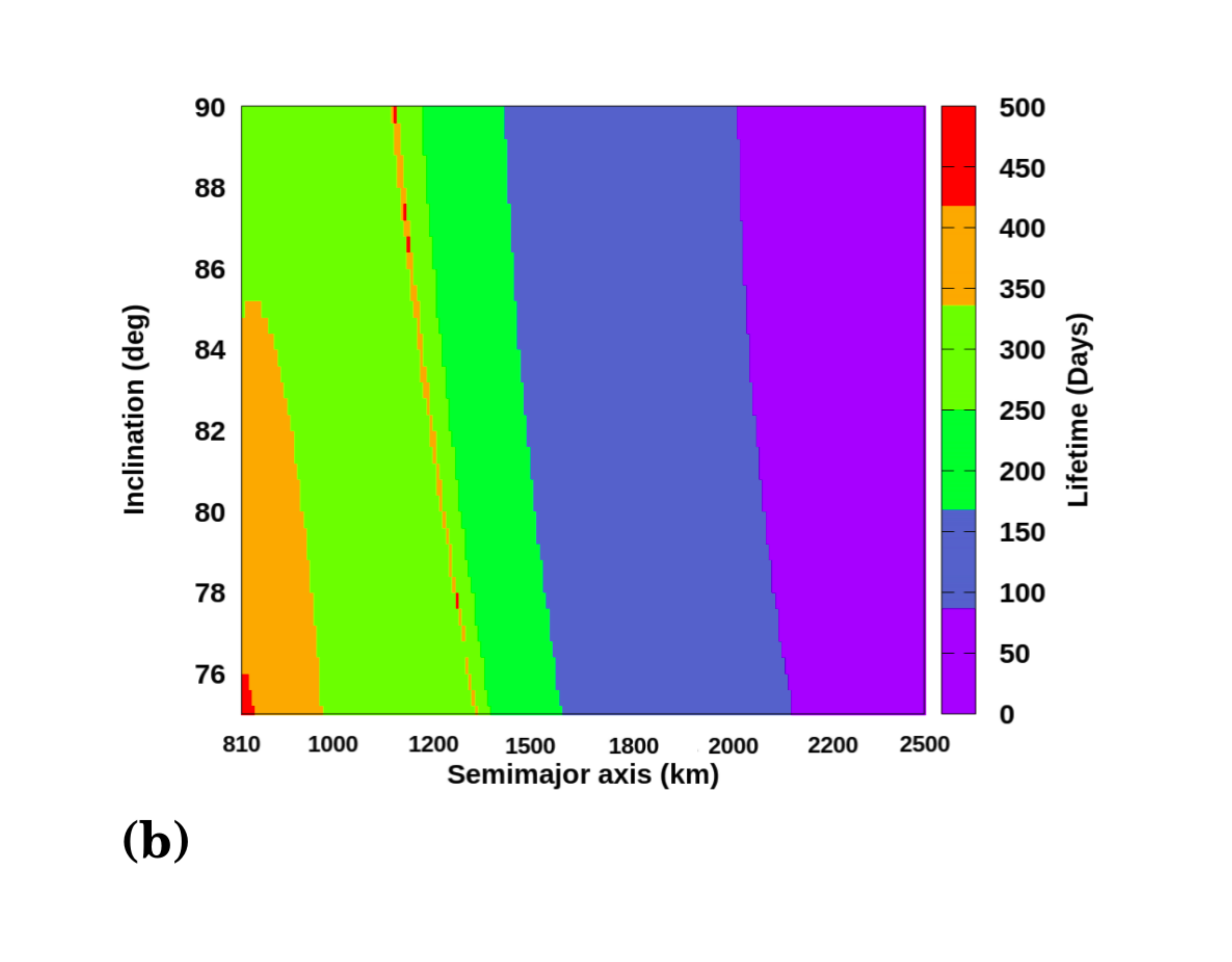}}\\
\fbox{\includegraphics[width=9.2cm]{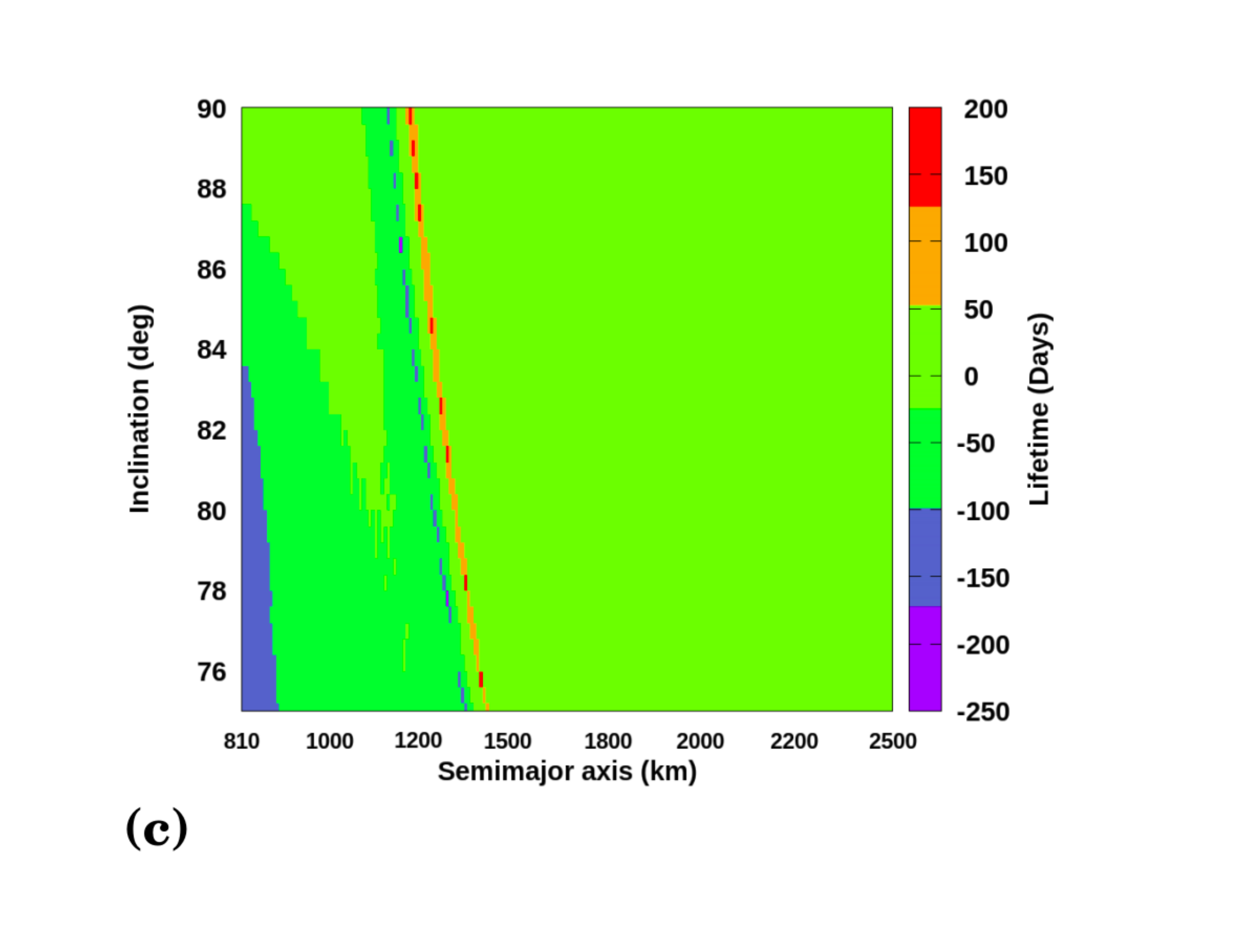}}
 \caption{Diagram of $a$ versus $I$ for $e=10^{-3}$: (\textbf{a}) considering only the effects from Uranus; (\textbf{b}) including
the third-body and the effects of $J_2$ and $C_{22}$ of Titania; (\textbf{c}) lifetime differences (third body $- J_2$ and $C_{22}$ of Titania). Initial values are \emph{a} = 810--1200~km, \emph{I} = 75--90\textdegree, $\omega=0^\circ$, and $\Omega=0^\circ$.}
\label{fig:2}
\end{figure}

For $ e = 10^{-3} $, shown in Figure \ref{fig:2}, our results, given in the diagram $a \times I$ when only the gravitational effect of Uranus is considered, are similar to those observed for $e = 0$ and $e = 10^{-4}$. As already noted, the terms $ J_2$ and $C_{22} $ of Titania are more important to increasing the lifetime of the probe, confirming that the effects caused by the gravity coefficients of Titania compensate for the effects caused by Uranus' gravitational attraction, as expected.

\vspace{-6pt}

The lifetime differences maps presented in Figure \ref{fig:2}c have two nearby regions where the changes in the lifetimes are expressive (regions in purple, red, and orange squares). In the region with purple squares, the lifetime for those orbits only considering the third body is up to 250 days. At the red dots, the probe's lifetime can reach up to 200 days when $ J_2, C_ {22} $ coefficients of Titania are considered.

Figure \ref{fig:2}b, including the gravity coefficients of Titania, presents a more extensive region with lifetimes of 350--500 days. This region is located very close to the surface of Titania (\emph{a} = 810--1000~km), where the terms $J_2$ and $C_{22}$ cancel the effects of the third body. Figure \ref{fig:2}a shows the numerical simulations where the gravitational effects of Uranus are considered. Orbits with a lifetime close to these values appear only on a small island, for all inclinations, and a semi-major axis closer to 1400~km.

In the regions where the probe remains around Titania for longer times, there is a balance between the perturbation caused by the third body and the perturbation due to the gravitational coefficients of Titania, causing the probe's lifetime to be extended. This balance is described in previous work \citep{Carvalho2012, Carvalho2018}, where it is shown that, when combined with $ J_2 $ and $ C_ {22} $ of the central body, this resultant perturbation acts as a ``protection mechanism'' responsible for softening the effects caused by the third body perturbation on the variation of the eccentricity.

\begin{figure}[H]
\begin{adjustwidth}{-\extralength}{0cm}
\fbox{\includegraphics[width=8.42cm]{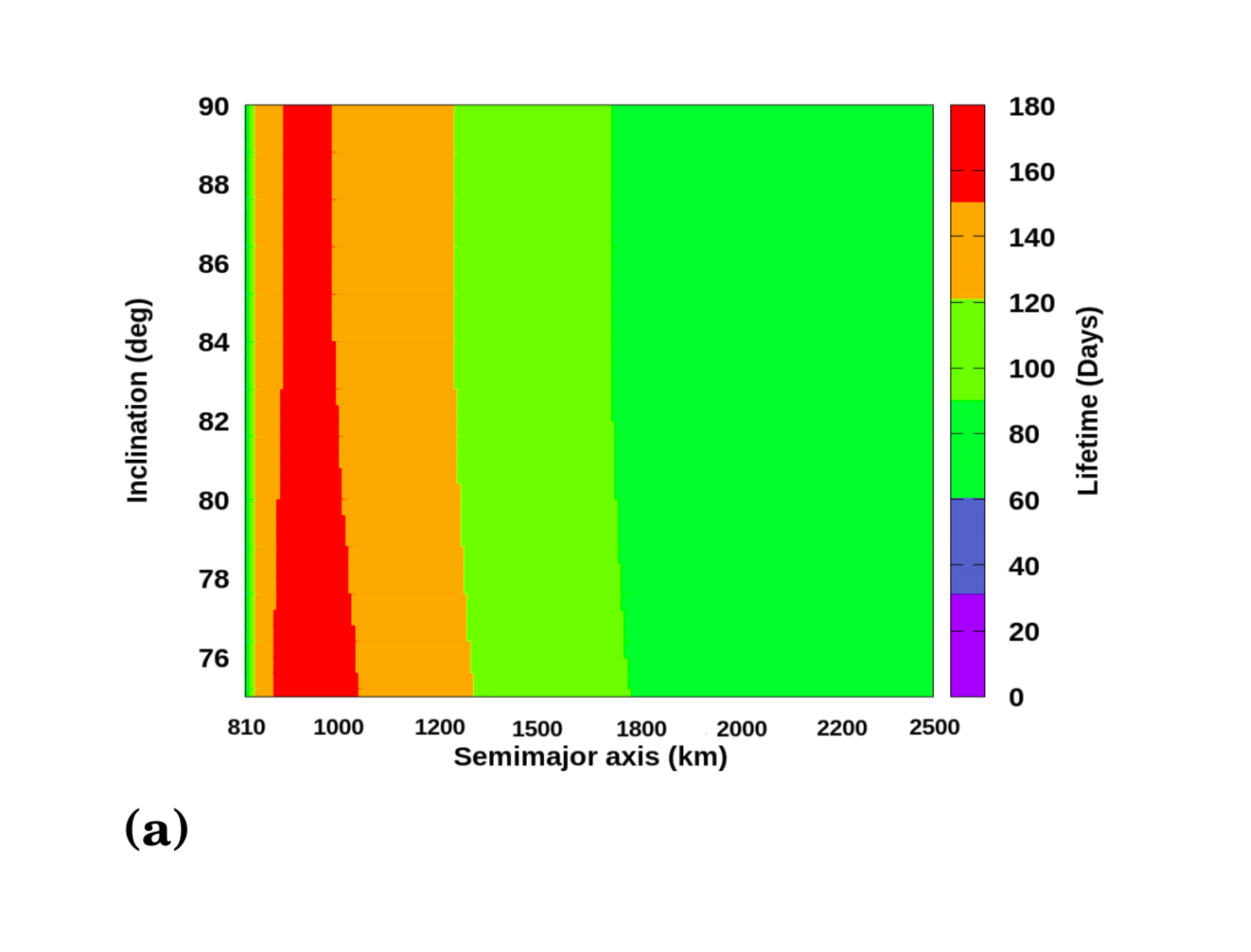}}
\fbox{\includegraphics[width=8.78cm]{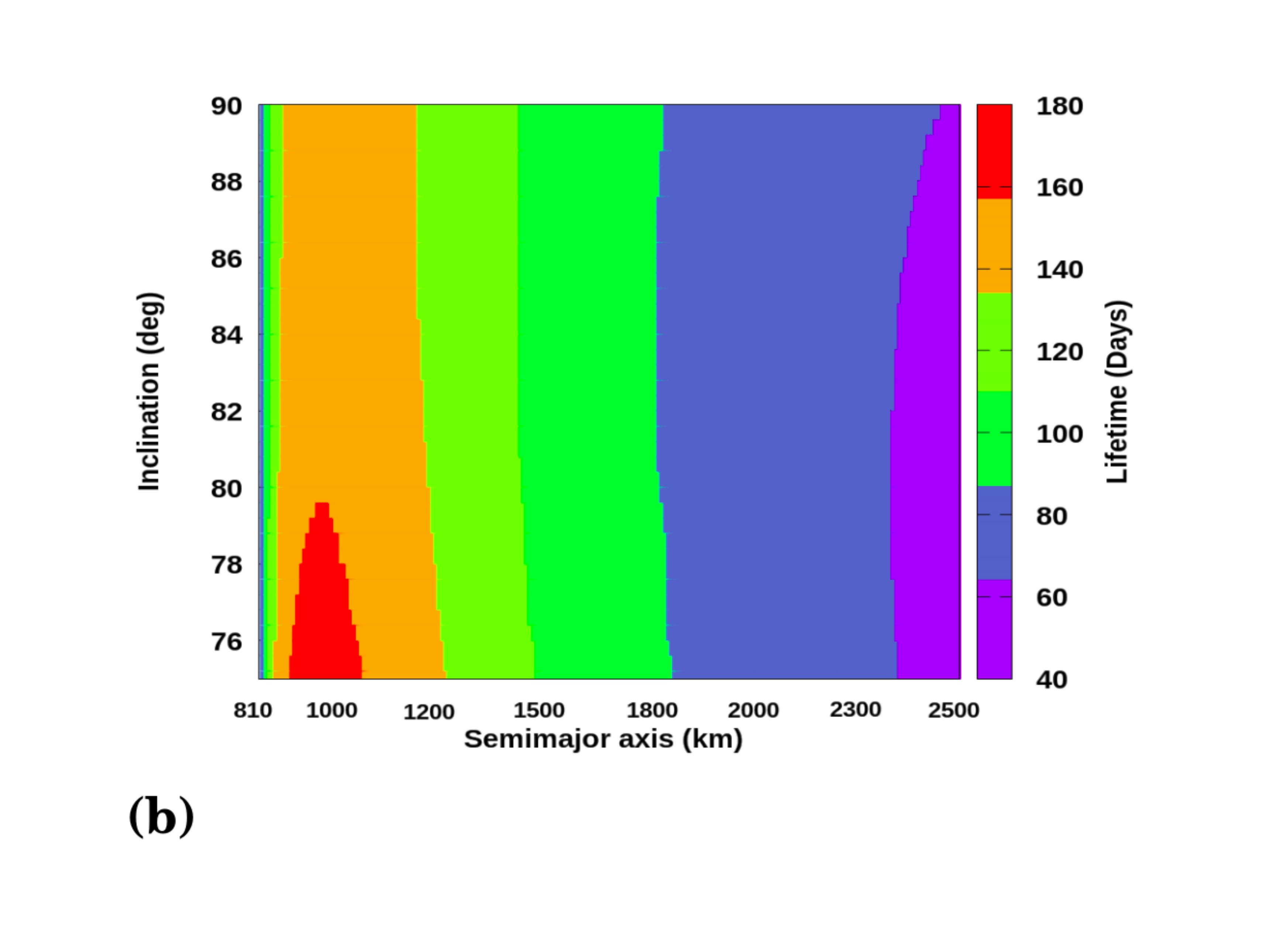}}
\end{adjustwidth}
 \caption{Diagram of $a$ versus $I$ for $e=10^{-2}$: (\textbf{a}) considering only the effects from Uranus; (\textbf{b}) including the third-body and the effects of $J_2$ and $C_{22}$ of Titania. Initial values are \emph{a} = 810--2500~km, \emph{I} = 75--90\textdegree, $\omega=0^\circ$, and $\Omega=0^\circ$.}
\label{fig:3}
\end{figure}

The results presented in Figure \ref{fig:3} are obtained for $e = 10^{-2}$. Note that there are no large differences between the results obtained when considering only the third body perturbation (a) and those when the $J_2$ and $C_{22}$ from Titania are added in the system (b). There is only a decrease in the region with lifetimes between 160 and 180 days, covering all the inclinations, when $J_2$ and $C_{22}$ of Titania are included. This region is located in the range \emph{a} = 900--1050~km and \emph{I} = 75--80\textdegree. 

For $ e = 10^{-1} $, the orbital duration maps show that the results are very close when all the perturbations are considered, and the behaviour is similar to the one obtained for $ e = 10^{-2} $. The difference is that for $ e = 10^{-1} $, the probe's maximum lifetime is much shorter, 60~days, but the regions of greater and lesser duration are very similar. Previous studies \citep{Prado2003, Gomes2016} found similar results, where more eccentric orbits have shorter lifetimes, as expected, as these orbits have smaller periapsis and, therefore, collide with the central body in less time. In addition, they have larger apoapsis, which increases the effect of the third body.

Previous work \cite{Cardoso2017,Cinelli2019,Shraddha2011} shows the relevance of the contribution of the argument of periapsis ($\omega$) and longitude of the ascending node ($\Omega$) in the duration of highly inclined orbits. In \cite{Cardoso2017}, the initial conditions capable of increasing the lifetime of the probe are $\omega=180^\circ$ and $\Omega=90^\circ$. In \cite{Cinelli2019}, the best values found for these angles are $\omega$ = 140.5--148\textdegree, $\omega$ = 321--327\textdegree,~and $\Omega = 165 ^\circ$. Given the above, we analyzed the regions with longer lifetimes, shown in Figures \ref{fig:1}--\ref{fig:3}, in a range of values of the periapsis argument and the node longitude. Within these regions, we chose the best values of $a$ and $I$ and built maps as a function of $\omega$ and $\Omega$ with the four values of eccentricities adopted earlier. For this analysis, we consider the perturbation due to the third body and also the gravity coefficients of Titania.

\begin{figure}[H]
\begin{adjustwidth}{-\extralength}{0cm}
\fbox{\includegraphics[width = 8.54cm]{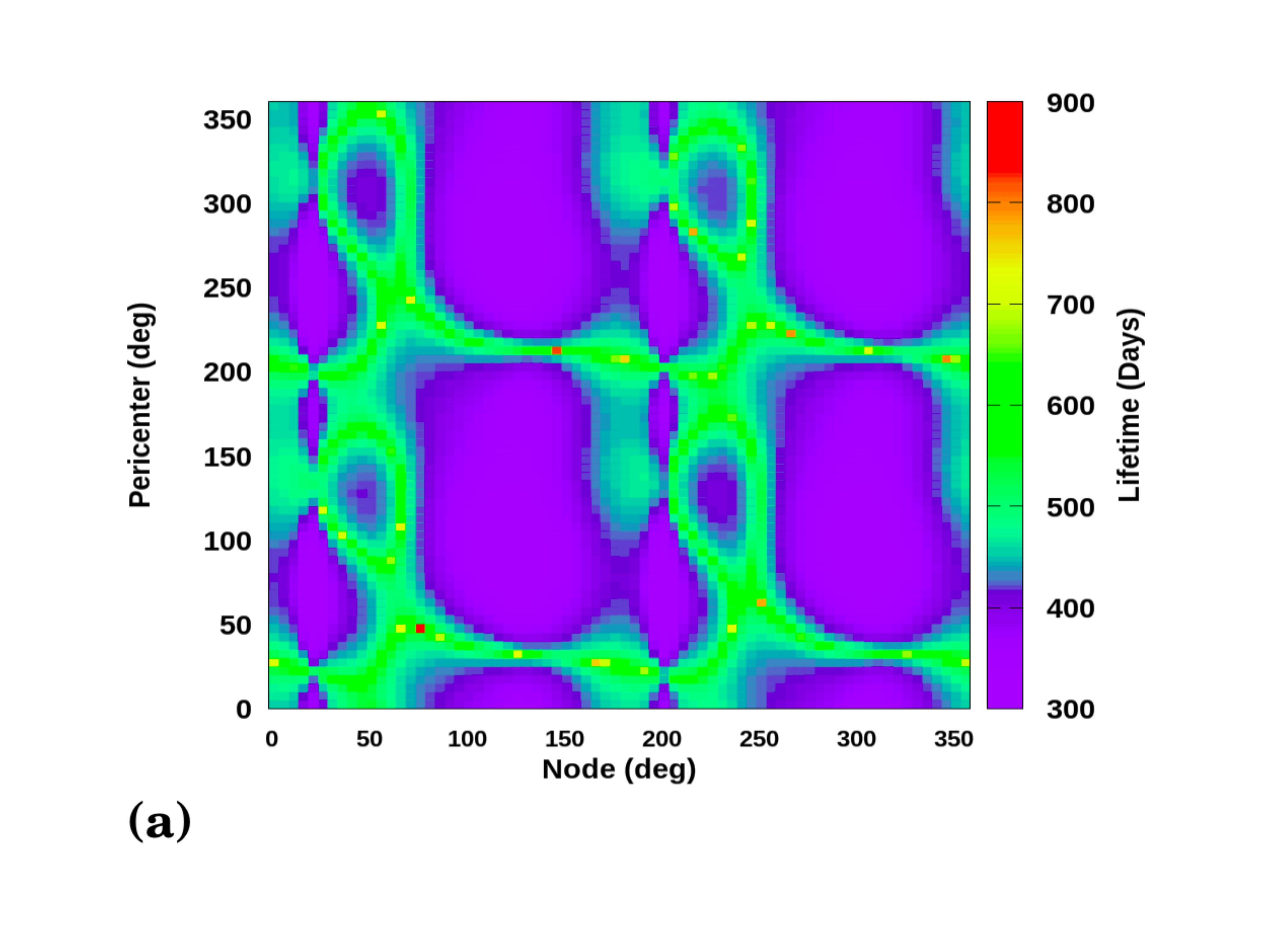}}  
\fbox{\includegraphics[width = 8.58cm]{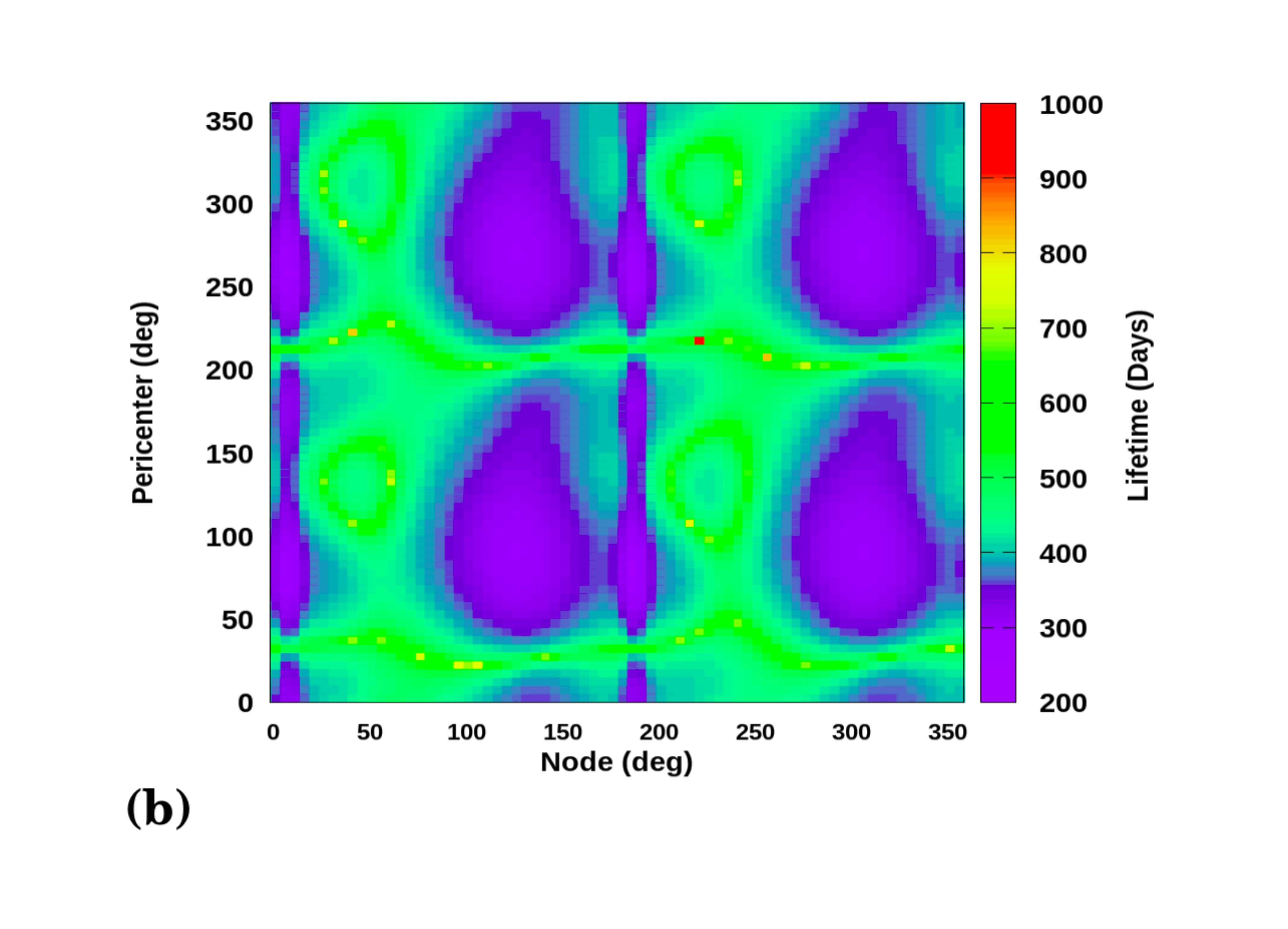}}\\
\fbox{\includegraphics[width = 8.57cm]{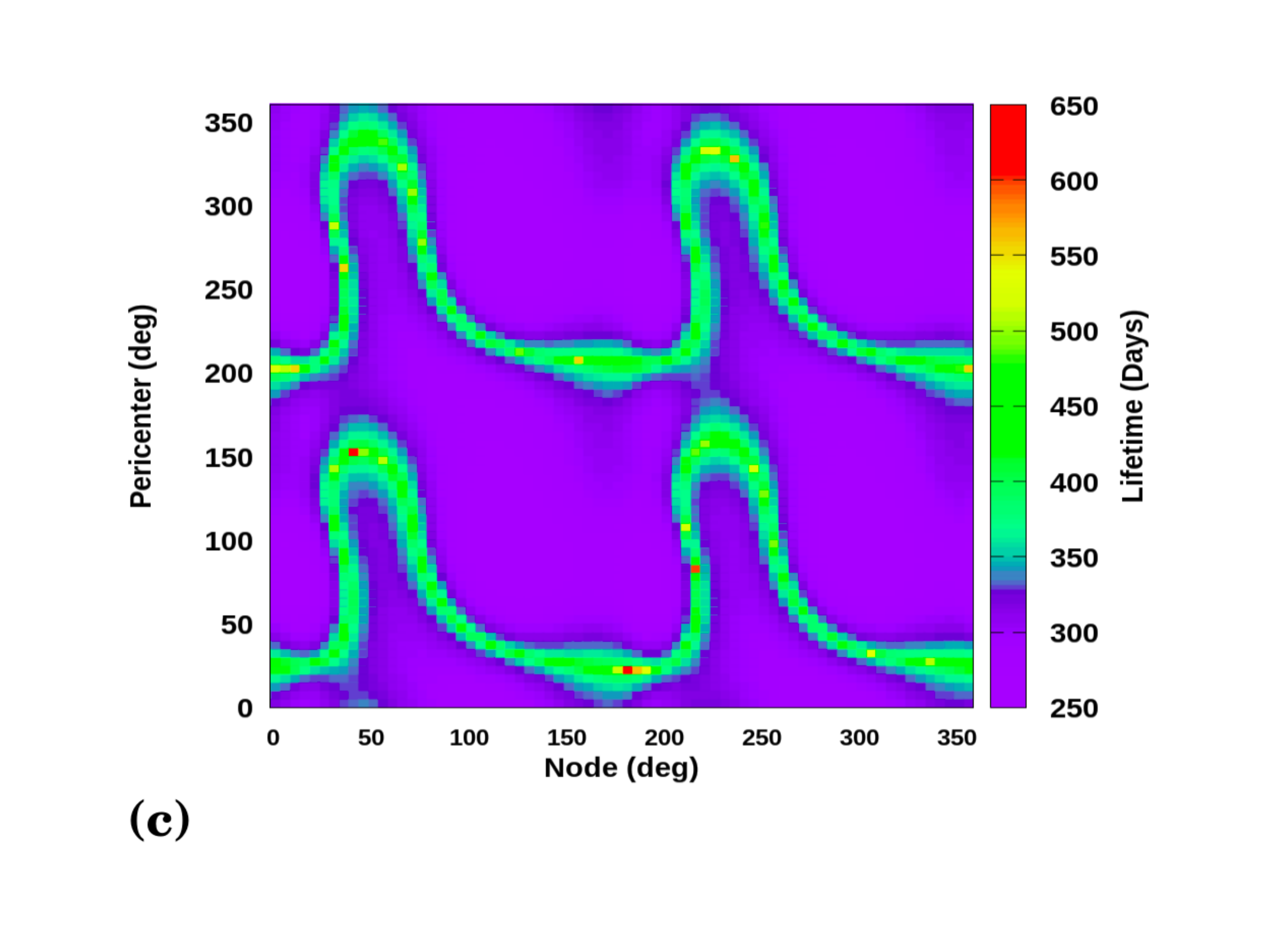}}\hspace{0.1cm}\fbox{\includegraphics[width = 8.57cm]{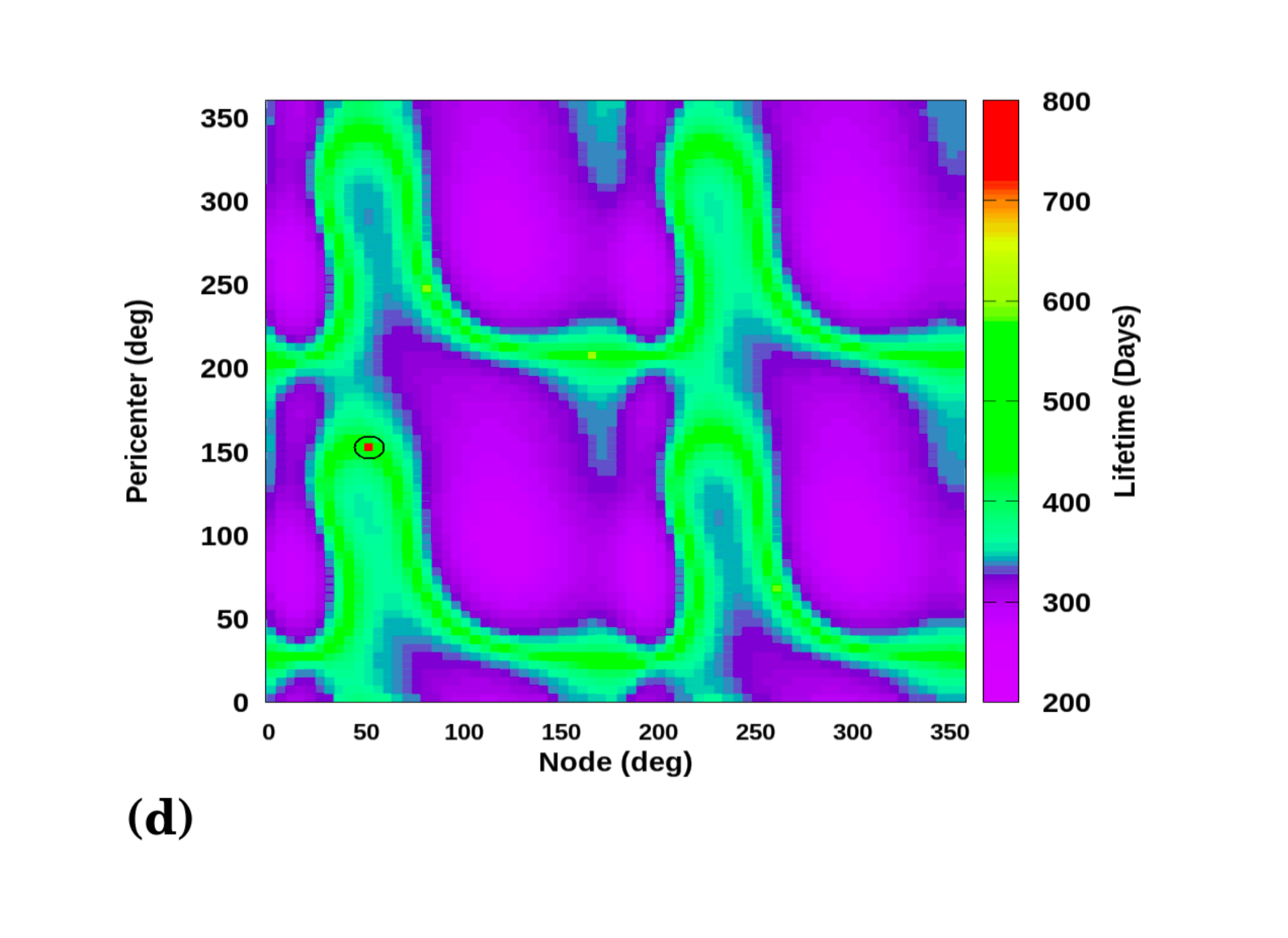}}
\end{adjustwidth}
\caption{Diagram of $\omega \times \Omega$ for $e=10^{-4}$ considering the effects of third-body and $J_2$ and $C_{22}$ of Titania: (\textbf{a}) $a=900$ km, $I=70^\circ$; (\textbf{b}) $a=900$ km, $I =80^\circ$; (\textbf{c}) $I=80^\circ$, $a=1058$ km; (\textbf{d}) $I=80^\circ$,  $a =1018$ km; $\omega$ = 0--360\textdegree, $\Omega$ = 0--360\textdegree.}
\label{fig:4}
\end{figure}

The first case to explore is for $e = 10^{-4}$. The results provided by the simulations showed that the longest orbits occurred for $a$ = 900--1058~km and $I$ = 70--80\textdegree. The result of this analysis can be seen in Figure \ref{fig:4}. 
 
We fixed a semi-major axis value, 900 km, and varied the inclination to $70^\circ$ (Figure \ref{fig:4}a) and $80^\circ$ (Figure \ref{fig:4}b). We then fixed the inclination at $80^\circ$ and vary the semi-major axis to 1018 km, (Figure \ref{fig:4}c) and 1058 km (Figure \ref{fig:4}d). The angles $\omega $ and $\Omega $ range from $0^\circ$ to $360^\circ $. 

Figure \ref{fig:4}a,b show similar regions for longer life orbits, $\omega$ = 0-- 150\textdegree, $\omega$ = 220--360\textdegree~and $\Omega$ = 0--100\textdegree, $\Omega$ = 200--300\textdegree. In Figure \ref{fig:4}a, where the considered inclination is $70^\circ$, the longer duration orbits have a lifetime of up to 900 days. As we increase the inclination by $10^\circ$, the lifetime increases by 100 days (Figure \ref{fig:4}b). This is due to the action of the gravitational coefficients $J_2$ and $C_{22}$, which are responsible for attenuating the effects caused by the third body for more inclined orbits.

The regions with longer duration orbits present in Figure \ref{fig:4}c,d, have values of $\omega$ = 25--150\textdegree, $\Omega$ = 50--100\textdegree,~and $\Omega$ = 150--200\textdegree. Orbits with a semi-major axis of 1058~km have a maximum lifetime of up to 650 days, whereas, in Figure \ref{fig:4}d, this time is 800 days for $a=1018$~km.

We chose the initial conditions of a specific point within these longer life regions to redo the simulations for $e=10^{-4}$. The chosen point is highlighted in Figure \ref{fig:4}d (black circle), and has values of $\Omega = 155^\circ$, $\Omega=55^\circ$, $a=1018$~km, and $I=80^\circ$. Figure \ref{fig:4}a,b present orbits with longer lifetimes compared to the point chosen in Figure \ref{fig:4}d. However, our main objective is to show that the adoption of these angles can increase the lifetime of the orbit even for regions with shorter orbits. In order to analyze the importance of these angles in the lifetimes, we made a new simulation with values of $\Omega = 155^\circ$, $\Omega = 55^\circ$ for $e = 10^{-4}$.

The insertion of the values $\omega$ and $\Omega$ in the new simulation for $e=10^{-4}$ increase the probe's life by approximately 78\%, a very large value. We extended the numerical integration for the two cases analyzed: third body and $J_2$ and $C_ {22}$ from Titania. However, the new results show distinct regions with more extended life for each perturbation considered. In the case of the gravitational effects of Uranus, a small island of orbits with a lifetime of 600--800 days appears with a semi-major axis ranging from $a = 810$ km to $900$ km and inclinations from $75^\circ$ to $78^\circ$. If we include the gravity coefficients of Titania, the maximum lifetime is approximately 600 days. It is located in the region \emph{a} = 810--1000~km and can be seen for all values of \emph{I}, increasing from \emph{I} = 78\textdegree~to 90\textdegree. Orbits with a lifetime of 800 days are found for a value of \emph{a} close to 900 km and \emph{I} = 78--90\textdegree.

The differences indicate that after the inclusion of the angles $ \omega $ and $ \Omega $, the gravity coefficients of Titania are more relevant for inclined orbits, \emph{I} = 80--90\textdegree. 

For \emph{a} = 875--950~km, the orbits reach lifetimes between 200 and 400 days. This lifetime is also experienced for orbits with a semi-major axis ranging from 950 km to 1050 km and \emph{I} = 75--80\textdegree, shown in the smaller island. From $ 810 $ km to 1100 km, the gravity coefficients of Titania can still extend the duration of the orbits up to 100 days for inclinations in the range of 80--90\textdegree. Before the inclusion of $ \omega = 55^\circ $ and $ \Omega =155^\circ $, the orbits that were most affected by the gravity coefficients of Titania had a maximum lifetime up to 60 days. Considering these specific values of those angles, this time has been extended to 400~days, a very large difference.

In the case of Uranus, the differences point out that, when $ \omega $ and $ \Omega $ angles are not zero, there is a deviation in the location of the orbits where their contribution was most significant. When $ \omega = \Omega =0^\circ$, the most inclined orbits were the most affected by the third body. After that there are orbits with inclinations between 75 and 76\textdegree~with lifetimes of 100--300~days.

\begin{figure}[H]
\begin{adjustwidth}{-\extralength}{0cm}
\fbox{\includegraphics[width = 8.55cm]{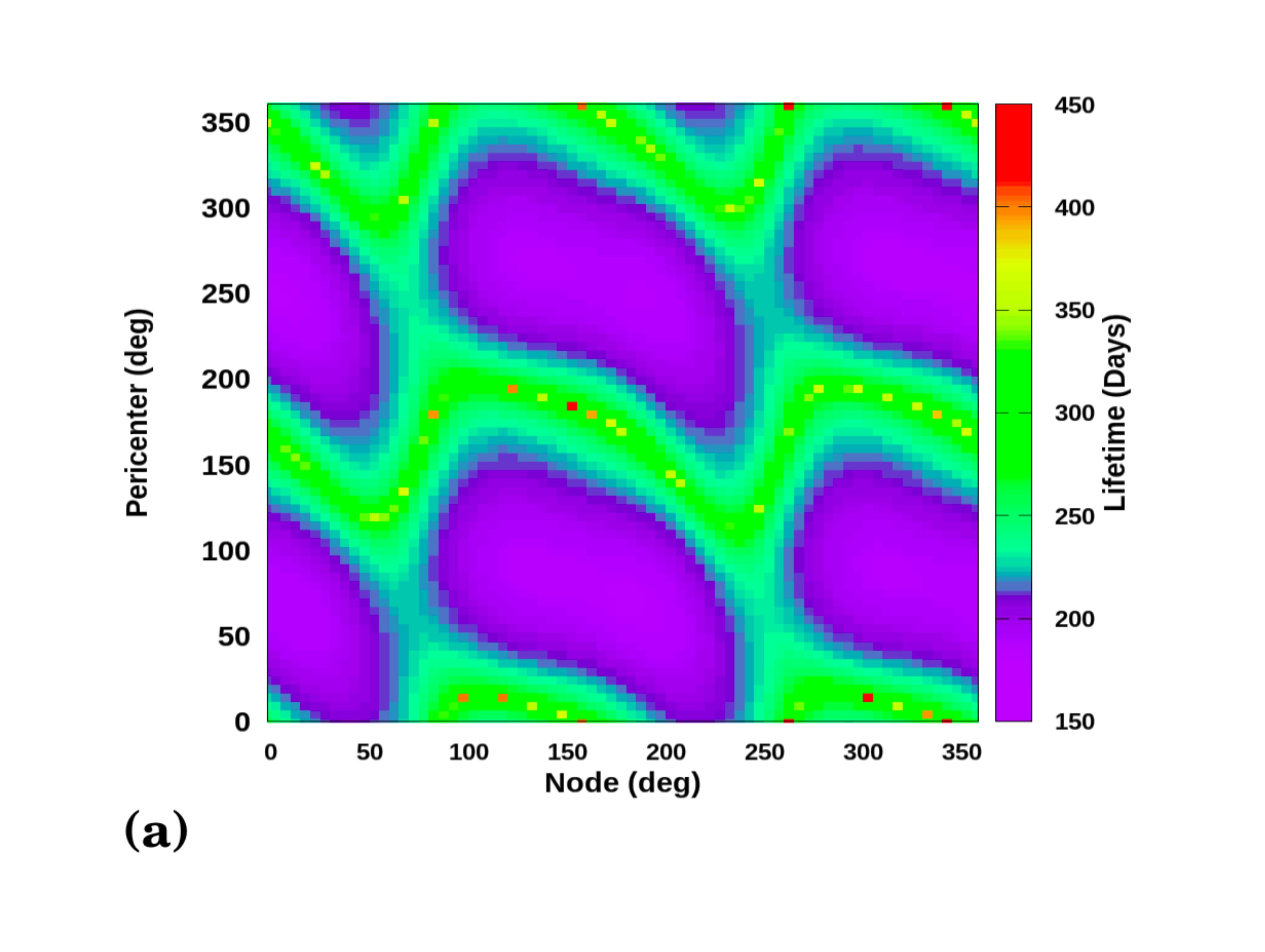}}  \fbox{\includegraphics[width = 8.66cm]{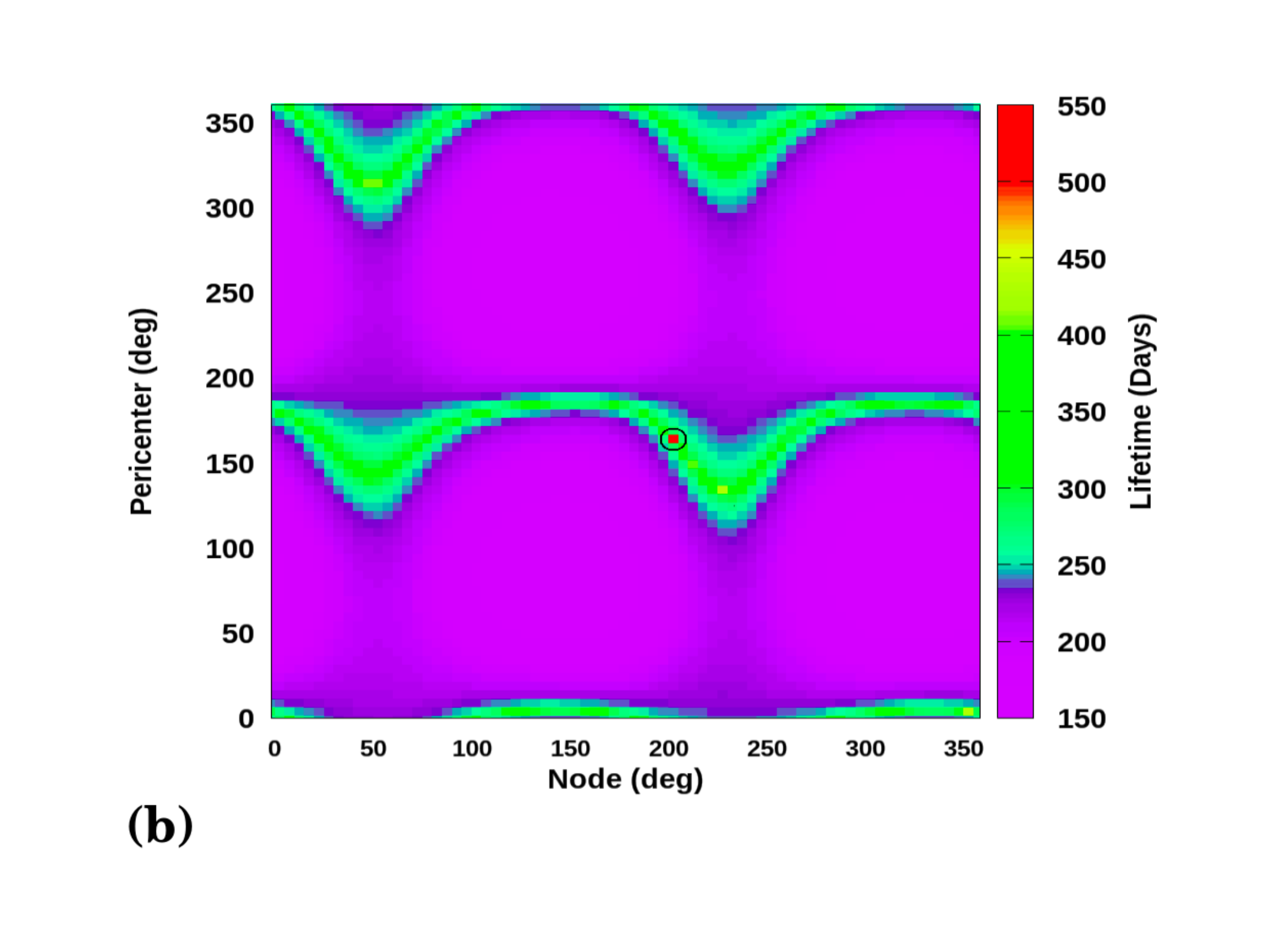}}\\
\fbox{\includegraphics[width = 8.55cm]{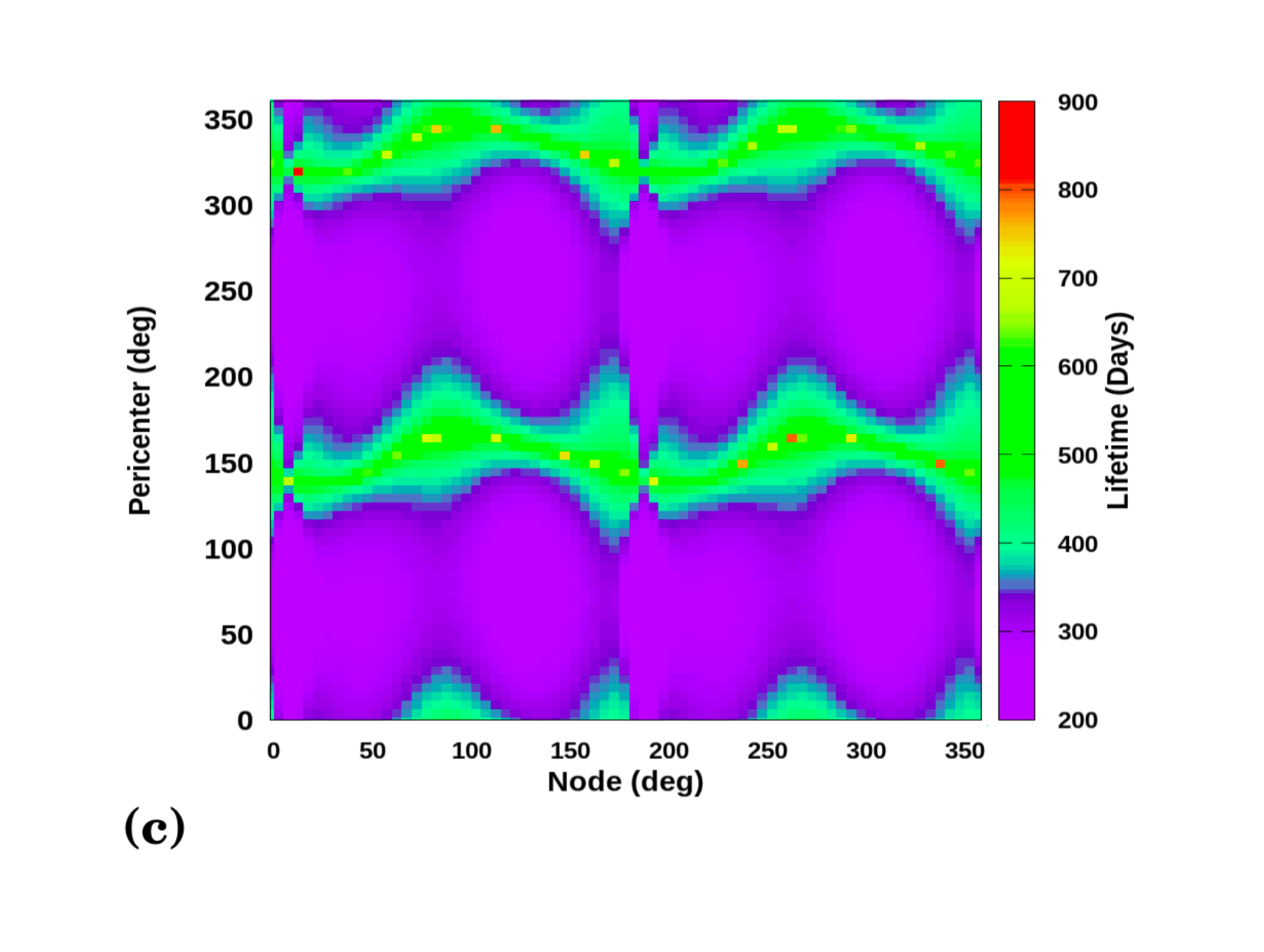}} 
\fbox{\includegraphics[width = 8.62cm]{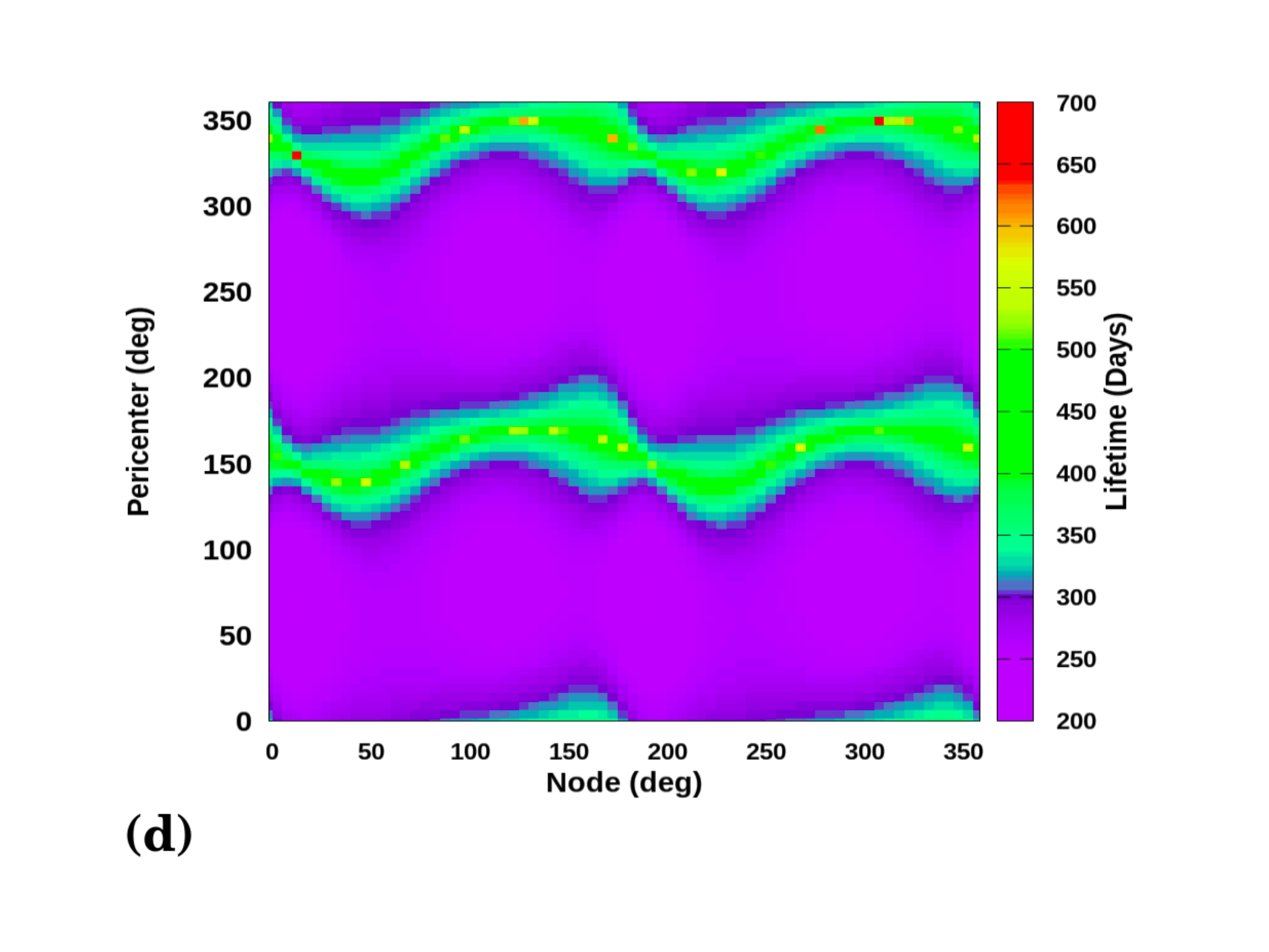}}
\end{adjustwidth}
\caption{Diagram of $\omega \times \Omega$ for $e=10^{-3}$ considering the effects of third-body and $J_2$ and $C_{22}$ of Titania: (\textbf{a}) $a=1265$ km, $I=70^\circ$;  (\textbf{b}) $a=1265$ km, $I =80^\circ$; (\textbf{c})  $I=80^\circ$, $a=834$ km;  (\textbf{d}) $I=80^\circ$, $a =1000$ km; $\Omega=0^\circ-360^\circ$, $\omega=0^\circ-360^\circ$}
\label{fig:5}
\end{figure}

For the case $e=10^{-3}$, the best initial conditions presented in Figure \ref{fig:2} point to a semi-major axis in the range of 834--1265~km and an inclination of 70--80\textdegree. In Figure \ref{fig:5} we present a diagram $\omega \times \Omega$ for some specific cases. In Figure \ref{fig:5}a, $a= 1265$~km and $I=70^\circ$, whereas in Figure \ref{fig:5}b, the initial conditions are $a= 1265$~km and $I=80^\circ$.

Figure \ref{fig:5}a shows large islands with orbits that reach 450 days. These islands are located at $\omega$ = 100--200\textdegree, $\omega$ = 300--360\textdegree, $\Omega=50^\circ-200^\circ$, and $\Omega$ = 250--360\textdegree. In the case of Figure~\ref{fig:5}b, where the inclination is $80^\circ$, the ``protection mechanism'' caused by the combination of the terms $J_2$ and $C_{22}$ increases the orbital duration to 550 days. However, these islands are much smaller and can be seen at $\omega$ = 125--180\textdegree, $\omega=300^\circ-360^\circ$ and $\Omega=50^\circ$ and $\Omega=250^\circ$.

Figure \ref{fig:5}c,d have the same inclination $80^\circ$, but different semi-major axes. In Figure~\ref{fig:5}c, with $a=834$ km, the maximum lifetime is almost 3 years. However, the regions are smaller compared to those seen in Figure \ref{fig:5}a,b. They are found with $\omega$ around $150^\circ$ and $ \Omega$ close to $325^\circ$. More distant orbits, with a semi-major axis equal to 1000 km, (Figure \ref{fig:5}d), it is noted that the lifetime decreases, and the values of $\omega$ and $\Omega$ for orbits with the longest duration are $100^\circ$ and $275^\circ$, respectively.

\begin{figure}[H]
\begin{adjustwidth}{-\extralength}{0cm}
\fbox{\includegraphics[width=9.2cm]{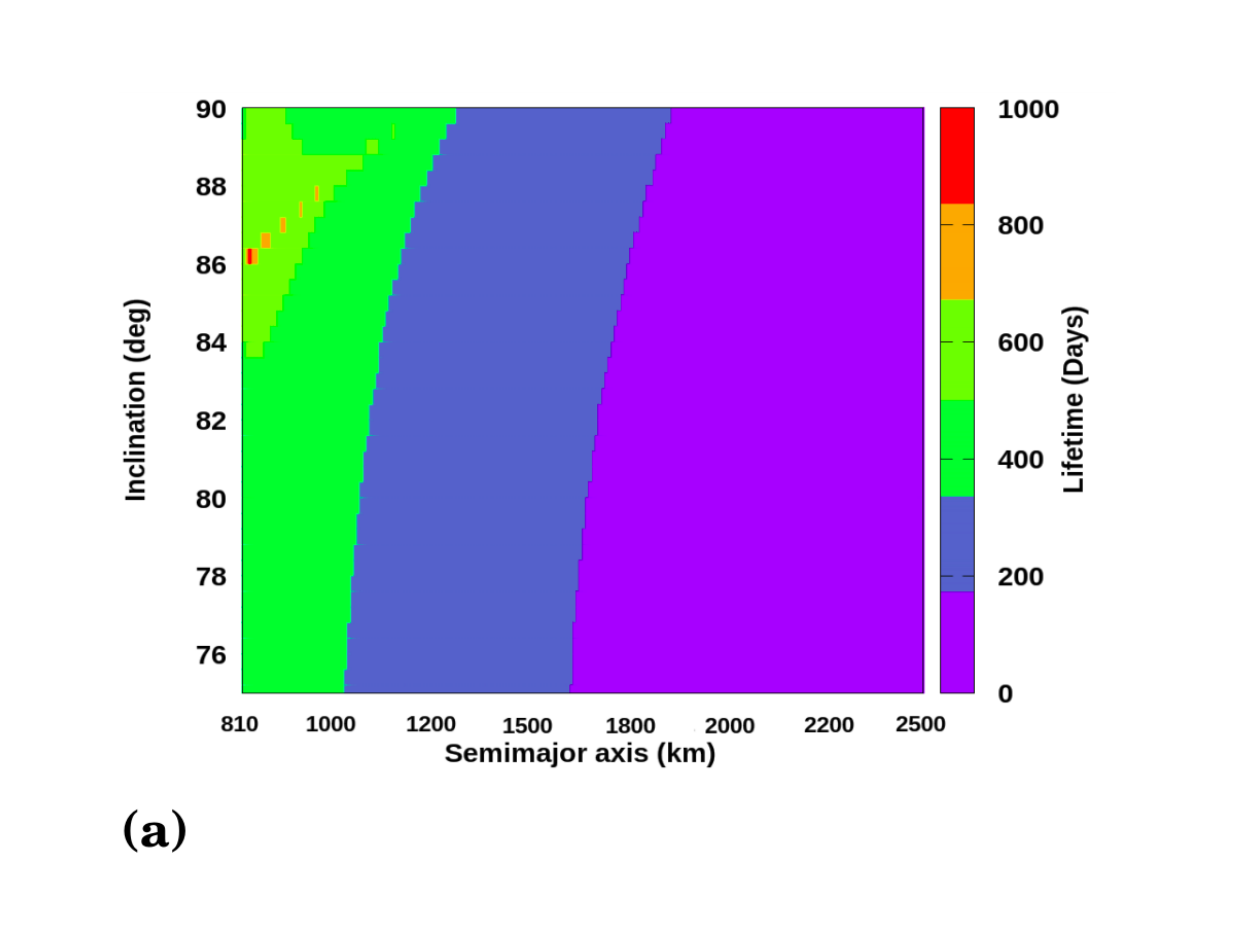}} 
\fbox{\includegraphics[width=9.7cm]{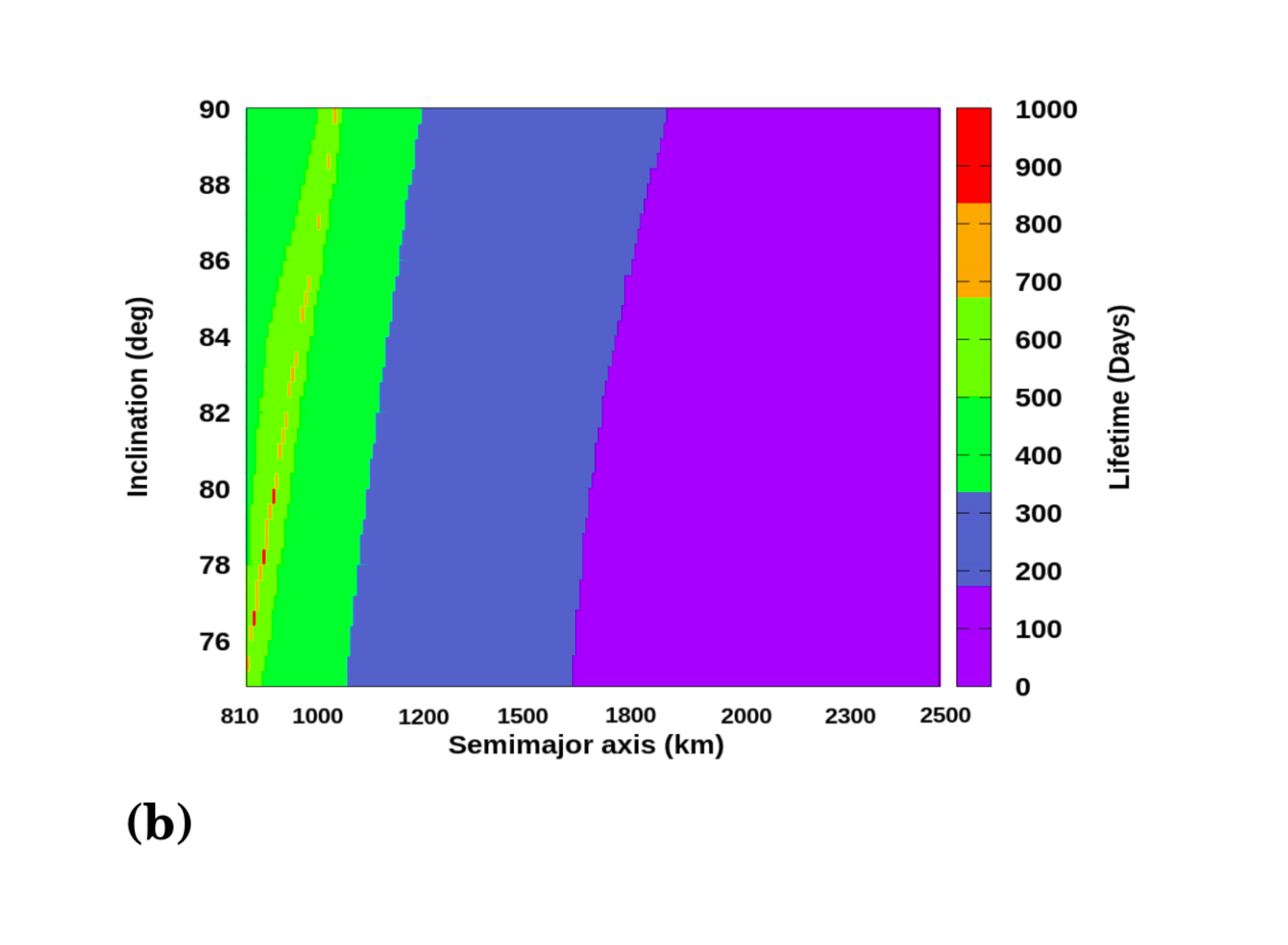}}
\fbox{\includegraphics[width=9.3cm]{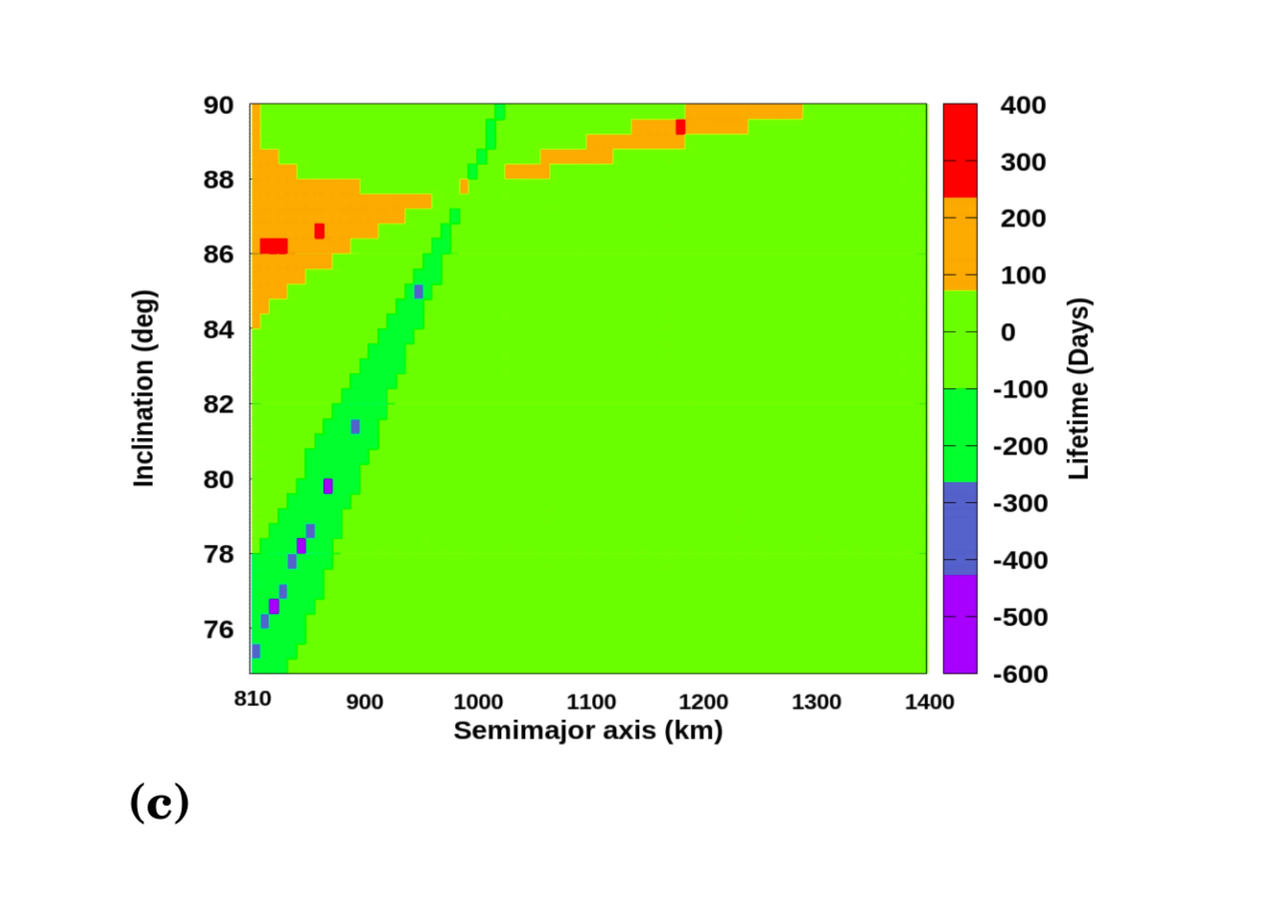}}
\end{adjustwidth}
 \caption{Diagram of $a$ versus $I$ for $e=10^{-3}$: (\textbf{a}) considering only the effects from Uranus; (\textbf{b}) including
the third-body and the effects of $J_2$ and $C_{22}$ of Titania; (\textbf{c}) lifetime differences (third body $- J_2$ and $C_{22}$ of Titania). Initial values are \emph{a} = 810--1400 km, \emph{I} = 75--90\textdegree, $\omega=165^\circ$, and $\Omega=205^\circ$.}
\label{fig:6}
\end{figure}

An initial condition (a point) with values of $\omega = 165^\circ $ and $ \Omega =205^\circ $ is highlighted (black circle) in Figure \ref{fig:5}b. The values of these angles were added in the new simulations for the initial conditions of Figure \ref{fig:2}, in order to investigate how they would affect the lifetime of the probe. The new investigation was carried out by considering only the perturbation of the third body and with the complete system, third body plus the $J_2$, $C_{22}$ coefficients of Titania. Shown in Figure \ref{fig:6}, it is observed that the use of the non-zero values for these angles in $e = 10^{-3}$ increases by 100\% of the lifetime of the probe in both cases. When both effects are included, the gravitational attraction of Uranus and the gravity coefficients of Titania, the use of non-zero values for $\omega$ and $\Omega$ also changed those regions where the lifetimes are longer. Orbits with a lifetime up to 600 days are found for inclinations ranging from $ 84^\circ $ to $ 90^\circ $ and $ a = 810 $ km to $ a = 1000 $ km (third-body effects).

When considering the gravity coefficients of Titania, orbits with this lifetime appear for an interval of semi-major axes equal to \emph{a} = 810--1050  km and $ I $ in the range 75--90\textdegree~(light green squares). 
The location of the orbits with a lifetime of approximately 400 days is practically the same for both cases, \emph{a} = 810--1000 km and \emph{I} = 75--90\textdegree~(green squares). The blue dot region has lifetimes in the range 200--300 days for all inclinations and with $ a $ in the range 1000--1600 km.

Figure \ref{fig:6}c shows that Uranus affects the lifetime of the more inclined orbits. This effect appears in a small orange island that goes from \emph{I} = 84--90\textdegree~for \emph{a} = 810--1300 km. In this region, the duration of the orbits is approximately 400 days, considering the lifetime of the probe affected by the gravity coefficients of Titania and Uranus. The effect of $ J_2 $ and $ C_{22} $ of Titania and the third-body acts on orbits with inclinations of 75--90\textdegree~and semi-major axes of 810--1000 km. The lifetimes of these orbits range from 100 to 
600 days compared to the case where only the effect of Uranus is considered; there are no significant changes for $a$ larger than 1400 km.

\begin{figure}[H]
\begin{adjustwidth}{-\extralength}{0cm}
\fbox{\includegraphics[width = 8.4cm]{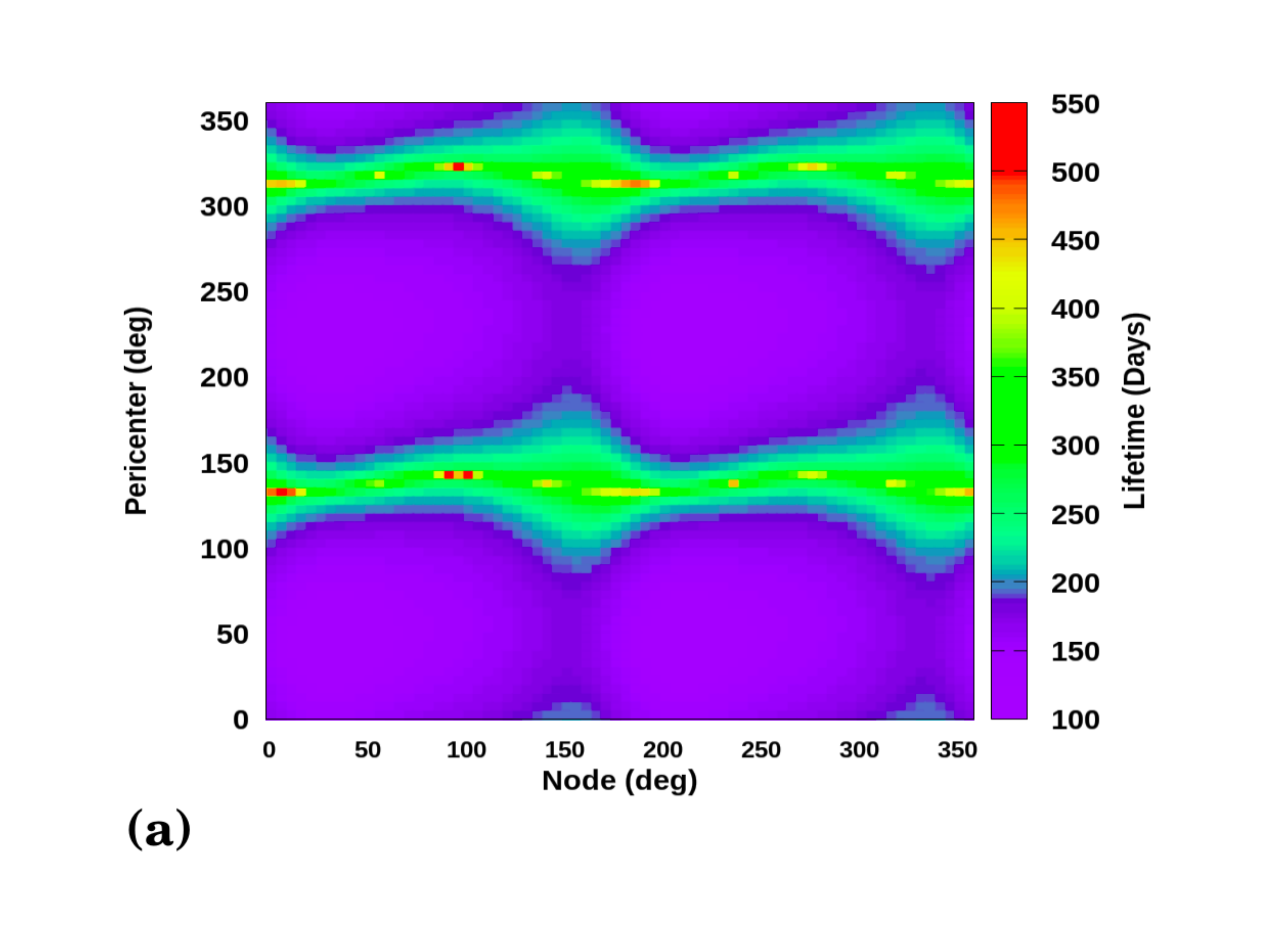}}  
\fbox{\includegraphics[width = 8.4cm]{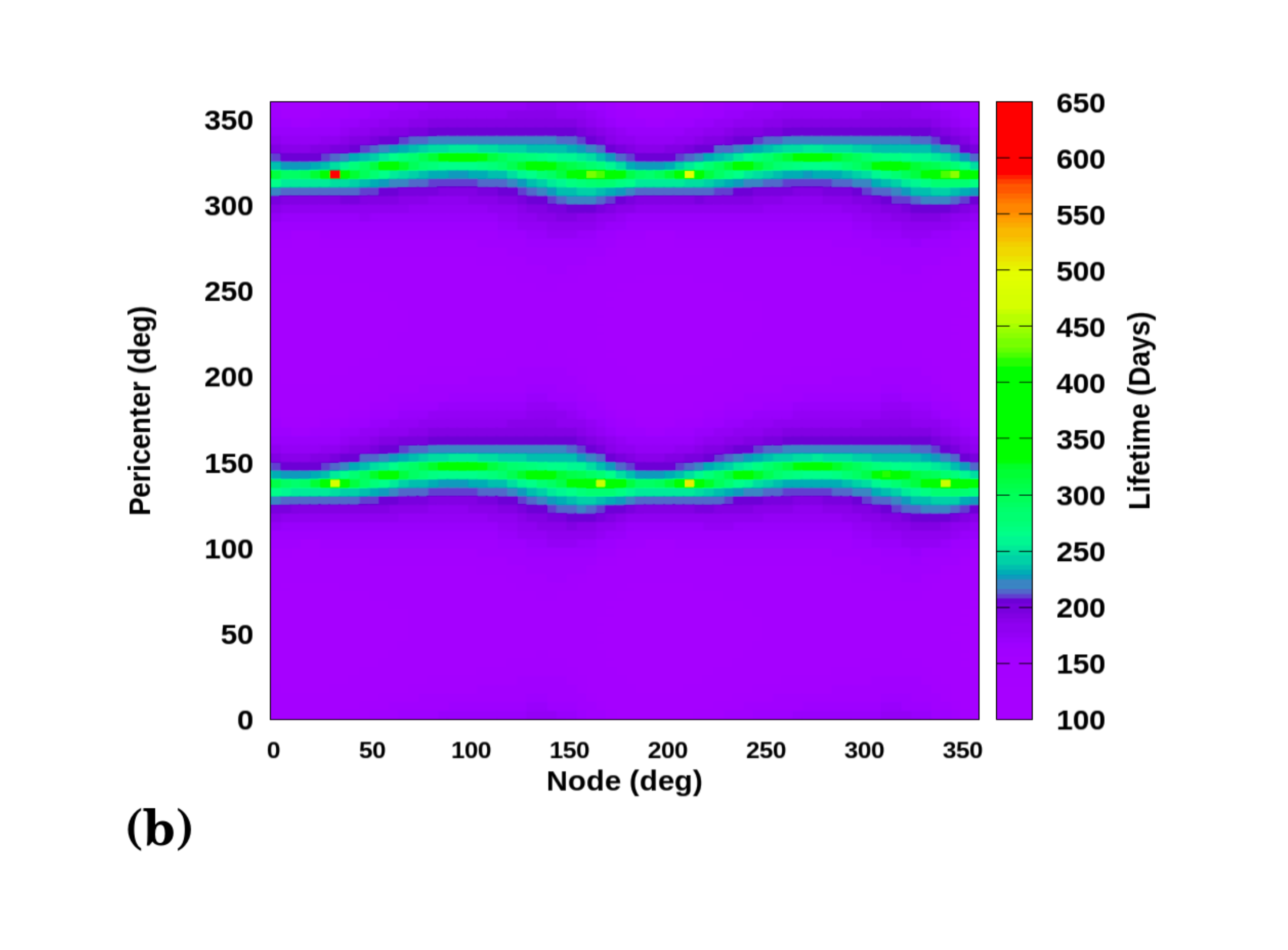}} \\
\fbox{\includegraphics[width =8.4cm]{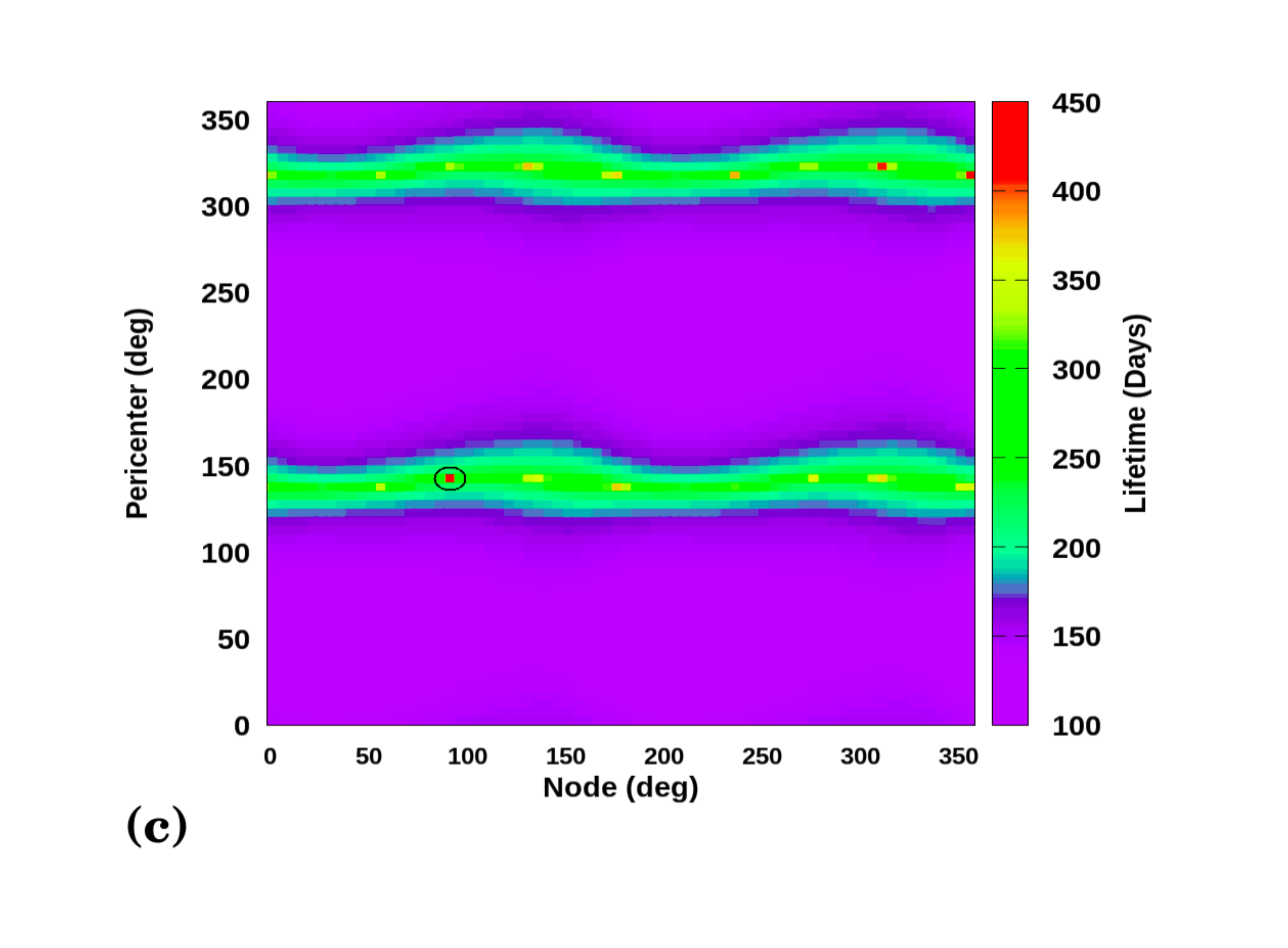}}  
\fbox{\includegraphics[width = 8.4cm]{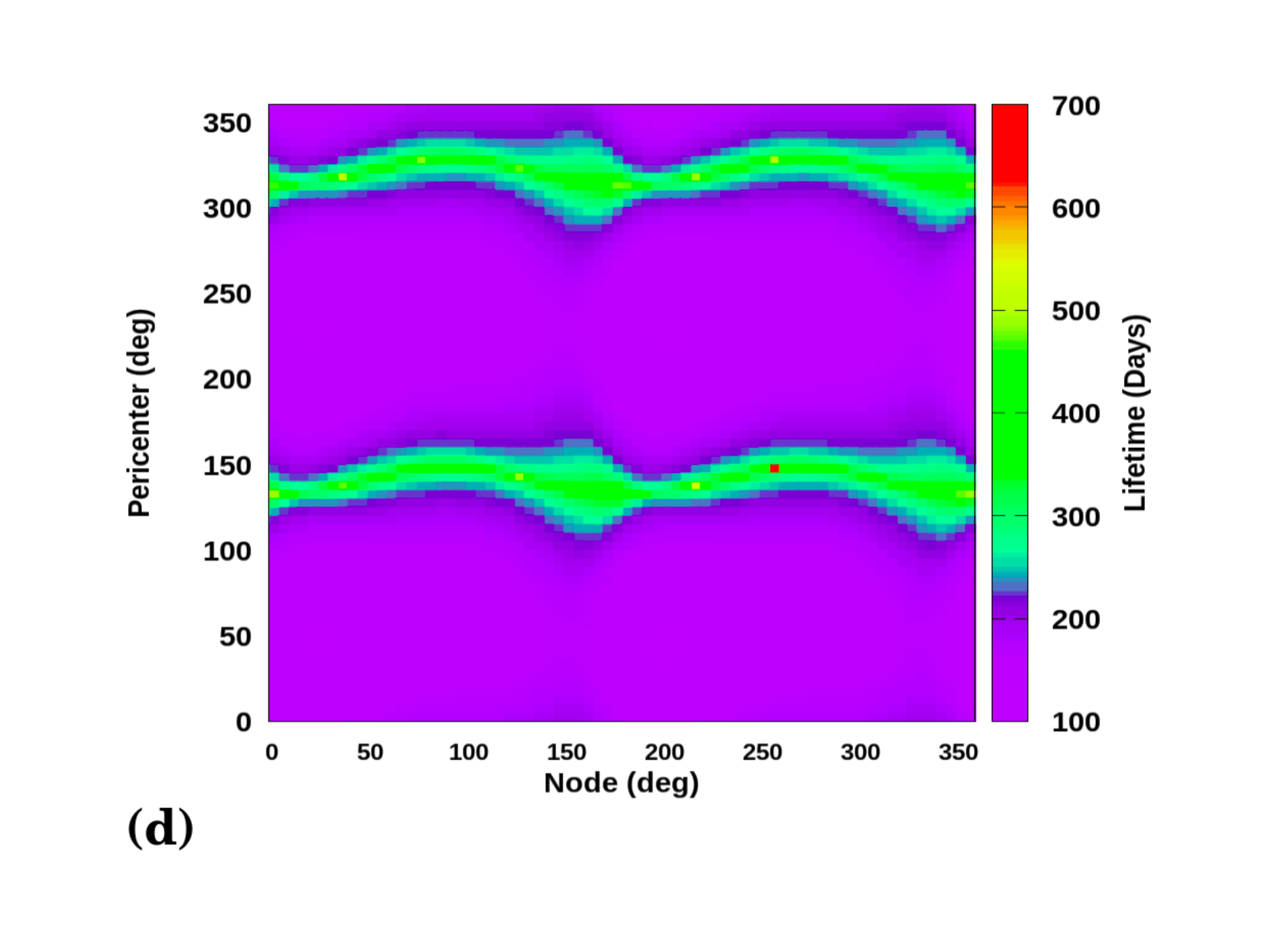}}
\end{adjustwidth}
\caption{Diagram of $\omega \times \Omega$ for $e=10^{-2}$ considering the effects of third-body and $J_2$ and $C_{22}$ of Titania: (\textbf{a}) $a=1000$ km, $I=70^\circ$;  (\textbf{b}) $a=1000$ km, $I=80^\circ$; (\textbf{c}) $I=80^\circ$, $a=1162$ km;  (\textbf{d})  $I=80^\circ$, $a=914$ km; $\omega$ = 0--360\textdegree, $\Omega$ = 0--360\textdegree.}
\label{fig:7}
\end{figure}

For $e=10^{-2}$, the longest-lived orbits are located in the range 900--1160 km and \emph{I} = 70--80\textdegree. Therefore, the initial conditions used to build the $\omega \times \Omega$ graphs are within this range of values. In Figure \ref{fig:7} we present four scenarios for analysis of these angles. In Figure~\ref{fig:7}a, the semi-major axis considered is 1000 km and the inclination is $70^\circ$, whereas in Figure~\ref{fig:7}b, $a$ also assumed the value 1000 km and $I = 70^\circ$.

There is a similarity between the regions of orbits with longer orbital duration shown in Figure \ref{fig:7}a,b. In both cases, the islands with the longest lifetime are located for values of $\Omega$ close to $150^\circ$, whereas for $\omega$ the best values are around $140^\circ$ and $325^\circ$. Despite the high value of the eccentricity, with the combination of Titania’s gravity terms with the third body perturbation, long-lived orbits can be found. In Figure \ref{fig:7}a the maximum lifetime is 550 days, whereas in Figure \ref{fig:7}b, where the inclination is greater, the duration of the orbits is up to 650~days.

The best orbits for $I=80^\circ$ (Figure \ref{fig:7}c,d), are for $\Omega = 150^\circ$ and for $\omega$ close to $140 ^\circ$ and $325^\circ$. Due to the proximity to the surface of Titania, the lifetime for the cases analyzed in Figure \ref{fig:7}d is longer, 700 days.

\begin{figure}[H]
\begin{adjustwidth}{-\extralength}{0cm}
 \fbox{\includegraphics[width=9.2cm]{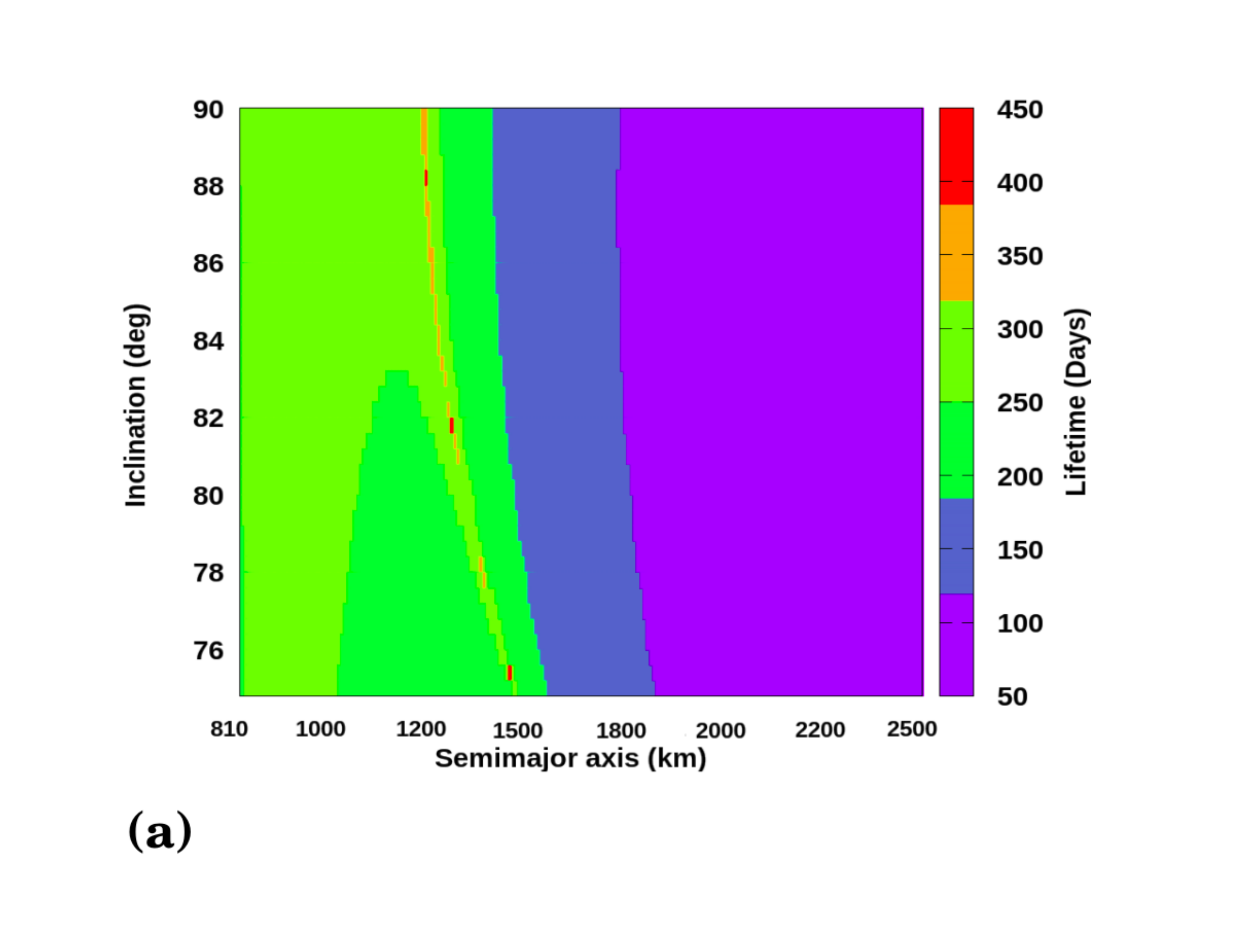}} 
 \fbox{\includegraphics[width=9.1cm]{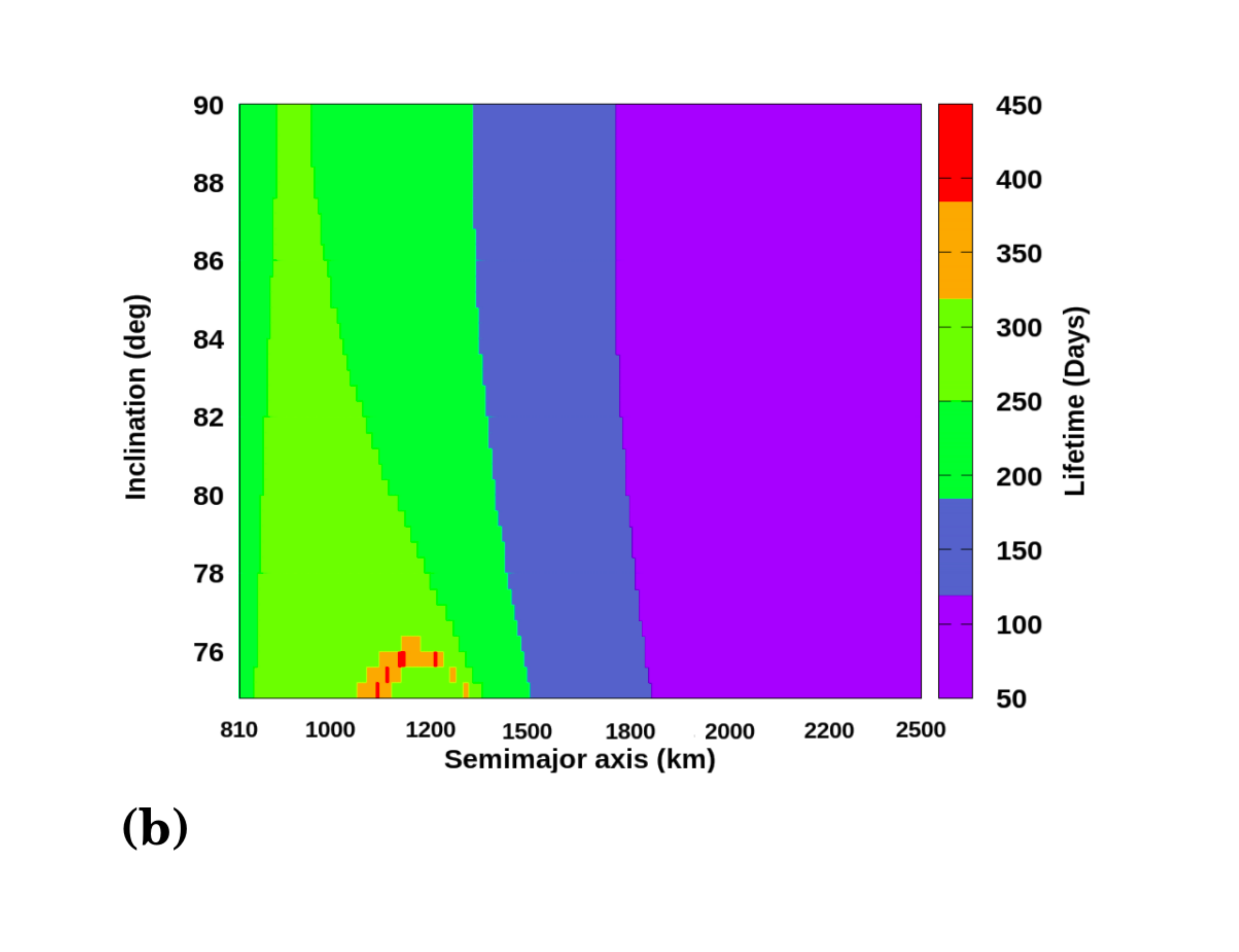}} 
\fbox{\includegraphics[width=9.2cm]{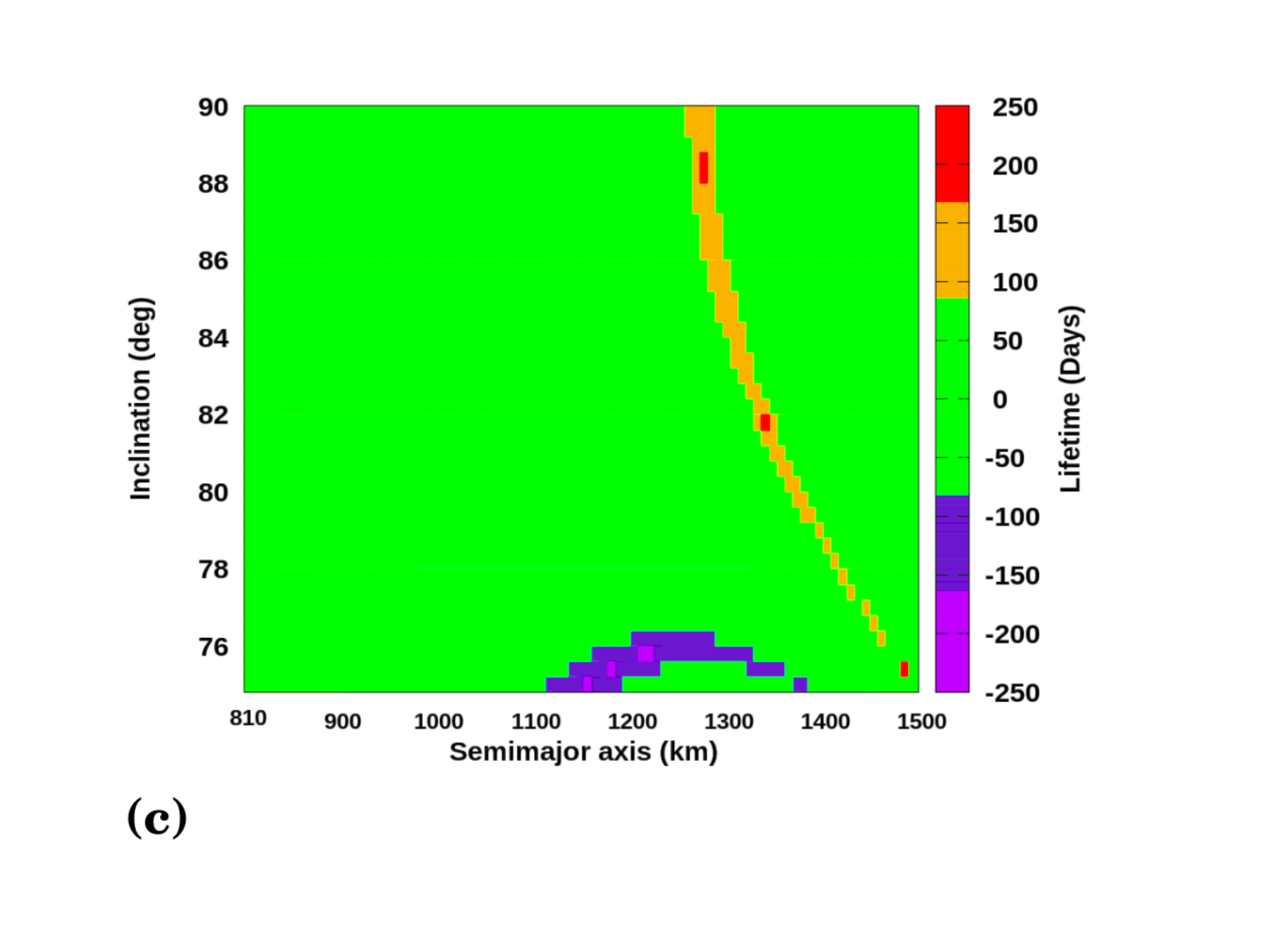}}
\end{adjustwidth}
\caption{Diagram of $a$ versus $I$ for $e=10^{-2}$: (\textbf{a}) considering only the effects from Uranus; (\textbf{b}) including third-body and effects from $J_2$ and, $C_{22}$ of Titania; (\textbf{c}) lifetime differences (third body $- J_2$ and $C_{22}$ of Titania). Initial values ares \emph{a} = 810--1500 km, \emph{I} = 75--90\textdegree, $\omega=145^\circ$, and $\Omega=95^\circ$.}
\label{fig:8}
\end{figure}

New numerical simulations for $e = 10^{-2}$ were made according to the initial condition highlighted in Figure \ref{fig:7}c (black circle), where the orbital elements are equal to $a$ = 1162~km, $I = 80^\circ$, $\omega = 145^\circ$, and $\Omega = 95^\circ$. These values for the argument of periapsis and the longitude of the ascending node were used in the new simulations with the same values of $a$, $e$, and $I$ used in Figure \ref{fig:3}. The results of the inclusion of these new values of $\omega$ and $\Omega$  are shown in Figure \ref{fig:8}. 

When considering $\omega$ and $\Omega$ non-zero, the increase of the lifetime of the probe is 2.5~times greater in comparison with $\omega=\Omega=0^\circ$. The lifetimes increased from 180 to 450~days.

In Figure \ref{fig:8}a, the orbits closer to the surface of Titania have lifetimes of 250--300 days with a semi-major axis in the range 810--1000 km. It occurs for all inclinations when we considered the perturbation of Uranus. When combining the effects of Uranus and the gravity coefficients of Titania, Figure \ref{fig:8}b, orbits with this lifetime occupy a slightly larger region, \emph{a} = 810--1350 km, also for all analyzed  values of $ I $. The islands with lifetimes up to 450 days are seen in both cases: when only the perturbation of Uranus is considered, and when the gravity coefficients of Titania are also taken into account. However, these orbits change locations according to each perturbation investigated. In Figure \ref{fig:8}a, these orbits can be seen from \emph{I} = 75--90\textdegree~and \emph{a} = 1300--1500 km, whereas in Figure \ref{fig:8}b, this lifetime is observed for orbits in the range \emph{a} = 1100--1400 km and  \emph{I} =75--76\textdegree~(orange and red dots). 

The orbits with  \emph{a} = 900--1450 km and \emph{I} =75--82\textdegree~have longer lifetimes when subjected to the effect of the gravity coefficients of Titania. The difference in the lifetimes is up to 50 days, in comparison with the situation where only the influence of the third body is considered. These orbits can be seen in Figure \ref{fig:8}c (green dots).

Numerical integration for $ e = 10^{-1}$ showed that the best values for $a$ and $I$ would be $900$ km and $ I =80^\circ $, respectively. We use these values and plot the lifetimes as a function of $ \omega \times \Omega $ varying these angles over the range 0--360\textdegree. The results of this analysis are not shown here in this work, as they are similar to the case $e=10^{-2}$. However, it is important to note that the adoption of specific values for $\omega$ and $\Omega$ also increased the orbit lifetime.

The results shown in Figure \ref{fig:6}b point out that the angles $\omega$ and $\Omega$ contribute to the increase in the lifetimes for orbits with high eccentricities, as in the case $ e = 10^{-1}$. 
With the values found for these angles, the duration of the orbits around Titania increased by two times when considering only the third body perturbation and more than eight times when the gravity coefficients of Titania were also included. In the analysis involving the contribution from Uranus gravitational attraction, the regions of greater and smaller duration remained similar, increasing only the maximum lifetime. 

However, exploring the scenario where $ J_2 $ and $ C_ {22} $ of Titania were added to the model, the orbits with zero and non-zero values of $ \omega $ and $ \Omega $ had very different maximum lifetimes as well as larger regions with longer lifetimes. Without the inclusion of specific values for these angles, the orbits with the longest lifetime were located at \emph{I} = 75--90\textdegree~and \emph{a} = 900--1200~km. In this interval, the maximum times reached by these orbits were 60 days. With the adoption of better values for $ \omega $ and $ \Omega $ in the new simulations, these orbits started to have lifetimes from 200 to 500 days, for \emph{a} = 900--1000 km and \emph{I} = 75--90\textdegree.

The analysis carried out in this section shows the importance of a different approach, first including variations in $\omega$ and $\Omega$, and then mapping $a \times I$ for the best values of $\omega$ and $\Omega$. In all cases investigated, the best values for $\omega$ can be found in two regions,  in the ranges 100--150\textdegree~and 27--360\textdegree. In the case of the ascending node longitude, the best values are close to 50--100\textdegree~and 250--360\textdegree. The orbits that had their lifetimes extended were those orbits closer to Titania's surface and with the greatest inclination. The reason for this is associated with the fact that lower and more inclined orbits are subject to the action of the zonal oblateness coefficient $J_2$, which, added to the term $C_{22}$ and the perturbation caused by the third body, promotes an equilibrium capable of increasing the duration of the orbit.

\section{Long-Duration Orbits}
\label{sec:element}

Previous studies of systems similar to the one presented here often use the double averaging method to find frozen orbits. According to the work of \cite{Carvalho2012}, in order to find a frozen orbit, the relations $\dfrac{de}{dt}=0$, $\dfrac{di}{dt}=0$, $ \dfrac{d\omega}{dt}=0$ must be satisfied. For such relationships to be met, it is common in the literature to define $\omega =\pi/2$ or $\omega=3\pi/2$, because, according to the Kozai--Lidov mechanism, under the presence of the third body, the argument of periapsis librate around these values as the eccentricity and inclination oscillate \cite{Kozai1962,Lidov1962}. However, when these values are fixed, $\dfrac{di}{dt}$ is non-zero due to the term $C_{22}$, which causes the inclination to increase significantly and hence the eccentricity. In other work, where the double mean model is used, it is possible to eliminate the short period terms that contain the $C_{22}$ coefficient and then obtain the necessary conditions to find the frozen orbits. 

In this work, we are not interested in finding frozen orbits, whose noted derivatives are equal to zero. Despite showing similarity to a frozen orbit, orbits that last longer are our main interest, independently of the value of these derivatives. This is because our goal is to analyze more realistic models and investigate the isolated effect of the term $C_{22}$ on the desired orbits.

\begin{figure}[H]
\includegraphics[width=10.3cm]{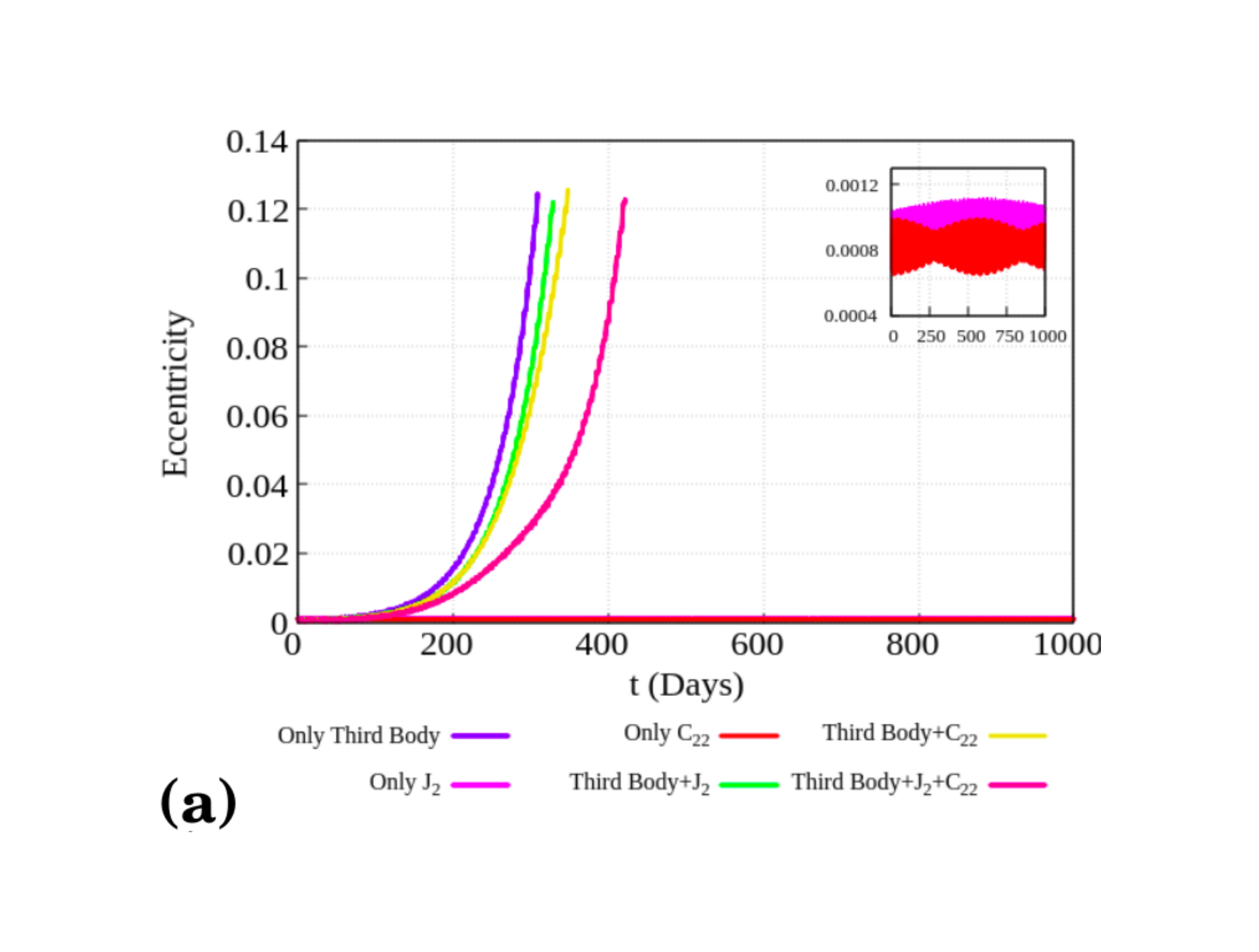}
\includegraphics[width=10.3cm]{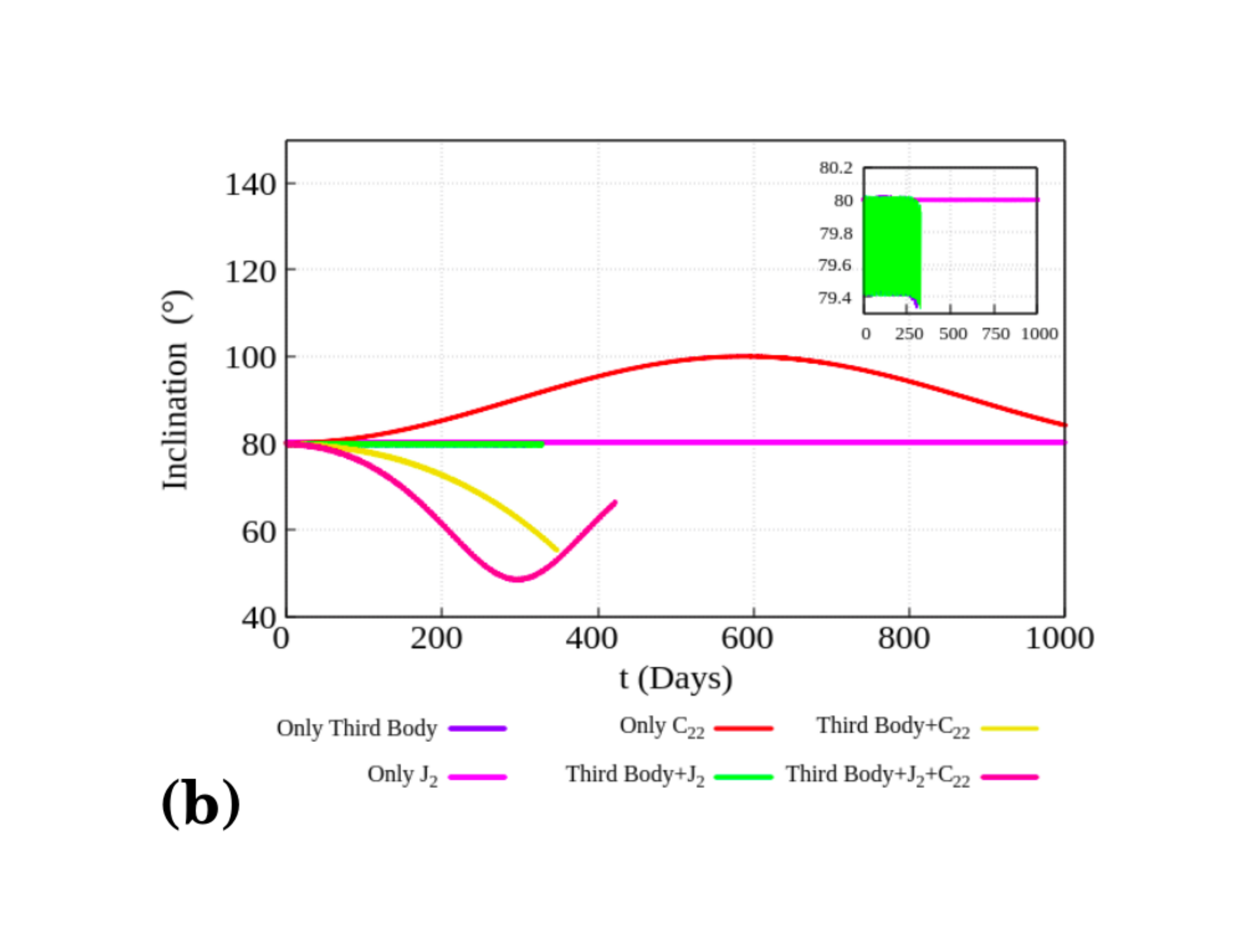}
 \caption{Diagram of $e \times t$ and $I \times t$ showing different perturbative effects: considering only the perturbations by Uranus, only $J_2$ from Titania, only $C_{22}$ from Titania, Uranus $+ J_2$ from Titania, Uranus $+ C_{22}$ from Titania, and Uranus $+ J_2 + C_{22}$ from Titania. Initial values are $a_i=900$ km, $e_i=10^{-3}$, $I_i= 80^\circ$, $\omega_i=\Omega_i=0^\circ$.}
\label{fig:9}
\end{figure}

\begin{figure}
\begin{adjustwidth}{-\extralength}{0cm}
 \includegraphics[width=8.9cm]{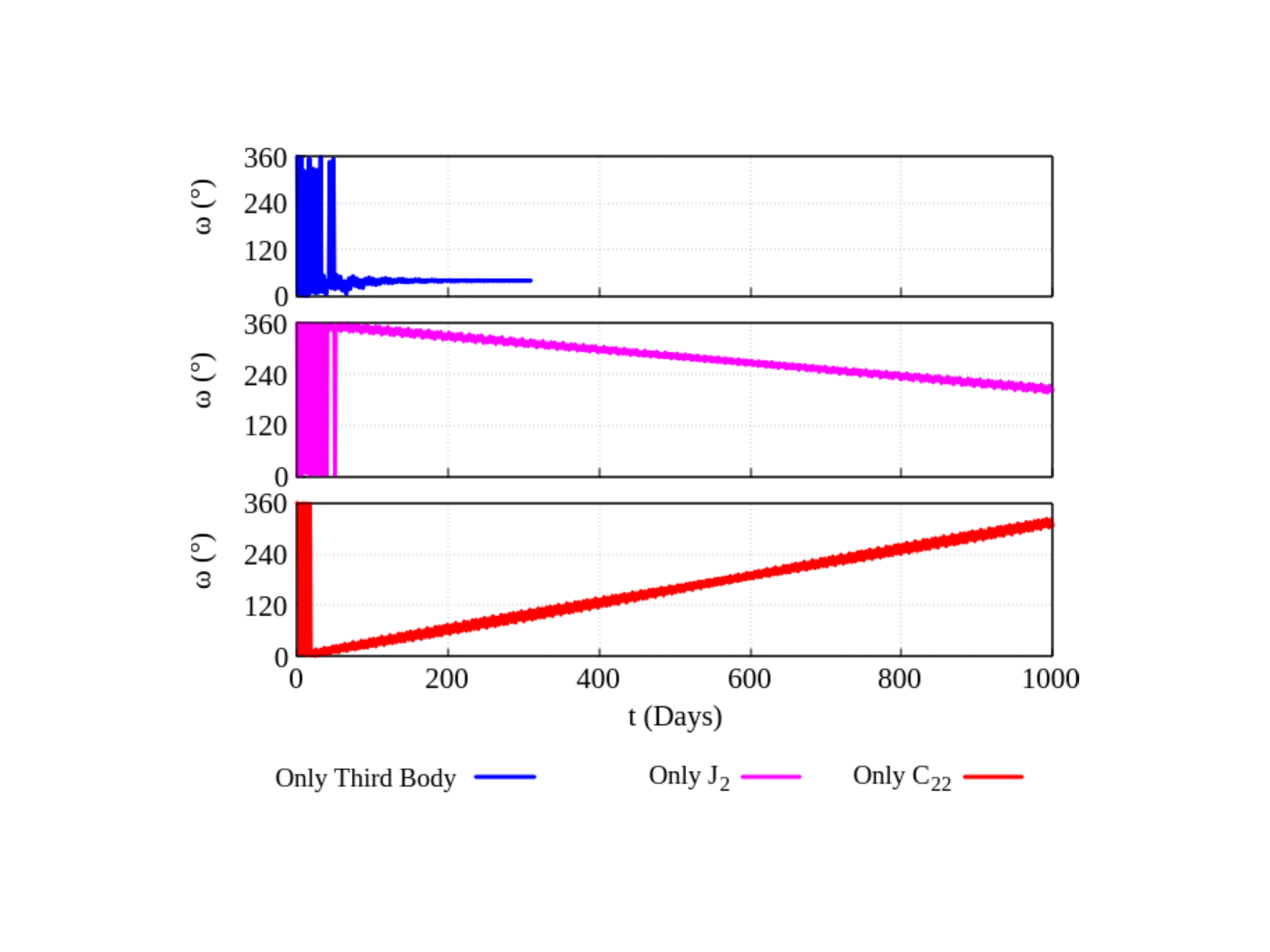}
 \includegraphics[width=8.9cm]{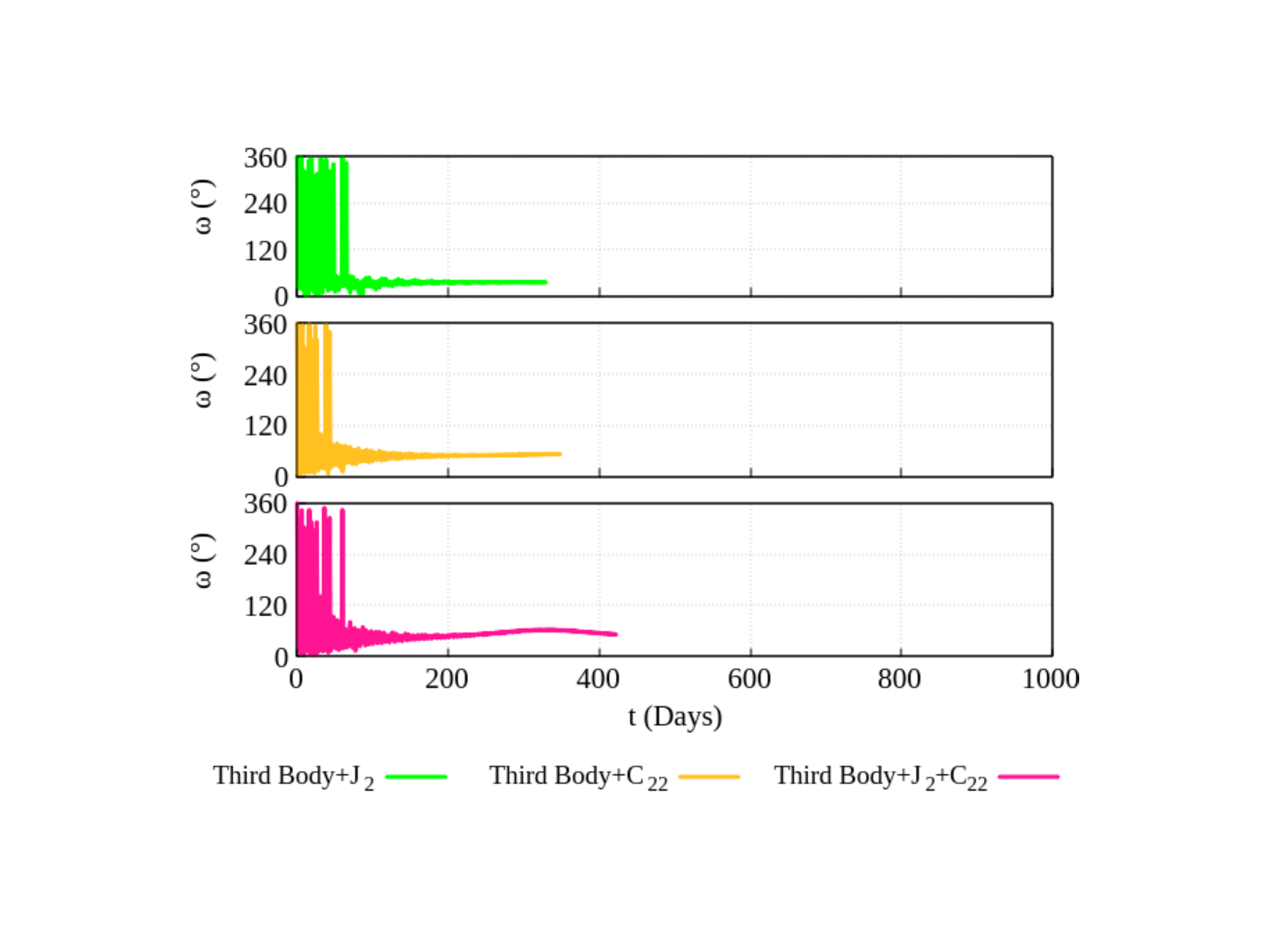}
\end{adjustwidth}
 \caption{Diagram of $\omega \times t$ showing different perturbative effects: considering only the perturbations by Uranus, only $J_2$ from Titania, only $C_{22}$ from Titania, Uranus $+ J_2$ from Titania, Uranus $+ C_{22}$ from Titania, and Uranus $+ J_2 + C_{22}$ from Titania. Initial values are $a_i=900$ km, $e_i=10^{-3}$, $I_i= 80^\circ$, $\omega_i=\Omega_i=0^\circ$.}
\label{fig:10}
\end{figure}

In this section, we present a study on the isolated effect of each perturbation on the evolution of some orbital elements of a probe for some specific cases. The perturbations considered here are: the gravitational effect of Uranus, the gravity coefficient $J_2$ from Titania, the ellipticity of Titania ($C_{22}$), Uranus + $J_2$ from Titania, Uranus + $C_{22}$ of Titania, and Uranus + $J_2$ + $C_{22}$ of Titania. This analysis is presented in Figures \ref{fig:9} and \ref{fig:10}, where the temporal evolution of the orbital elements $e$, $I$, and $\omega$ are shown. In all cases analyzed, the numerical integration stops when the probe collides with the surface of Titania.

Figure \ref{fig:9}a shows the temporal evolution of the eccentricity under the effect of these perturbations. It can be observed that the effect of the $J_2$ and $C_{22}$ gravity coefficients of Titania are small compared to the effect caused by the third body. This is seen even when the other perturbations are added. When $J_2$ and $C_{22}$ are included in the system, the lifetime increases, as already noted; however, the eccentricity continues to increase due to the presence of the third body. 

In the circular restricted three body problem, the Kozai-Lidov effect is common in some cases. This phenomenon is directly related to the presence of the third body, which, as shown in Figure \ref{fig:9}, causes the eccentricity of the probe to reach high values in a short period of time. The Kozai--Lidov mechanism is still capable of making the argument of periapsis librate around a constant value. These values are usually $90^\circ$ or $270^\circ$. However, in cases where the eccentricity grows rapidly, as is the case of the present work, a region of libration is not found for the diagram $e \times \omega$ (e.g., see \citep{Carvalho2012}). When $ J_2$ and $C_ {22} $ of Titania are included, there is a balance between the effects caused by the gravity coefficients of Titania and those caused by the Kozai--Lidov effect. This balance causes a slow increase in the eccentricity and, consequently, the lifetime of the orbit increases.

Regarding the evolution of the inclination, shown in Figure \ref{fig:9}b, the third body and the $J_2$ coefficient do not cause large variations in the probe inclination. The variation caused by $J_2$ is small and does not exceed $1^\circ$, although the lifetime is 1000 days. When Uranus is considered, the variation is small and the lifetime is just 300 days. When only Uranus and $J_2$ are taken into account, it is noted that the presence of the third body reduces lifetime compared to the case when only the $J_2$ coefficient was considered.

The effect responsible for causing large variations in the inclination is the tesseral coefficient $C_{22}$. When we analyze the dynamics only assuming $C_{22}$, the inclination increases from $80^\circ$ to $100^\circ$ degrees and the lifetime is 1000 days. In all cases in which $C_{22}$ is considered, the variation of $I$ is large. This can be seen in the cases Uranus $+ C_{22}$ and Uranus $+ J_2 + C_{22}$, in which the variation of $I$ can reach $30^\circ$. This effect caused by $C_{22}$ is expected, as shown in previous work \citep{Carvalho2012,Tzirt2009,Tzirt2010,Saedeleer2006}.

In the analysis of the argument of periapsis, presented in Figure \ref{fig:10}, it is possible to note that, as well as in the evolution of the eccentricity shown in Figure \ref{fig:9}a, the third body is responsible for causing a greater perturbation in this element. The behaviour of $\omega$ is similar in all cases where Uranus is added. The argument undergoes a great variation in the first days, with oscillations ranging from $0^\circ$ to $360^\circ$. After a certain time, the argument of periapsis begins to librate around $\sim$50\textdegree~for a short period of time. An expected behaviour for $\omega$ would be its libration around $\pi/2$ or $3\pi/2$, as described by Kozai--Lidov \cite{Kozai1962,Lidov1962}. However, as the short period terms that contain the $C_{22}$ coefficient were not eliminated, the inclination grows very quickly and, as the eccentricity depends on the slope, it also reaches high values in a few days. 

When only the effects of $J_2$ of Titania are assumed, the argument of periapsis reaches $360^\circ$ and then decays to approximately $240^\circ$ until a collision occurs in 1000 days. When only the perturbation due to $C_{22}$ is considered, the lifetime of the argument of periapsis is also 1000 days; however, the argument of periapsis increases to approximately $360^\circ$ before the collision.

\begin{figure}[H]
\begin{adjustwidth}{-\extralength}{0cm}
  \includegraphics[width=8.9cm]{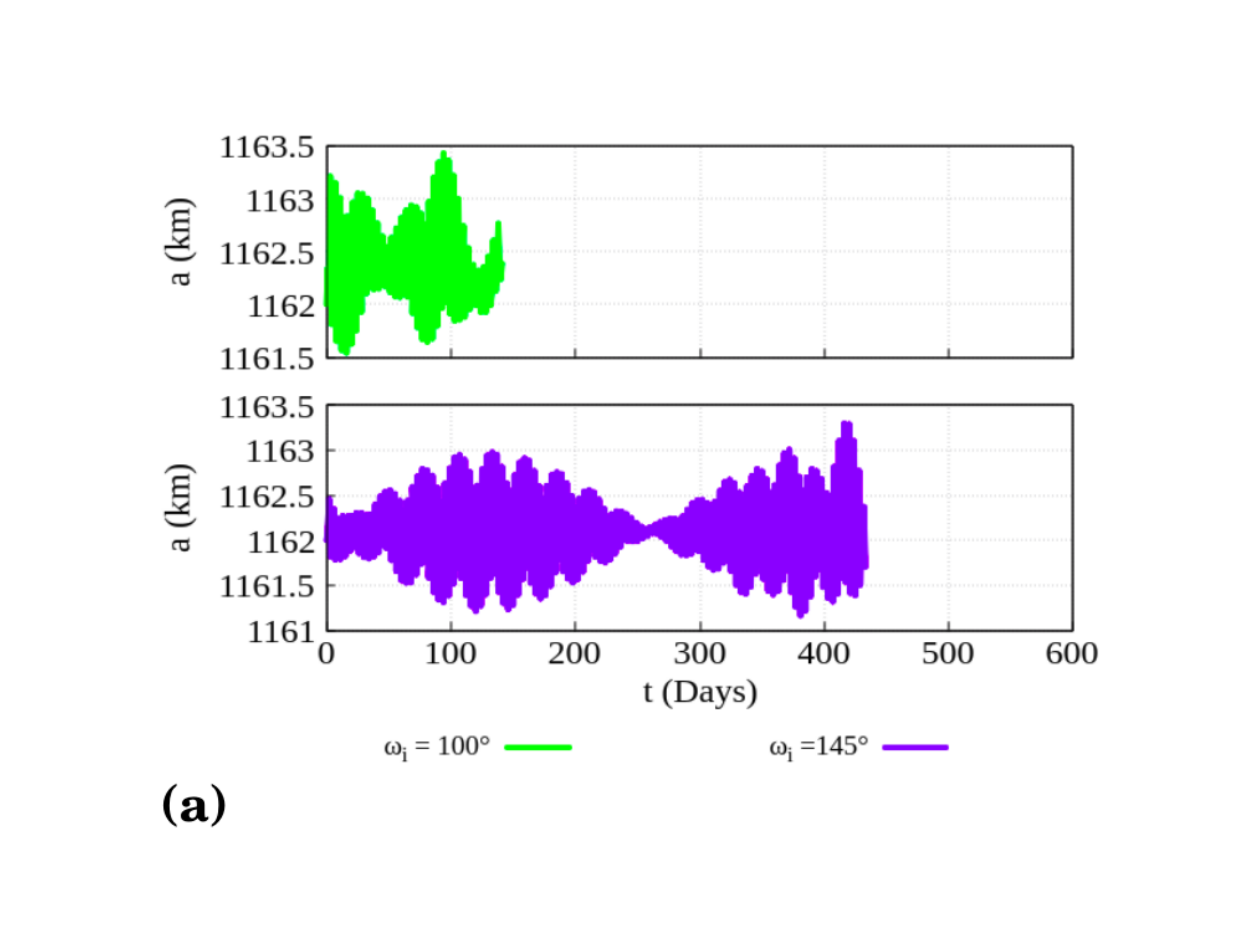}
  \includegraphics[width=8.9cm]{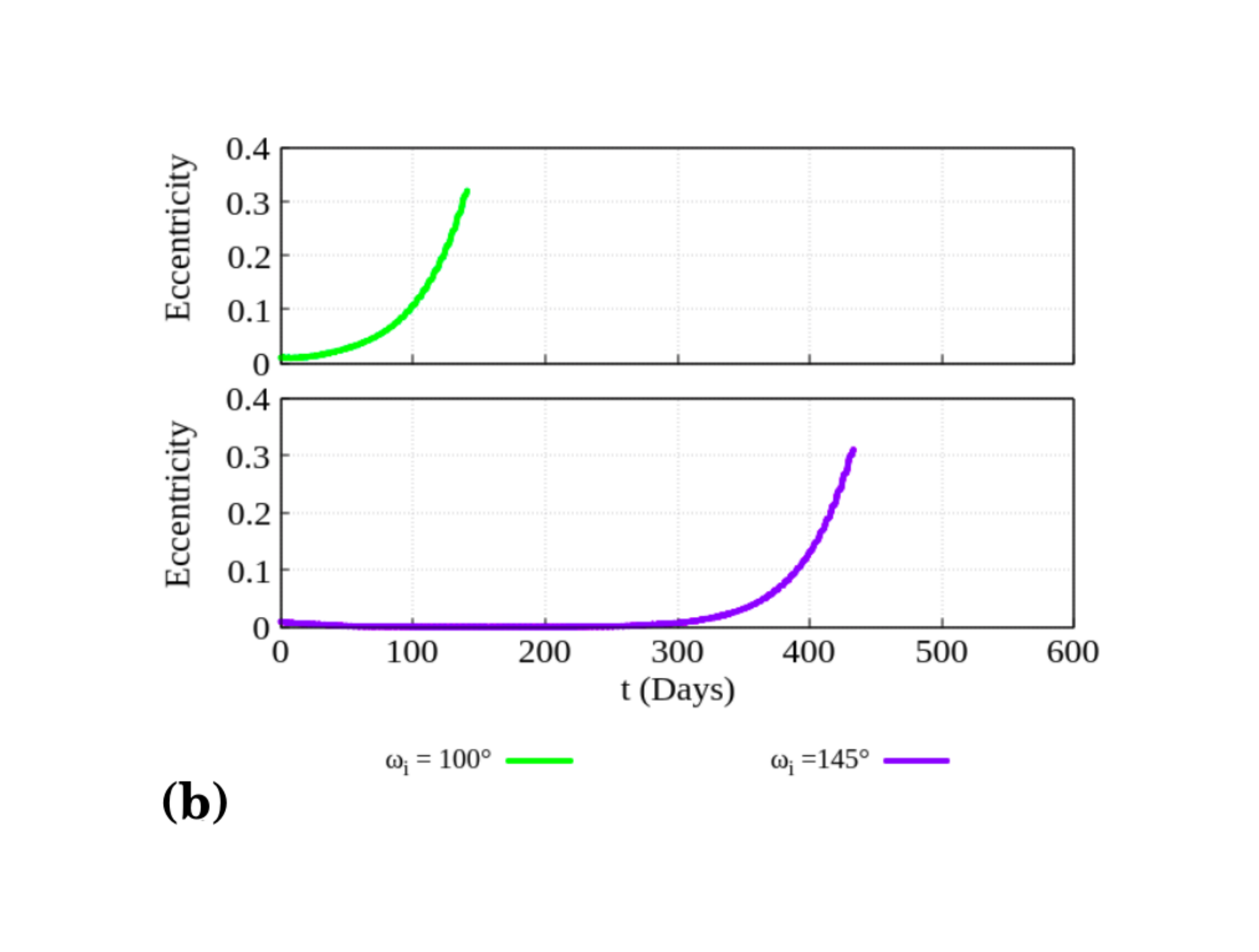}\\
\includegraphics[width=8.9cm]{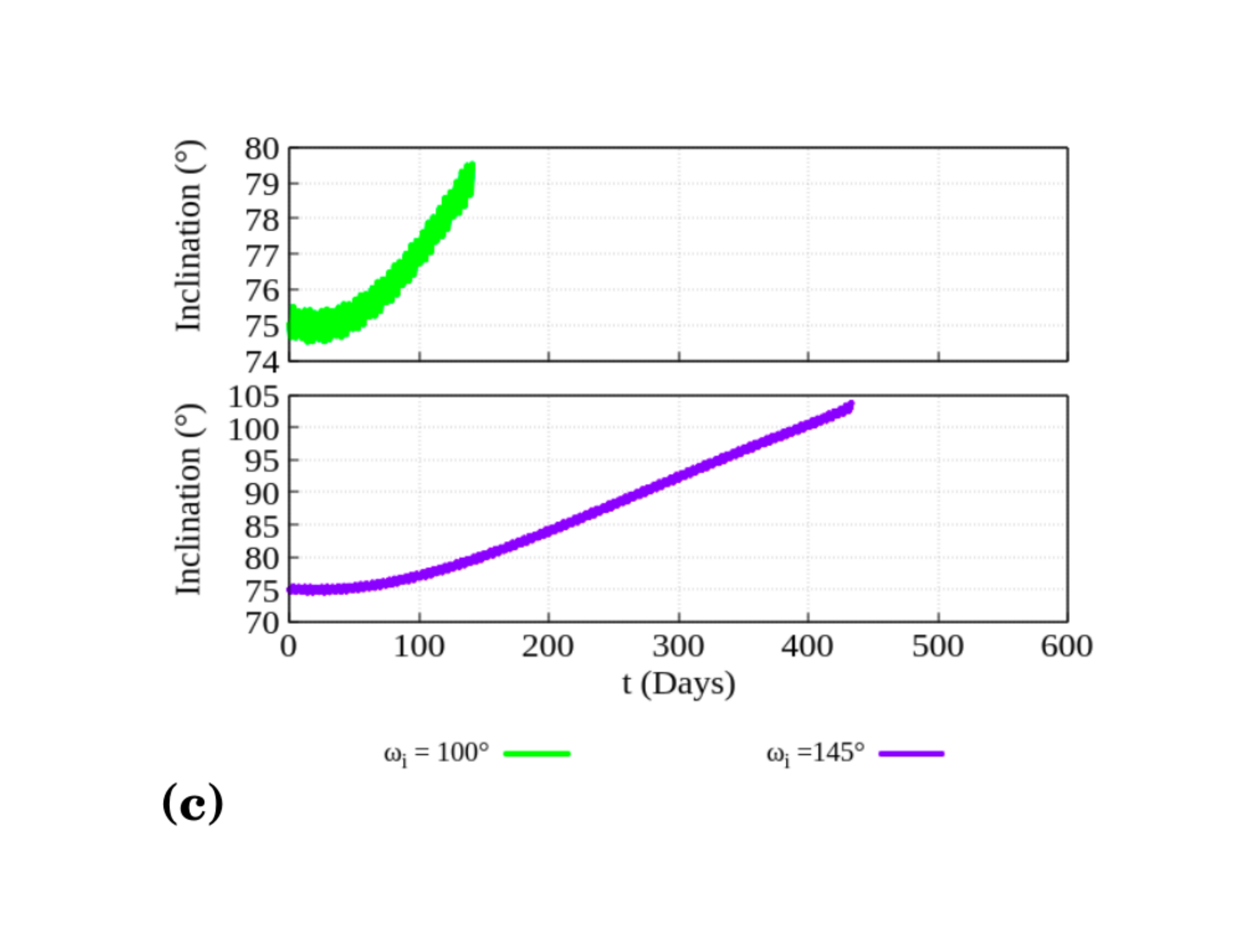}\\
 \includegraphics[width=8.9cm]{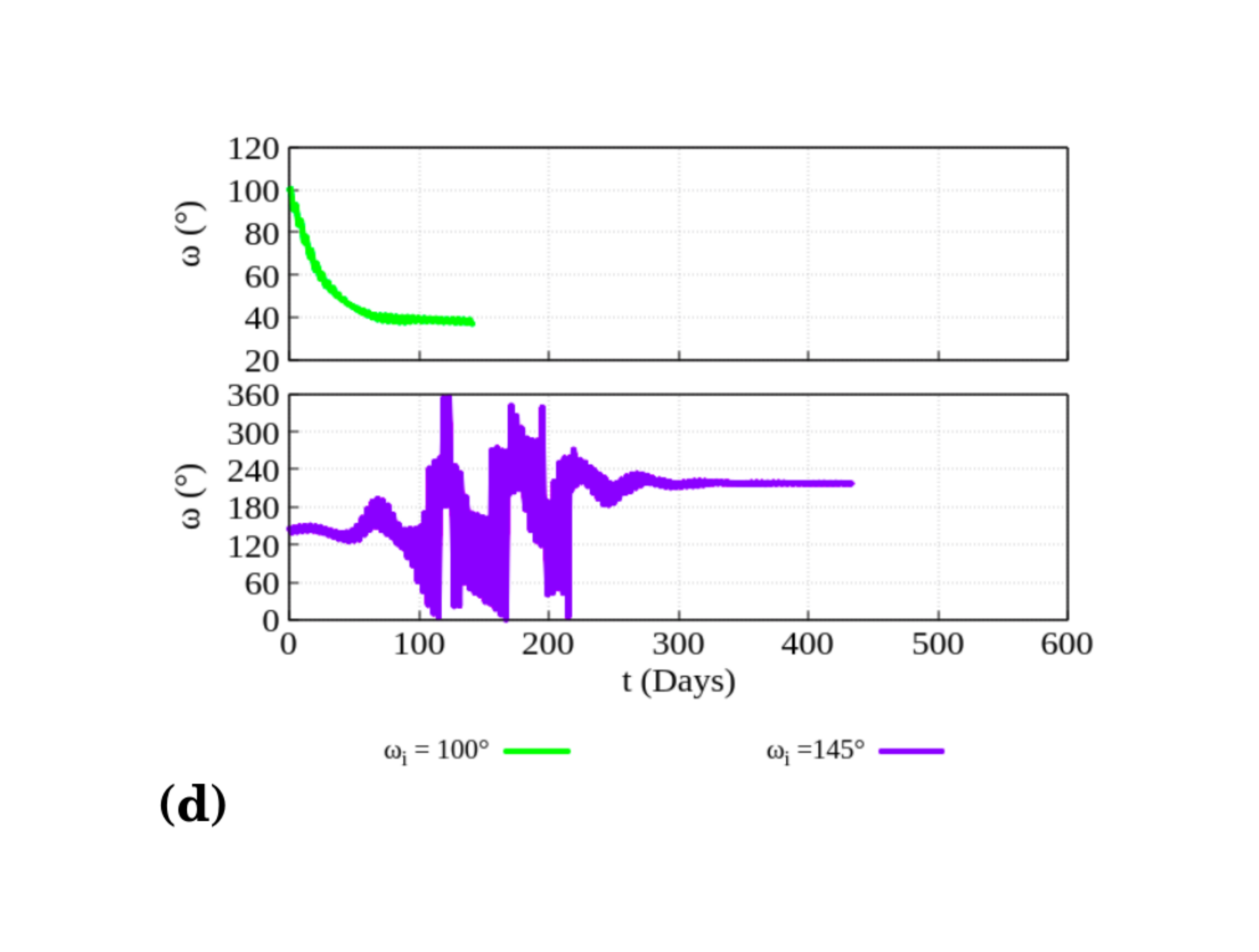}
 \includegraphics[width=8.9cm]{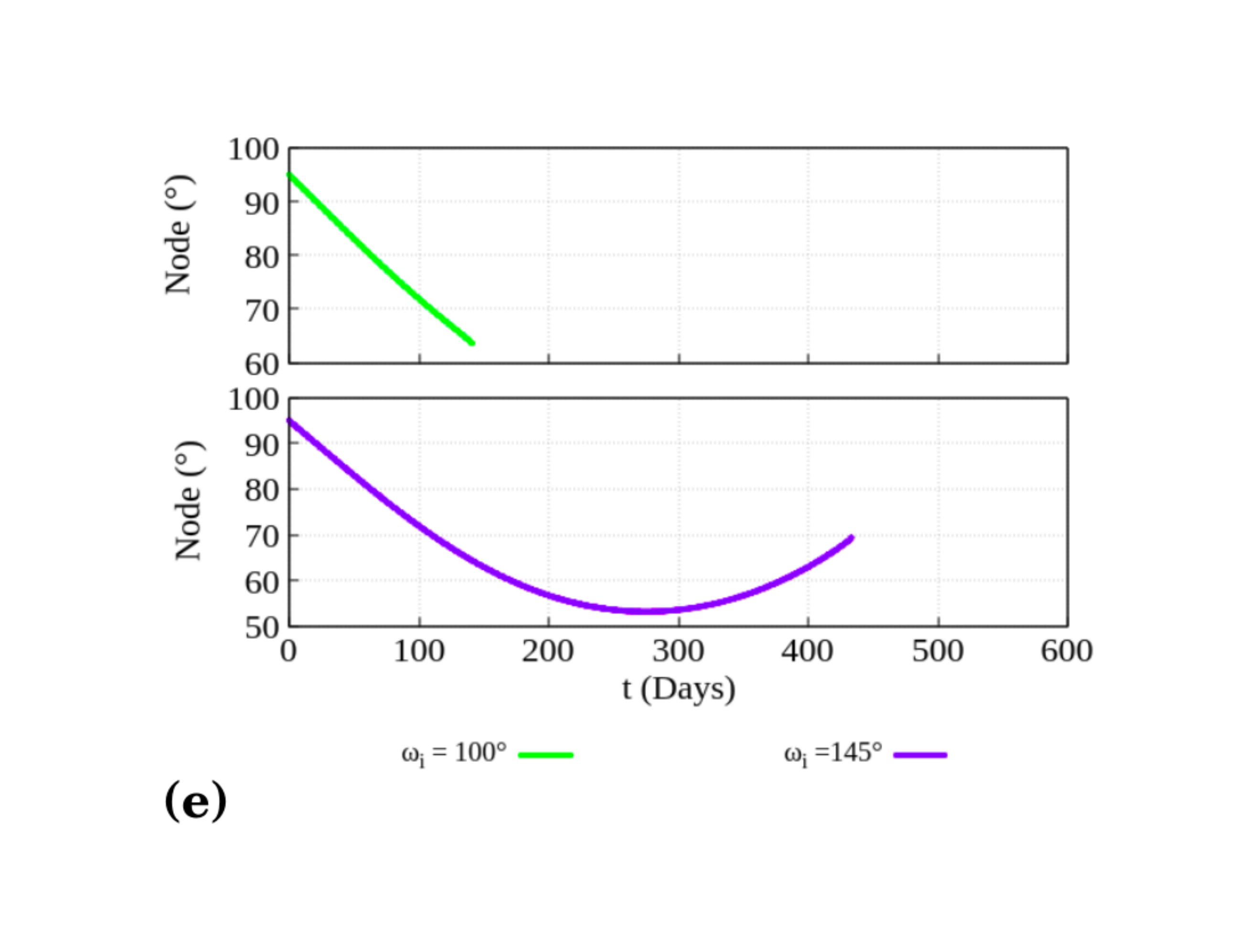}
\end{adjustwidth}
 \caption{Temporal evolution of the orbital elements considering $\omega=100^\circ$ and $\omega=145^\circ$: (\textbf{a}) $a \times t$, (\textbf{b}) $e \times t$, (\textbf{c}) $I \times t$, (\textbf{d}) $\omega \times t$, and (\textbf{e}) $\Omega \times t$. Initial conditions are $a_i=1162$ km, $e_i=10^{-2}$, $I_i=75^\circ$, $\Omega_i=95^\circ$. In these figures the effects considered are the perturbation by Uranus, $+J_2$ and $C_{22}$ of Titania.}
\label{fig:11}
\end{figure}

The relevance of the $\omega$ and $\Omega$ angles have already been evident in the lifetime maps shown in Figures \ref{fig:1}--\ref{fig:8}. In Figure \ref{fig:11}, an analysis of the influence of these angles on the orbital elements of two orbits is presented. The orbital elements analyzed in Figure \ref{fig:11} have different values of $\omega $. The first adopted value of $\omega$ and the other values of the orbital elements were taken from the point highlighted in Figure \ref{fig:7} (black circle). 

The second orbit has the same initial conditions as the highlighted initial condition, except $\omega$. The value of $\omega$ for the highlighted point is $145^\circ$. The use of this value is able to significantly increase the lifetimes of the orbits. For the sake of comparison, a value of $\omega$ just below the circled point was chosen, $\omega=100^\circ$.

In all cases shown in Figure \ref{fig:11}, the value of $\omega=145^\circ$ causes an increase in the lifetimes of the probe. The variation of the semi-major axis, shown in Figure \ref{fig:11}a, is very small in both cases. For $\omega=100^\circ$ and $\omega=145^\circ$, the variations are no more than 1 km. For the eccentricity, shown in Figure \ref{fig:11}b, the value of $\omega=145^\circ$ attenuates its rapid growth. It remains almost constant for 300 days and only then begins to grow, reaching a maximum of 0.3 in 430 days. Using $\omega=100^\circ$, the eccentricity increases much faster, and in just 135~days, the collision with the surface of Titania occurs.

The evolution of the inclination, shown in Figure \ref{fig:11}c, presents a variation of $30^\circ$ when $\omega =145^\circ$, but has a long lifetime, 430 days. Using $\omega =100^\circ$, the variation is only $5^\circ$, but the collision happens in a few days.

In the case of the evolution of the argument of periapsis, Figure \ref{fig:11}d, the behaviour is similar to the one shown in Figure \ref{fig:11} in the cases where the third body was considered. For both $\omega =100^\circ $ and $\omega=145^\circ$, the argument of periapsis oscillates around a constant value for some time. When we consider $\omega=100^\circ$, the argument decays and librates around $40^\circ$ for approximately 80 days. For $\omega=145^\circ$, the argument of periapsis circulates for about 200 days and then proceeds to librate around $238^\circ$ for almost 230 days. This behaviour is expected due to the aforementioned Kozai--Lidov mechanism.

In Figure \ref{fig:11}e, the evolution of the ascending node is presented for $\omega=100^\circ$ and $\omega=145^\circ$. For $\omega=100^\circ $, the ascending node decays almost $25^\circ $ in 140 days, whereas for $\omega=145^\circ$ this decay is almost $45^\circ$ during the first 250 days of integration. After that, the node starts to describe a periodic behaviour until the collision occurs.

In order to gather all the best orbits investigated, a summary of the best lifetimes for each initial condition and for each case analyzed is presented in Table \ref{tab:3}. This table presents the best initial conditions for the five values of the eccentricity studied in this work. In addition to the results for $\omega=\Omega=0^\circ$, the results considering $\omega$ and $\Omega$ different from $0^\circ$ are also presented. The table shows the best initial conditions obtained for long-duration orbits, with these conditions given by: $a=826$~km, $e=10^{-3}$, $I=86.2^\circ$, $\omega=165^\circ$, and $\Omega= 205^\circ$ considering only the third body perturbation. For the system, third-body $+ J_2 ,C_{22}$ of Titania, we have $a=874$~km, $e=10^{-3}$, $I=80^\circ$, $\omega= 165^\circ$, and $\Omega= 205^\circ$. In both cases, the orbits have a lifetime of 1000 days.

\begin{table}[H]
  \caption{Best lifetime for each system analyzed.}
 \label{tab:3}
 \begin{adjustwidth}{-\extralength}{0cm}
\newcolumntype{C}{>{\centering\arraybackslash}X}
\begin{tabularx}{\fulllength}{>{\centering}m{5cm}CCCCCC}
\toprule
 {\bf System} &  \boldmath{$a_i$} {\bf (km)} & \boldmath{$e_i$} & \boldmath{$I_i~(^\circ)$} & \boldmath{$\omega_i~(^\circ)$} & \boldmath{$\Omega_i~(^\circ)$} & {\bf Optimum  Lifetime (Days)}   \\
\midrule
Third-body                              & 858 & $0$ & 75 & 0 & 0 & 365  \\

Third-body + $J_2$,$C_{22}$ of Titania & 842  & $0$ & 75  & 0 & 0 & 450 \\

Third-body                             & 842  & $10^{-4}$ & 75  & 0 & 0 & 405  \\

Third-body + $J_2$,$C_{22}$ of Titania & 842  & $10^{-4}$ & 75  & 0 & 0 & 440 \\

Third-body                             & 1282 & $10^{-3}$ & 88.2  & 0 & 0 & 435  \\

Third-body + $J_2$,$C_{22}$ of Titania & 1242  & $10^{-3}$ & 86.6  & 0 & 0 & 500 \\

Third-body                             & 978  & $10^{-2}$ & 75  & 0 & 0 & 160  \\

Third-body + $J_2$,$C_{22}$ of Titania & 922  & $10^{-2}$ & 75.8  & 0 & 0 & 180 \\

Third-body                             & 1178  & $10^{-1}$ & 75  & 0 & 0 & 57  \\

Third-body + $J_2$,$C_{22}$ of Titania & 1266  & $10^{-1}$ & 75  & 0 & 0 & 57  \\

Third-body                              & 826  & $10^{-4}$ & 75  & 155 & 55 & 720  \\
 
Third-body + $J_2$,$C_{22}$ of Titania & 922  & $10^{-4}$ & 82.2  & 155 & 55 & 800  \\

Third-body                              & 826  & $10^{-3}$ & 86.2  & 165 & 205 & 1000  \\
 
Third-body +  $J_2$,$C_{22}$ of Titania & 874  & $10^{-3}$ & 80  & 165 & 205 & 1000  \\

Third-body                              & 1282  & $10^{-2}$ & 88.2  & 144.9 & 95 & 450  \\
 
Third-body + $J_2$,$C_{22}$ of Titania & 1186  & $0^{-2}$ & 75  & 145 & 95 & 450  \\

Third-body                              & 970  & $10^{-1}$ & 89.8  & 150 & 115 & 120  \\
 
Third-body + $J_2$,$C_{22}$ of Titania & 938  & $10^{-1}$ & 88.6  & 150 & 115 & 520  \\
\bottomrule
\end{tabularx}
\end{adjustwidth}
\end{table}

\begin{figure}[H]
\begin{adjustwidth}{-\extralength}{0cm}
\includegraphics[width=6.1cm]{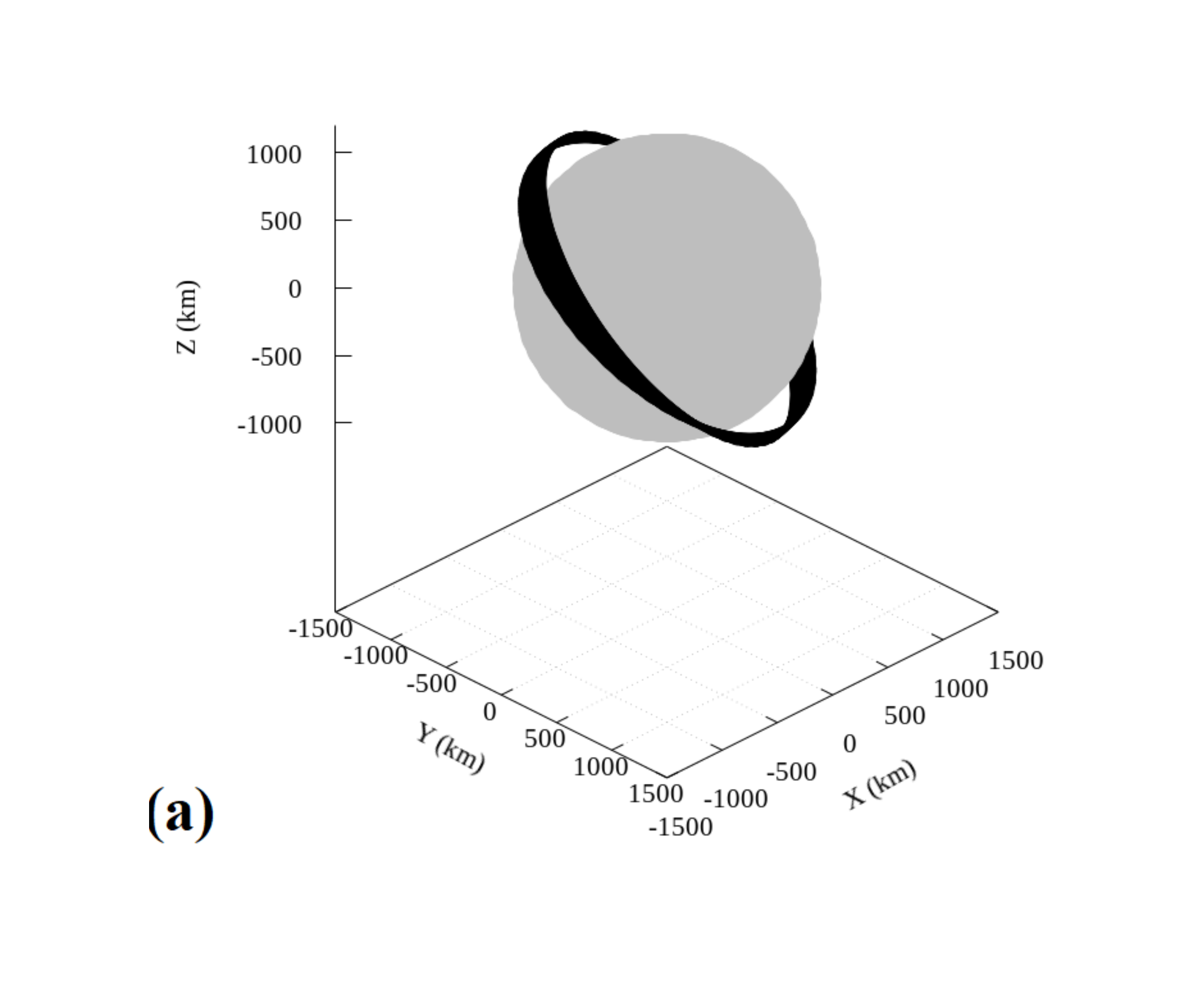}
\includegraphics[width=6.1cm]{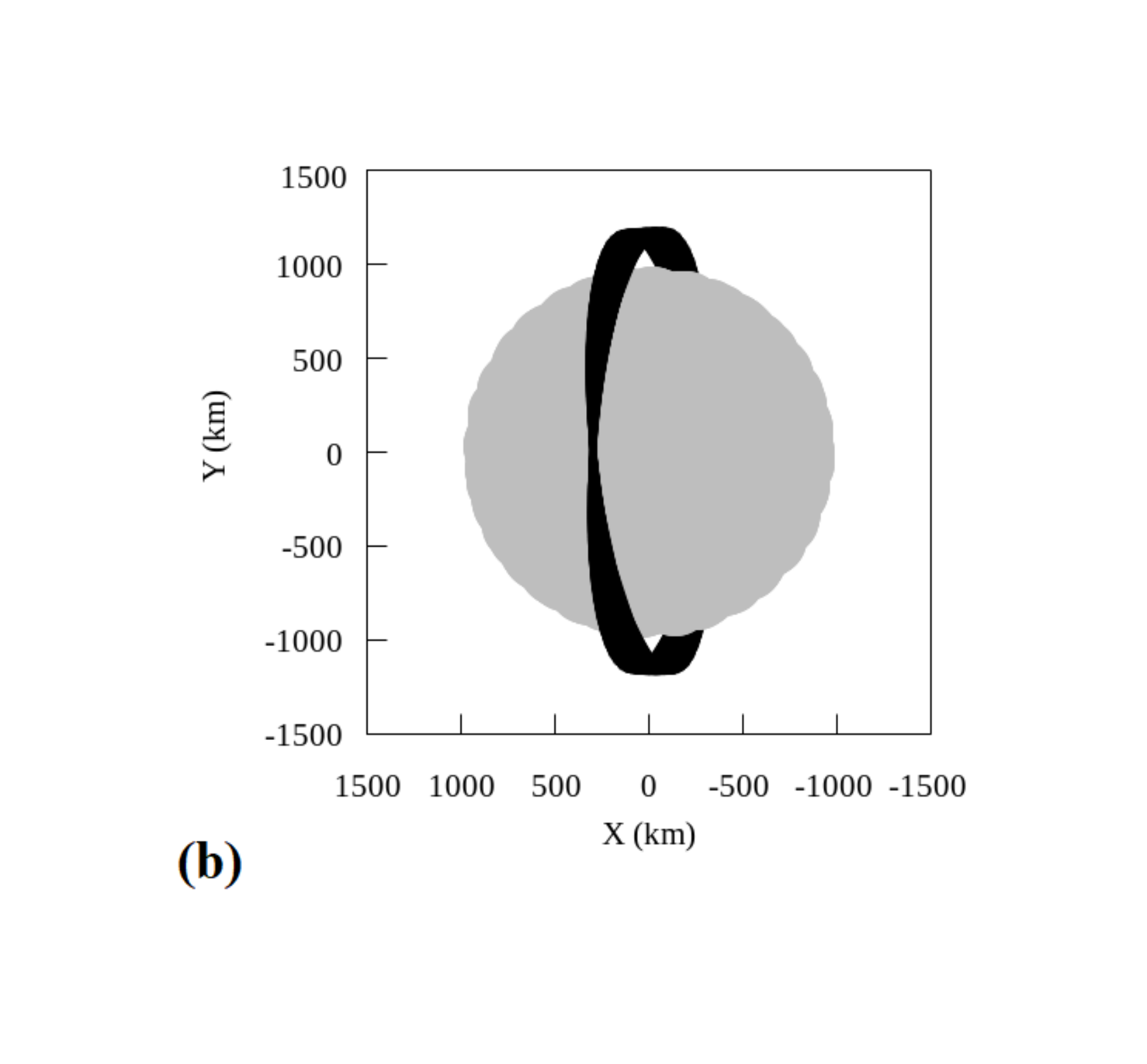}
\includegraphics[width=6.1cm]{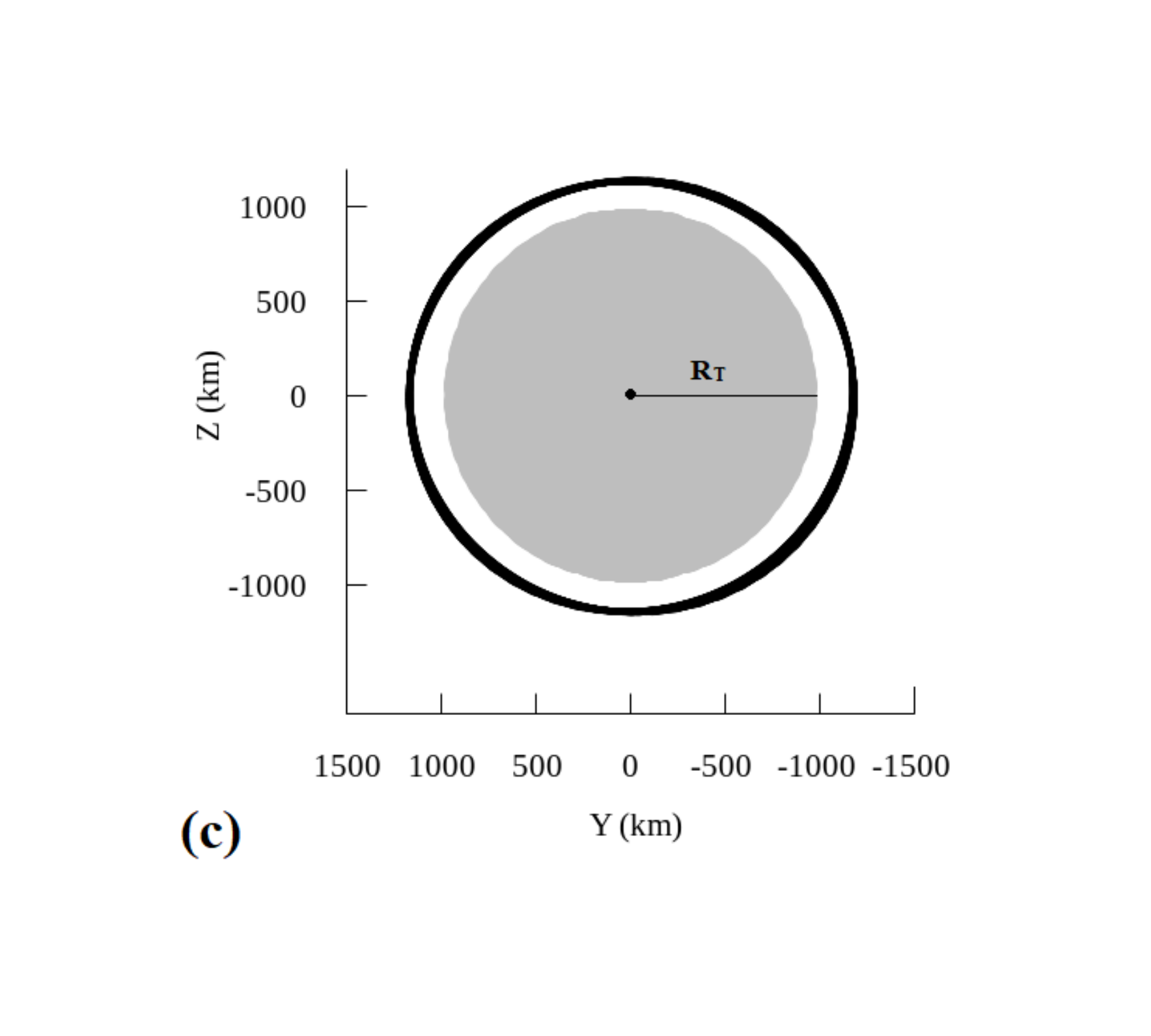}
\end{adjustwidth}
 \caption{Trajectory numerically simulated for 991 days, considering the complete system, the gravitational effects of the third body (Uranus), and Titania’s gravitational coefficients ($J_2$ and $C_{22}$). (\textbf{a}) Projection of the orbit in three dimensions ($xyz$). (\textbf{b}) Projection of the orbit in the plane ($xy$). \linebreak (\textbf{c}) Projection of the orbit in the plane ($yz$). The initial conditions are: $a=898$~km, $e=10^{-3}$, $I= 81.4^\circ$, $=140^\circ$, and $\Omega= 225^\circ$.}
\label{fig:12}
\end{figure}

In Figure \ref{fig:12} we present different projections of one of the orbits with the best initial conditions found in this work. The trajectory was numerically simulated for 991 days, considering the complete system, the gravitational effects of the third body (Uranus), and Titania’s gravitational coefficients ($J_2$ and $C_{22}$). The initial conditions are: $a=898$~km, $e=10^{-3}$, $I=81.4^\circ$, $\omega=140^\circ$, and $\Omega= 225^\circ$.

\section{Analyzing the Gravitational Coefficients of Titania}
\label{sec:grav}

\begin{figure}[H]
\begin{adjustwidth}{-\extralength}{0cm}
\centering 
\includegraphics[width=8.5cm]{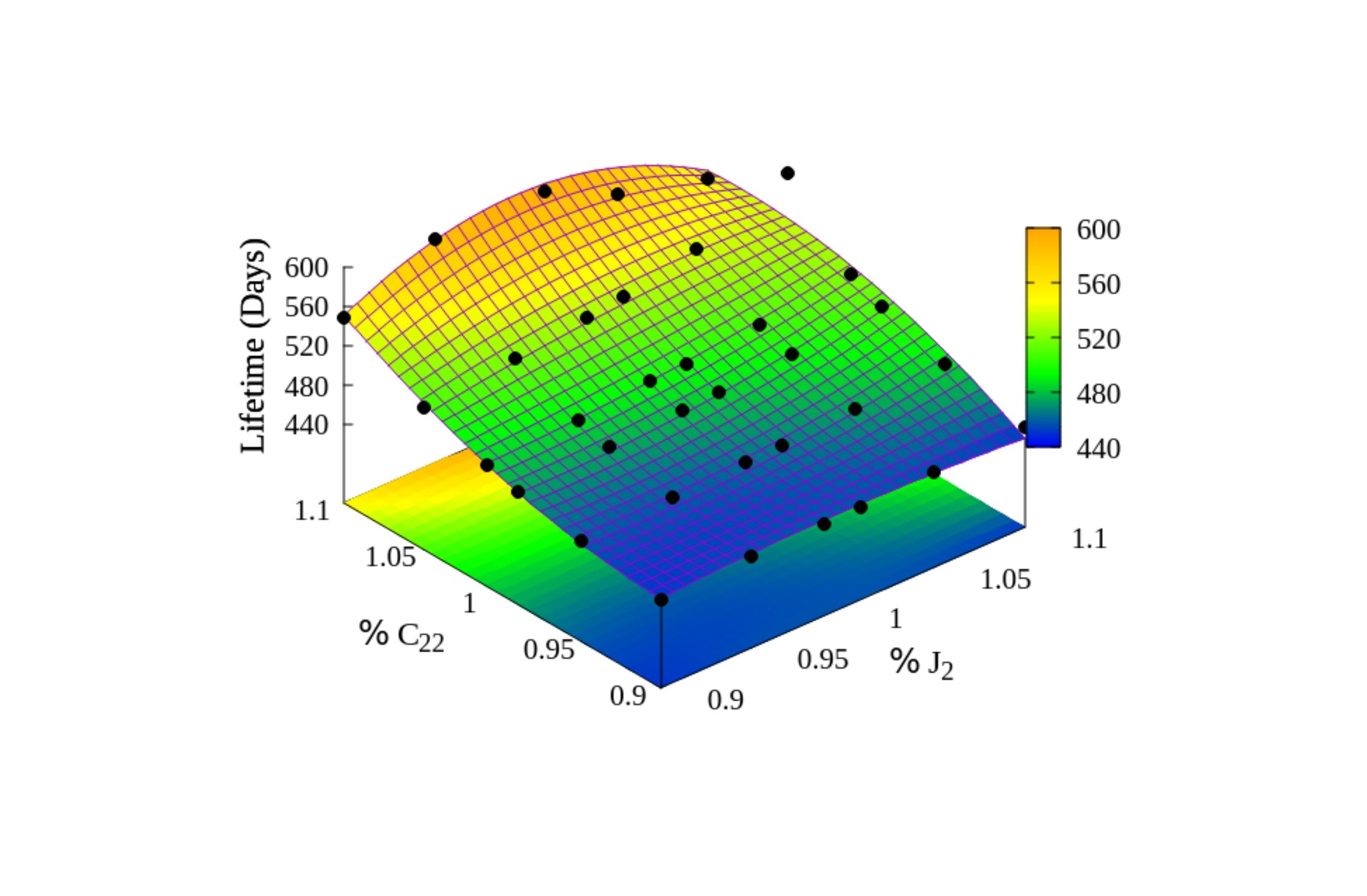}
\includegraphics[width=8.5cm]{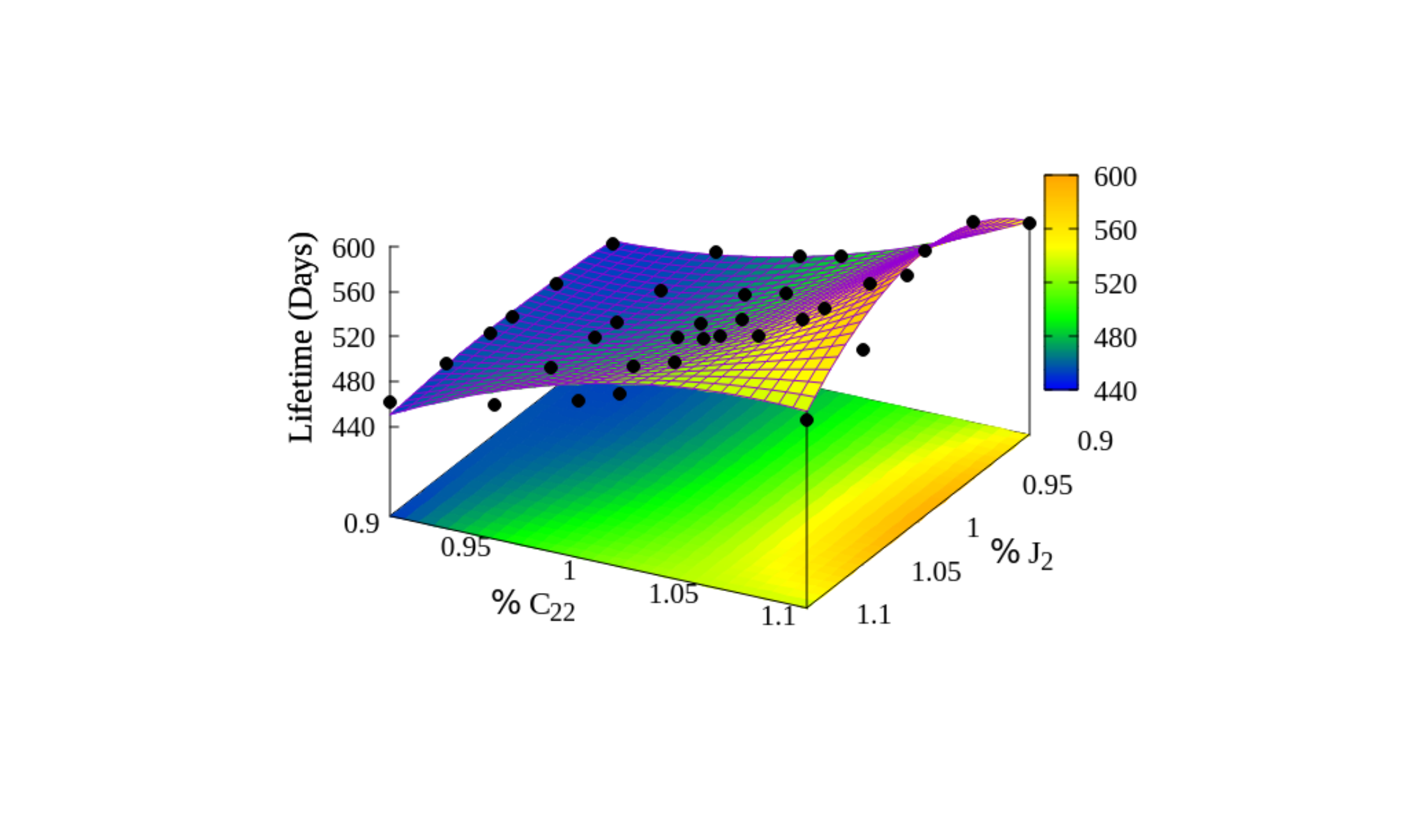}
\end{adjustwidth}
 \caption{Response surface for multiple regression with three variables. This figure represents the most complete model in two different angles, with $R^2=0.92$. This model is represented by \linebreak Equation~\eqref{eq:11}. Initial conditions are $a=1075$ km, $e=10^{-3}$, $I=76^\circ$, $\omega=0^\circ$, and $\Omega=0^\circ$.}
\label{fig:13}
\end{figure}

In order to analyze how possible errors in the values of Titania's gravitational coefficients can affect the probe's lifetime, our next results present a detailed analysis of possible error values in both coefficients. We also present multiple regression models in order to find a relationship as a function of the coefficients $J_2$ and $C_{22}$ capable of predicting error in the probe's lifetime. Figure \ref{fig:13} shows the response surface found for the more complete regression model that matches the equation.

We investigate a range from $-10\%$ to $10\%$ for $J_2$ and $C_{22}$. To find the noted function, we consider only the case where the lifetime variation is significant compared to the cases obtained from \cite{Chen2014}. The initial conditions for the regression were taken from the percentages analyzed for $e = 10^{-3}$, as it shows the largest increasing in the lifetime. We present four~multiple regression models for different values of the coefficient of determination $R^2$. The models and their fit coefficients are presented in Equations \eqref{eq:8}--\eqref{eq:11}.

\begin{equation}
 Y= 1230 +1469J_2 -684J_2^2 -3635C_{22} +2115C_{22}^2, \, \  (R^2=0.84)
\label{eq:8}
\end{equation}

 \begin{equation}
 \begin{split}
 Y=&24139 - 44921J_2+ 22699 J_2^2 -26544C_{22}\\ &+2115C_{22}^2 +4639J_2^2C_{22}-23382J_2^2C_{22}, \,\,\,\,\,\,\,\,\,\,\,\,\ (R^2=0.87)
\end{split}
\label{eq:9}
\end{equation}\\
 \begin{equation}
 \begin{split}
Y= & 48401-69183J_2+22699J_2^2 -75272C_{22}+26480C_{22}^2\\
& +95118J_2^2C_{22} -23383J_2^2C_{22}-24364J_2C_{22}^2, \,\,\,\,\,\,\,\,\  (R^2=0.89)
\end{split}
\label{eq:10}
\end{equation}\\
\begin{equation}
 \begin{split}
 Y=&-348975+728920J_2-376353J_2^2+722831C_{22}\\
 & -372572C_{22}^2-1507821J_2C_{22}+778087J_2^2C_{22}\\
& +777105C_{22}^2J_2+((-400735\cdot(J_2^2))\cdot(-400735(C_{22}^2)) \, \  (R^2=0.92)
 \end{split}
 \label{eq:11}
\end{equation}
 where $Y$ is the lifetime dependent variable, and $J_2$ and $C_{22}$ are independent variables. It is important to highlight that in the values of $J_2$ and $C_{22}$ were added percentages ranging from \textminus10\% to +10\% with respect to the nominal value.

Equation \eqref{eq:8} presents a relationship for the lifetime as a function the percentages of $J_2$ and $C_{22}$. For this model (Model 1), the coefficient of determination $R^2$ has a value of 0.84. Although the value of $R^2$ is not as close as to 1 and the model does not present an interaction term between $J_2$ and $C_{22}$, the function provides a good approximation of the lifetime as a function of the percentage values of Titania's gravitational coefficients.

For Equations \eqref{eq:9}--\eqref{eq:11}, the interactions between the percentages of the coefficients are analyzed. With a coefficient $R ^ 2 = 0.92 $, Equation~\eqref{eq:11} (Model 4) presents a better precision for estimating the lifetime as a function of $ J_2 $ and $ C_ {22} $. The graphical representation of this model is shown in Figure \ref{fig:13}.

\section{Final Comments}
\label{sec:final}

In this work we study the dynamics of orbits around the natural satellite Titania, the largest moon in the Uranus system. The gravitational attraction due to Uranus is considered, as well as the gravity coefficients $J_2$ and $C_{22}$ of Titania. Titania is at the center of the system, and Uranus is in an elliptical orbit and acts as the third body to perturb the motion of the space probe around Titania.

Through a set of numerical simulatons of the equations of motion, we build lifetime maps for different values of eccentricities of the orbit of the space probe. The construction of these maps allowed us to find values of $a$, $e $, $I$, $\omega$, and $\Omega$ capable of increasing the lifetime of the probe. We also constructed lifetime ``difference maps'' in order to analyze the regions where the effect of the third body and the gravity coefficients of Titania were in equilibrium, allowing longer life for the space probe.

We also present an analysis of the importance of the angles $\omega$ and $\Omega$ in the lifetime of the probe. In addition, we studied the response of each effect on the orbital element.

Our results showed that, for the altitudes adopted in this work, the oblateness coefficients $J_2$ and $J_4$ of Uranus do not affect the orbits of the space probe. We also showed that the orbits with a longer lifetime had a semi-major axis close to 900 km, whereas the best eccentricity is between $10^{- 3}$ and $10^{- 4}$ when considering $\omega=\Omega=0^\circ$. The results also point out that the inclusion of specific values of $\omega$ and $\Omega$ are extremely important in increasing the lifetime of the probe.  In all cases, the increase in the lifetime was more than 50\% after adopting specific values for these two angles. The best values for $\omega$ are around 145--165\textdegree~and for $\Omega$ around 55--205\textdegree. In some cases, including these values resulted in the probe lifetime increasing up to eight times compared to cases with $\omega=\Omega=0^\circ$.

Our main objective is to find long-duration orbits around Titania. After analyzing the results including the values of $\omega$ and $\Omega$, we found ideal initial conditions for an orbit lasting up to 1000 days. This orbit has initial orbital elements given by: $a=874$~km, $e=10^{-3}$, $I=80^\circ$, $\omega=165^\circ$, and $\Omega=205 ^\circ$.

Analyzing the effect of each perturbation, we show that the third body is responsible for causing a greater perturbation in the eccentricity and argument of periapsis, whereas the ellipticity coefficient $C_{22}$ of Titania causes great oscillations in the inclination. The zonal coefficient $J_2$, when acting alone, has little influence on the orbital elements; however, when added to the other perturbations, it helps to promote a balance between third body and gravity coefficients, and then we have the so-called ``protection mechanism'' prolonging the lifetimes of the orbits.

The gravitational coefficients of Titania have not yet been used to investigate orbits close to its surface. Thus, we present an original study on the sensitivity of the lifetime as a function of some possible errors attributed to the values of $J_2$ and $C_{22}$. Our results show that the hypothetical errors in the values of Titania's gravitational coefficients can greatly increase or decrease the lifetime of the orbit, especially when the eccentricity is equal to $10^{-3}$. To try to predict the lifetime as a function of these errors, we present multiple regression models for the lifetime as a function of the coefficients $J_2$ and $C_{22}$. We found a function with a value of $R^2$ equals to $0.92$, capable of providing a good approximation of the lifetime for values of $J_2$ and $C_{22}$ between $-10\%$ and $+10\%$ of the values given in \cite{Chen2014}.

The results presented in this paper can help plan missions to Titania, which can bring important scientific data, in particular considering that the Uranus system has been studied mostly from Earth.





\vspace{6pt} 



\authorcontributions{Conceptualization, J.X., A.B.P., and S.G.W.; methodology, J.X. and A.A.; formal analysis, J.X., A.B.P., S.G.W., and A.A.; writing—preparation of original draft, J.X., A.B.P., and S.G.W.; writing—review and editing, J.X., A.B.P., and S.G.W.; visualization, J.X., A.B.P., and S.G.W.; supervision, A.B.P. and S.G.W. All authors read and agreed with the published version of the manuscript.}

\funding{Improvement Coordination Higher Education Personnel---Brazil (CAPES)---Financing Code 001. The Center for Mathematical Sciences Applied to Industry (Ce-MEAI), funded by FAPESP (grant 2013/07375-0) and the project 2016/23542-1 from FAPESP. CNPq (Proc 313043/2020-5). This paper has been supported by the RUDN University Strategic Academic Leadership Program.}

\institutionalreview{Not applicable.}

\informedconsent{Not applicable.}

\dataavailability{All data generated or analyzed during this study are included in this published article in the form of figures.} 

\acknowledgments{
The authors thank Improvement Coordination Higher Education Personnel---Brazil (CAPES)---Financing Code 001. The Center for Mathematical Sciences Applied to Industry (Ce-MEAI), funded by FAPESP (grant 2013/07375-0) and the project 2016/23542-1 from FAPESP. SMGW thanks CNPq (Proc 313043/2020-5) for the financial support. This paper has been supported by the RUDN University Strategic Academic Leadership Program.}

\conflictsofinterest{The authors declare that they have no conflicts of interest with research institutions, professionals, researchers, and/or financial supports.} 

\ethicalapproval{The submitted work is original and has not been published elsewhere in any form or language. The work presents the results of a single study. The results are presented clearly, honestly, and without fabrication, falsification, or inappropriate data manipulation. No data, text, or theories by others are presented as if they were the authors' own.}

\begin{adjustwidth}{-\extralength}{0cm}

\reftitle{References}

\end{adjustwidth}

\begin{thebibliography}{999}

\bibitem[{Jarmak} \em{et~al.}(2020){Jarmak}, {Leonard}, {Akins}, {Dahl},
  {Cremons}, {Cofield}, {Curtis}, {Dong}, {Dunham}, {Journaux}, {Murakami},
  {Ng}, {Piquette}, {Girija}, {Rink}, {Schurmeier}, {Stein}, {Tallarida},
  {Telus}, {Lowes}, {Budney}, and {Mitchell}]{Jarmak2020}
{Jarmak}, S.; {Leonard}, E.; {Akins}, A.; {Dahl}, E.; {Cremons}, D.R.;
  {Cofield}, S.; {Curtis}, A.; {Dong}, C.; {Dunham}, E.T.; {Journaux}, B.;
  et~al.
\newblock {QUEST: A New Frontiers Uranus orbiter mission concept study}.
\newblock {\em Acta Astronaut.} {\bf 2020}, {\em 170},~6--26.
\newblock
  https://doi.org/{\changeurlcolor{black}\href{https://doi.org/10.1016/j.actaastro.2020.01.030}{\detokenize{10.1016/j.actaastro.2020.01.030}}}.

\bibitem[Cartwright \em{et~al.}(2021)Cartwright, Beddingfield, Nordheim, Elder,
  Castillo-Rogez, Neveu, Bramson, Sori, Buratti, Pappalardo, Roser, Cohen,
  Leonard, Ermakov, Showalter, Grundy, Turtle, and Hofstadter]{Cartwright2021}
Cartwright, R.J.; Beddingfield, C.B.; Nordheim, T.A.; Elder, C.M.;
  Castillo-Rogez, J.C.; Neveu, M.; Bramson, A.M.; Sori, M.M.; Buratti, B.J.;
  Pappalardo, R.T.;  et~al.
\newblock The science case for spacecraft exploration of the Uranian
  satellites: Candidate ocean worlds in an ice giant system.
\newblock {\em arXiv} {\bf 2021},  arXiv:2105.01164.

\bibitem[Hofstadter \em{et~al.}(2019)Hofstadter, Simon, Atreya, Banfield,
  Fortney, Hayes, Hedman, Hospodarsky, Mandt, Masters, Showalter, Soderlund,
  Turrini, Turtle, Reh, Elliott, Arora, and Petropoulos]{Hofstadter2019}
Hofstadter, M.; Simon, A.; Atreya, S.; Banfield, D.; Fortney, J.J.; Hayes, A.;
  Hedman, M.; Hospodarsky, G.; Mandt, K.; Masters, A.;  et~al.
\newblock Uranus and Neptune missions: A study in advance of the next Planetary
  Science Decadal Survey.
\newblock {\em Planet. Space Sci.} {\bf 2019}, {\em 177},~104680.
\newblock
  https://doi.org/{\changeurlcolor{black}\href{https://doi.org/https://doi.org/10.1016/j.pss.2019.06.004}{\detokenize{10.1016/j.pss.2019.06.004}}}.

\bibitem[Cartwright \em{et~al.}(2015)Cartwright, Emery, Rivkin, Trilling, and
  Pinilla-Alonso]{Cartwright2015}
Cartwright, R.; Emery, J.; Rivkin, A.; Trilling, D.; Pinilla-Alonso, N.
\newblock Distribution of CO2 ice on the large moons of Uranus and evidence for
  compositional stratification of their near-surfaces.
\newblock {\em Icarus} {\bf 2015}, {\em 257},~428--456.

\bibitem[Cartwright \em{et~al.}(2018)Cartwright, Emery, Pinilla-Alonso, Lucas,
  Rivkin, and Trilling]{Cartwright2018}
Cartwright, R.J.; Emery, J.P.; Pinilla-Alonso, N.; Lucas, M.P.; Rivkin, A.S.;
  Trilling, D.E.
\newblock Red material on the large moons of Uranus: Dust from the irregular
  satellites?
\newblock {\em Icarus} {\bf 2018}, {\em 314},~210--231.
\newblock
  https://doi.org/{\changeurlcolor{black}\href{https://doi.org/https://doi.org/10.1016/j.icarus.2018.06.004}{\detokenize{10.1016/j.icarus.2018.06.004}}}.

\bibitem[Fletcher \em{et~al.}(2020)Fletcher, Helled, Roussos, Jones, Charnoz,
  André, Andrews, Bannister, Bunce, Cavalié, Ferri, Fortney, Grassi, Griton,
  Hartogh, Hueso, Kaspi, Lamy, Masters, Melin, Moses, Mousis, Nettleman,
  Plainaki, Schmidt, Simon, Tobie, Tortora, Tosi, and Turrini]{Fletcher2020}
Fletcher, L.N.; Helled, R.; Roussos, E.; Jones, G.; Charnoz, S.; André, N.;
  Andrews, D.; Bannister, M.; Bunce, E.; Cavalié, T.;  et~al.
\newblock Ice Giant Systems: The scientific potential of orbital missions to
  Uranus and Neptune.
\newblock {\em Planet. Space Sci.} {\bf 2020}, {\em 191},~105030.
\newblock
  https://doi.org/{\changeurlcolor{black}\href{https://doi.org/https://doi.org/10.1016/j.pss.2020.105030}{\detokenize{10.1016/j.pss.2020.105030}}}.

\bibitem[{Gomes} and {Domingos}(2016)]{Gomes2016}
{Gomes}, V.; {Domingos}, R.d.C.
\newblock {Studying the lifetime of orbits around Moons in elliptic}.
\newblock {\em JO---Comput. Appl. Math.} {\bf 2016}, {\em
  35},~653--661.
\newblock
  https://doi.org/{\changeurlcolor{black}\href{https://doi.org/10.1007/s40314-015-0258-8}{\detokenize{10.1007/s40314-015-0258-8}}}.

\bibitem[{Cardoso} \em{et~al.}(2017){Cardoso}, {Carvalho}, {Prado}, and
  {Vilhena de Moraes}]{Cardoso2017}
{Cardoso}, J.; {Carvalho}, J.P.S.; {Prado}, A.F.B.A.; {Vilhena de Moraes}, R.
\newblock {Lifetime maps for orbits around Callisto using a double-averaged
  model}.
\newblock {\em Astrophys. Space Sci.} {\bf 2017}, {\em 362},~227.
\newblock
  https://doi.org/{\changeurlcolor{black}\href{https://doi.org/10.1007/s10509-017-3200-2}{\detokenize{10.1007/s10509-017-3200-2}}}.

\bibitem[{Carvalho} \em{et~al.}(2012){Carvalho}, {Elipe}, {Vilhena de Moraes},
  and {Prado}]{Carvalho2012}
{Carvalho}, J.P.S.; {Elipe}, A.; {Vilhena de Moraes}, R.; {Prado}, A.F.B.A.
\newblock {Low-altitude, near-polar and near-circular orbits around Europa}.
\newblock {\em Adv. Space Res.} {\bf 2012}, {\em 49},~994--1006.
\newblock
  https://doi.org/{\changeurlcolor{black}\href{https://doi.org/10.1016/j.asr.2011.11.036}{\detokenize{10.1016/j.asr.2011.11.036}}}.

\bibitem[Carvalho \em{et~al.}(2012)Carvalho, Mourão, Elipe, Vilhena~de Moraes,
  and Prado]{Carvalho2012_2}
Carvalho, J.; Mourão, D.; Elipe, A.; Vilhena~de Moraes, R.; Prado, A.
\newblock Frozen orbits around Europa.
\newblock {\em Int. J. Bifurc. Chaos Appl. Sci. Eng.} {\bf 2012}, {\em 22}, 12502409.
\newblock
  https://doi.org/{\changeurlcolor{black}\href{https://doi.org/10.1142/S0218127412502409}{\detokenize{10.1142/S0218127412502409}}}.

\bibitem[{Nie} and {Gurfil}(2018)]{Tao2018}
{Nie}, T.; {Gurfil}, P.
\newblock {Lunar frozen orbits revisited}.
\newblock {\em Celest. Mech. Dyn. Astron.} {\bf 2018}, {\em
  130},~61.
\newblock
  https://doi.org/{\changeurlcolor{black}\href{https://doi.org/10.1007/s10569-018-9858-0}{\detokenize{10.1007/s10569-018-9858-0}}}.

\bibitem[Elipe and Lara(2003)]{Lara2003}
Elipe, A.; Lara, M.
\newblock Frozen Orbits About the Moon.
\newblock {\em Pre-Publ. Sem. Mat. ``García de
  Galdeano"} {\bf 2003}, {\em 26}, 200320.
\newblock
  https://doi.org/{\changeurlcolor{black}\href{https://doi.org/10.2514/2.5064}{\detokenize{10.2514/2.5064}}}.

\bibitem[{Abd El-Salam} and {Abd El-Bar}(2016)]{Abdelsalam2016}
{Abd El-Salam}, F.; {Abd El-Bar}, S.
\newblock Families of frozen orbits of lunar artificial satellites.
\newblock {\em Appl. Math. Model.} {\bf 2016}, {\em
  40},~9739--9753.
\newblock
  https://doi.org/{\changeurlcolor{black}\href{https://doi.org/https://doi.org/10.1016/j.apm.2016.06.036}{\detokenize{10.1016/j.apm.2016.06.036}}}.

\bibitem[Carvalho \em{et~al.}(2018)Carvalho, Cardoso~dos Santos, Prado, and
  Moraes]{Carvalho2018}
Carvalho, J.; Cardoso~dos Santos, J.; Prado, A.; Moraes, R.
\newblock Some characteristics of orbits for a spacecraft around Mercury.
\newblock {\em Comput. Appl. Math.} {\bf 2018}, {\em
  37},~267–281.
\newblock
  https://doi.org/{\changeurlcolor{black}\href{https://doi.org/10.1007/s40314-017-0525-y}{\detokenize{10.1007/s40314-017-0525-y}}}.

\bibitem[{De Saedeleer} and Henrard(2006)]{Saedeleer2006}
{De Saedeleer}, B.; Henrard, J.
\newblock The combined effect of J2 and C22 on the critical inclination of a
  lunar orbiter.
\newblock {\em Adv. Space Res.} {\bf 2006}, {\em 37},~80--87.

\bibitem[{Tzirti} \em{et~al.}(2009){Tzirti}, {Tsiganis}, and
  {Varvoglis}]{Tzirt2009}
{Tzirti}, S.; {Tsiganis}, K.; {Varvoglis}, H.
\newblock {Quasi-critical orbits for artificial lunar satellites}.
\newblock {\em Celest. Mech. Dyn. Astron.} {\bf 2009}, {\em
  104},~227--239.
\newblock
  https://doi.org/{\changeurlcolor{black}\href{https://doi.org/10.1007/s10569-009-9207-4}{\detokenize{10.1007/s10569-009-9207-4}}}.

\bibitem[{Kozai}(1962)]{Kozai1962}
{Kozai}, Y.
\newblock {Secular perturbations of asteroids with high inclination and
  eccentricity}.
\newblock {\em  Astron. J.} {\bf 1962}, {\em 67},~591--598.
\newblock
  https://doi.org/{\changeurlcolor{black}\href{https://doi.org/10.1086/108790}{\detokenize{10.1086/108790}}}.

\bibitem[Lidov(1962)]{Lidov1962}
Lidov, M.
\newblock The evolution of orbits of artificial satellites of planets under the
  action of gravitational perturbations of external bodies.
\newblock {\em Planet. Space Sci.} {\bf 1962}, {\em 9},~719--759.
\newblock
  https://doi.org/{\changeurlcolor{black}\href{https://doi.org/https://doi.org/10.1016/0032-0633(62)90129-0}{\detokenize{10.1016/0032-0633(62)90129-0}}}.

\bibitem[Naoz(2016)]{Naoz2016}
Naoz, S.
\newblock The Eccentric Kozai-Lidov Effect and Its Applications.
\newblock {\em Annu. Rev. Astron. Astrophys.} {\bf 2016}, {\em
  54},~441--489.
\newblock
  https://doi.org/{\changeurlcolor{black}\href{https://doi.org/10.1146/annurev-astro-081915-023315}{\detokenize{10.1146/annurev-astro-081915-023315}}}.

\bibitem[Naoz \em{et~al.}(2017)Naoz, Li, Zanardi, de~El{\'{\i}}a, and
  Sisto]{Naoz_2017}
Naoz, S.; Li, G.; Zanardi, M.; de~El{\'{\i}}a, G.C.; Sisto, R.P.D.
\newblock The Eccentric Kozai{\textendash}Lidov Mechanism for Outer Test
  Particle.
\newblock {\em  Astron. J.} {\bf 2017}, {\em 154},~18.
\newblock
  https://doi.org/{\changeurlcolor{black}\href{https://doi.org/10.3847/1538-3881/aa6fb0}{\detokenize{10.3847/1538-3881/aa6fb0}}}.

\bibitem[{Tzirti} \em{et~al.}(2010){Tzirti}, {Tsiganis}, and
  {Varvoglis}]{Tzirt2010}
{Tzirti}, S.; {Tsiganis}, K.; {Varvoglis}, H.
\newblock {Effect of 3rd-degree gravity harmonics and Earth perturbations on
  lunar artificial satellite orbits}.
\newblock {\em Celest. Mech. Dyn. Astron.} {\bf 2010}, {\em
  108},~389--404.
\newblock
  https://doi.org/{\changeurlcolor{black}\href{https://doi.org/10.1007/s10569-010-9313-3}{\detokenize{10.1007/s10569-010-9313-3}}}.

\bibitem[{Paskowitz} and {Scheeres}(2006)]{Scheeres2006}
{Paskowitz}, M.E.; {Scheeres}, D.J.
\newblock {Design of Science Orbits About Planetary Satellites: Application to
  Europa}.
\newblock {\em J. Guid. Control Dyn.} {\bf 2006}, {\em
  29},~1147--1158.
\newblock
  https://doi.org/{\changeurlcolor{black}\href{https://doi.org/10.2514/1.19464}{\detokenize{10.2514/1.19464}}}.

\bibitem[Lara and Russell(2006)]{Lara2006}
Lara, M.; Russell, R.
\newblock On the design of a science orbit about Europa.
\newblock {\em Adv. Astronaut. Sci.} {\bf 2006}, {\em 124}. 


\bibitem[Carvalho \em{et~al.}(2010)Carvalho, Vilhena~de Moraes, and
  Prado]{Carvalho2010}
Carvalho, J.; Vilhena~de Moraes, R.; Prado, A.
\newblock Some orbital characteristics of lunar artificial satellites.
\newblock {\em Celest. Mech. Dyn. Astron.} {\bf 2010}, {\em
  108},~371--388.
\newblock
  https://doi.org/{\changeurlcolor{black}\href{https://doi.org/10.1007/s10569-010-9310-6}{\detokenize{10.1007/s10569-010-9310-6}}}.

\bibitem[{Chambers}(1999)]{Chambers1999}
{Chambers}, J.E.
\newblock {A hybrid symplectic integrator that permits close encounters between
  massive bodies}.
\newblock {\em Mon. Not. R. Astron. Soc.} {\bf 1999}, {\em 304},~793--799.
\newblock
  https://doi.org/{\changeurlcolor{black}\href{https://doi.org/10.1046/j.1365-8711.1999.02379.x}{\detokenize{10.1046/j.1365-8711.1999.02379.x}}}.

\bibitem[{French} and {Showalter}(2012)]{French2012}
{French}, R.S.; {Showalter}, M.R.
\newblock {Cupid is doomed: An analysis of the stability of the inner uranian
  satellites}.
\newblock {\em Icarus} {\bf 2012}, {\em 220},~911--921. 
\newblock
  https://doi.org/{\changeurlcolor{black}\href{https://doi.org/10.1016/j.icarus.2012.06.031}{\detokenize{10.1016/j.icarus.2012.06.031}}}.

\bibitem[{Chen} \em{et~al.}(2014){Chen}, {Nimmo}, and {Glatzmaier}]{Chen2014}
{Chen}, E.M.A.; {Nimmo}, F.; {Glatzmaier}, G.A.
\newblock {Tidal heating in icy satellite oceans}.
\newblock {\em Icarus} {\bf 2014}, {\em 229},~11--30. \linebreak
\newblock
  https://doi.org/{\changeurlcolor{black}\href{https://doi.org/10.1016/j.icarus.2013.10.024}{\detokenize{10.1016/j.icarus.2013.10.024}}}.

\bibitem[{Prado}(2003)]{Prado2003}
{Prado}, A.F.
\newblock {Third-Body Perturbation in Orbits Around Natural Satellites}.
\newblock {\em J. Guid. Control Dyn.} {\bf 2003}, {\em
  26},~33--40.
\newblock
  https://doi.org/{\changeurlcolor{black}\href{https://doi.org/10.2514/2.5042}{\detokenize{10.2514/2.5042}}}.

\bibitem[Cinelli \em{et~al.}(2019)Cinelli, Ortore, and Circi]{Cinelli2019}
Cinelli, M.; Ortore, E.; Circi, C.
\newblock Long Lifetime Orbits for the Observation of Europa.
\newblock {\em J. Guid. Control Dyn.} {\bf 2019}, {\em 42}, 123--135.

\bibitem[Gupta(2011)]{Shraddha2011}
Gupta, S.
\newblock Effect of Altitude, Right Ascension of Ascending Node and Inclination
  on Lifetime of Circular Lunar Orbits.
\newblock {\em Int. J. Astron. Astrophys.} {\bf 2011},
  {\em 1}, 155--163.
\newblock
  https://doi.org/{\changeurlcolor{black}\href{https://doi.org/10.4236/ijaa.2011.13020}{\detokenize{10.4236/ijaa.2011.13020}}}.

\end{thebibliography}
\end{document}